\documentclass{aa}  

\usepackage{graphicx,epsfig,amsmath,amssymb,color,stmaryrd,multirow,makecell}
\usepackage{txfonts}
\bibpunct{(}{)}{;}{a}{}{,}
\newcommand{\sound}{{\mathrm{s}}}

\newcommand{\Ms}{{\mathrm{M}_\odot}}

\newcommand{\pc}{{\mathrm{pc}}}
\newcommand{\cc}{{\mathrm{cm^{-3}}}}
\newcommand{\gcc}{{\mathrm{g~cm^{-3}}}}

\begin{document} 

   \title{Stellar mass spectrum within massive collapsing clumps \\
  I. Influence of the initial conditions}

   \titlerunning{Stellar mass spectrum within collapsing clumps}

   \author{
          Yueh-Ning Lee \inst{1,2}
          \and
          Patrick Hennebelle\inst{1,2,3}%\fnmsep\thanks{}
          }
   \institute{IRFU, CEA, Universit\'{e} Paris-Saclay, F-91191 Gif-sur-Yvette, France\\
              \email{yueh-ning.lee@cea.fr}
         \and
                Universit\'{e} Paris Diderot, AIM, Sorbonne Paris Cit\'{e}, CEA, CNRS, F-91191 Gif-sur-Yvette, France
         \and
             LERMA (UMR CNRS 8112), Ecole Normale Sup\'{e}rieure, 75231 Paris Cedex, France\\
             \email{patrick.hennebelle@lra.ens.fr }
%             %\thanks{}
             }
   \date{Received 06 July 2017; accepted 18 October 2017}

% \abstract{}{}{}{}{} 
% 5 {} token are mandatory
 
  \abstract
   {Stars constitute the building blocks of our Universe, and their formation  
is an astrophysical problem of great importance.
   }
   {We aim to understand the fragmentation of massive molecular star-forming clumps
and the effect of initial conditions, namely the density and the level of turbulence, on the 
resulting distribution of stars. For this purpose, we conduct numerical experiments in which we
systematically vary the initial density over four orders of magnitude and the turbulent velocity over a factor ten. 
In a companion paper, we investigate the dependence of this distribution on the gas thermodynamics.}
   {We performed a series of hydrodynamical numerical simulations using adaptive mesh refinement,
     with special attention to numerical convergence. 
      We also adapted an  existing analytical 
     model to the case of collapsing clouds by employing a density probability distribution function (PDF) $\propto \rho^{-1.5}$
     instead of a lognormal distribution. 
      }
   { Simulations and analytical model both show two support regimes, 
a thermally dominated regime and a turbulence-dominated regime. 
%   a thermally dominated and a \textcolor[rgb]{1,0.501961,0}{turbulent dominated} \LEt{this is ungrammatical, do you mean "turbulence-dominated
%regime"? in analogy, it probably should be "temperature-dominated
%regime", not "thermally dominated", correct?}regime. 
For the first regime, we infer that $d N / d\log M \propto M ^ 0$, while for the second 
regime, we obtain  $d N / d\log M \propto M ^ {-3/4}$ . This is valid 
up to about ten times the mass of the first Larson core, as explained in the companion paper, leading to a peak 
of the mass spectrum at $\sim 0.2~\Ms$. 
From this point, the mass spectrum  decreases with decreasing mass 
except for the most diffuse clouds, where disk fragmentation leads to the formation 
of objects down to the mass of the first Larson core, that is,
to a few $10^{-2}~\Ms$. 
}
   {Although the mass spectra we obtain for the most compact clouds qualitatively resemble the 
observed initial mass function, the distribution exponent  is shallower
than the expected Salpeter exponent of $-1.35$. Nonetheless, we observe a possible transition 
toward a slightly steeper value that is broadly compatible with the 
Salpeter exponent for masses above a few solar masses. This change in behavior is associated 
with the change in density PDF, which switches from a power-law to
a lognormal distribution. Our results suggest that while gravitationally induced fragmentation could play an important 
role for low masses, it is likely the turbulently induced fragmentation that leads to the Salpeter exponent.
      }


   \maketitle

%________________________________________________________________________________

%------------------INTRODUCTION------------------------
\section{Introduction}
The formation of stars inside a cluster depends on local as well as on global conditions. 
Since the majority of stars forms inside clusters \citep[e.g.,][]{Lada03}, 
clarifying these effects will lead to an advanced understanding of the star formation processes. 
It is commonly accepted that the high-mass end of the initial mass function (IMF) follows some power law $dN/d\log M \propto M^{\Gamma}$, 
where \citet{Salpeter55} derived that $\Gamma \simeq -1.35$ with the field star mass spectrum. 
The shape of the IMF around  and below solar masses is quite different. 
While \citet{kroupa2001} suggested a power-law behavior 
with lower values of $\Gamma$ at lower masses \citep[see also][]{hillenbrand2004}, \citet{chabrier2003} proposed a lognormal 
distribution that peaks at about 0.2-0.3 $\Ms$.  
\citet{Bastian10} compiled IMFs observed in clusters and reported strong variations of $\Gamma$ between -0.4 and -2 \citep[see also, e.g.,][]{moraux2007}, 
and this value is, furthermore, sensitive to the mass range considered. 

Many numerical studies have been dedicated to the understanding of the IMF. 
An important difficulty arises from the need 
to resolve sufficiently small spatial scales while
at the same time considering relatively large spatial scales that are necessary for adequate statistics. 
Earlier cluster formation simulations \citep[e.g.,][]{Bate03, Bate05a, Bate05b, Clark08,Offner08, Offner09}, although valuable in the pioneering illustration of some IMF physics, 
are limited in either statistics or resolution and practically do not produce well-defined IMFs. 
More recently, thanks to the increase in computing power, larger simulations that provided more reliable statistics have been performed. 
\citet{Bonnell03} simulated an isothermal $1000~\Ms$ cloud of 1 pc diameter, with Mach $\sim 10$, 
and produced an IMF that was broadly consistent with $\Gamma \sim -1$.
\citet{Bate09a,Bate12} simulated $500~\Ms$ clouds of 0.404 radius with Mach number $\mathcal{M}=13.7$, with either a polytropic equation of state (eos) or radiation hydrodynamics. 
The mass spectra  they presented  have $\Gamma$ slightly higher than -1 in the mass range $0.1-3~\Ms$ in both cases. 
\citet{Maschberger10,Maschberger14} typically found $\Gamma \gtrsim -1$ in sub-clusters of a $10^4~\Ms$ simulation inside a cylinder of 10 pc length and 3 pc diameter \citep{Bonnell08,Bonnell11}. 
These simulations used a piecewise polytropic eos, with the gas being isothermal at density $5.5 \times 10^{-19}-5.5 \times 10^{-15}~\gcc$.
\citet{Krumholz11} simulated $1000~\Ms$ clouds at $2.4 \times 10^5~\cc$ and studied the effect of radiation hydrodynamics compared to the isothermal condition. 
Their mass spectra also showed $\Gamma > -1$ in the mass range $0.1- 2~\Ms$, either with or without radiation. 
\citet{Girichidis11} started with various density profiles for $100~\Ms$ molecular clouds at a density $4.6 \times 10^5~\cc$, with $\mathcal{M}\sim 3.5$, 
and generally found $\Gamma \sim -1$. 
When a density profile $\rho \propto r^{-2}$ is employed, an initial Mach number $\mathcal{M} <13$ leads to the formation of a single star. 
\citet{BallesterosParedes15} simulated a piece of isothermal molecular clouds of $1000~\Ms$ in a $1~\pc$ box for several Mach numbers and obtained an IMF compatible with $\Gamma \sim -1$, whose shape is established quite early. 
When we examine the mass spectra of these studies in detail, 
statistical fluctuations remain, and it is not obvious how the value of $\Gamma$ is determined. 
It is common to see slopes that are shallower at the intermediate-mass range than the canonical -1.35 inferred by \citet{Salpeter55}.

This work aims to examine the stellar mass spectrum shape as a result of the initial conditions by covering a wide range of initial parameters. 
We perform a series of numerical simulations of cluster formation inside molecular clouds of $1000~\Ms$ with varying initial conditions, namely density 
and turbulence level. 
To overcome the above-mentioned computational obstacles, 
a compromise is made between the need for high resolution and the need for sufficient statistics to obtain 
meaningful mass spectra.  
The first Larson core scale (a few AU) is resolved to ensure
a good description of the gravitational fragmentation.
Sink particles are employed as a subgrid modeling of stars. 
The choice of the cloud mass ($1000~\Ms$) allows us to produce a statistically significant cluster. 
With this simple setup, we show that the stellar distribution is environment dependent and is universal only under certain circumstances. 

The molecular gas was initially confined in a sphere of radius $0.042 - 1 ~\pc$, 
with the corresponding number density between $10^3~\cc$ and $10^7~\cc$. 
While noting that this configuration is rather dense compared to general molecular clouds, 
the very wide density range allows us to illustrate the influence of the initial density.
To isolate the effect of initial density, we intentionally left out  magnetic field, cooling, 
radiative transfer, and all stellar feedback effects, while representing the thermodynamics of the gas with a simple smoothed two-slope polytropic eos. 
These numerical experiments, therefore, are not intended to represent fully realistic molecular clouds.

In the second section, we present the numerical setup and the initial conditions. 
The results are qualitatively described in the third section. 
The fourth section is devoted to the stellar mass spectra inferred from the various simulations, 
while in the fifth section, the model of 
\citet{HC08} is adapted to the case of a collapsing cloud, and we compare theory to simulation
results. 
The sixth section discusses the results, and the seventh concludes the paper. 

%------------------NUMERICS------------------------
\section{Simulations}

\subsection{Initial conditions}
The simulation box was initialized with a spherical molecular cloud of mass $M=1000 ~\Ms$ and a 
 density profile $\rho(r) = \rho_0/\left[1+\left(r/r_0\right)^2\right]$, where $r$ is the distance to the cloud center, and
 $\rho_0$ and $r_0$ are the density and size of the central plateau, respectively.
The density contrast between the cloud center and edge was a factor
of ten, 
and the radius consequently was $3r_0$.
The simulation box was twice the size of the cloud, 
and the surrounding space was patched with diffuse medium of density $\rho_0/100$. 
The turbulence was initialized from a Kolmogorov spectrum with random phases. 
The seed was the same for all simulations, 
while the amplitude was scaled to match the assigned turbulent energy level and box size. 
The temperature, $T$, is given by a smoothed two-slope polytropic eos, 
such that the gas was at 10 K at low density and followed the dependence $T \propto \rho^{2/3}$ at a number density higher than $10^{10}~\cc$. 

Two effects of the cloud initial conditions were studied: the density, and the turbulence level.
We performed a series of simulations by varying the compactness of the cloud, 
defined by the ratio between the free-fall time and the sound-crossing time $t_\mathrm{ff}/t_\mathrm{sc} = 0.15,~0.1,~0.05,~\text{and}~0.03$. 
The four setups increased in density, while the virial parameter remained the same. 
This was done by fixing the ratio between the free-fall time and the turbulence-crossing
%\textcolor[rgb]{1,0.501961,0}{turbulent-crossing} \LEt{again,
%please check for correct nomenclature here, is this "turbulence-crossing
%time"? please also check throughout for correctness}
time $t_\mathrm{ff}/t_\mathrm{tc} = 1.1$, 
which is close to virial equilibrium. 
On the other hand, we also set $t_\mathrm{ff}/t_\mathrm{tc} = 1.5,~0.5,~0.3,~\text{and}~0.1,$ while keeping $t_\mathrm{ff}/t_\mathrm{sc} = 0.05$. 
The level of turbulence is important since  turbulence provides dynamical support against self-gravity and creates local overdensities through shocks. 
The simulation parameters are listed in Table~\ref{table_compactness}. 
  The number density $n=\rho/ (\mu m_{\rm p})$, where $\mu=2.33$ is the mean molecular weight and $m_{\rm p}$ is the atomic hydrogen mass, is presented in the table instead of the volumetric density $\rho$. 

\begin{table*}[t!]
\caption{Simulation parameters. The cloud mass is $1000 ~\Ms$. The size of the cloud decreases through runs A, B, C, and D, and the density increases correspondingly. The number 1 in the label denotes that the relaxation run was performed at the first refinement level. The plus and minus signs denote higher/lower maximum refinement level with respect to the canonical run. We list in the columns the run label, the ratio between free-fall time and turbulence-crossing time, the ratio between free-fall time and sound-crossing time, the initial cloud radius, the central density, the  density PDF peak (as illustrated in Fig.~\ref{fig_pdf}) after relaxation, the initial Mach number before relaxation, the base grid refinement level, the maximum refinement level, the physical resolution, and the relaxation duration.}
\label{table_compactness}
\centering
\begin{tabular}{l l l c c c c c c c c }
\hline\hline
Label   & $t_\mathrm{ff}/t_\mathrm{tc}$  & $t_\mathrm{ff}/t_\mathrm{sc}$   & Radius (pc)& $n_0 (\mathrm{cm}^{-3})$ & $n_{\rm pdf peak} (\mathrm{cm}^{-3})$ & Mach & $l_\mathrm{min}$ &  $l_\mathrm{max}$ & resolution (AU) \vspace{.5mm} & relaxation (kyr)\\
\hline
A1    & 1.1 & 0.15   &  0.75   &  8.2 $\times10^4$ & $\simeq 2 \times 10^2$   &7&  8  & 14& 38 & 88\\%& -- &\\
A1+  &      &           &           &              &                    &&  8  & 15& 19 & \\%& -- &\\      
A1++  &      &           &           &              &                    &&  8  & 16& 9 & \\%& -- &\\      
\hline
B1    & 1.1 & 0.10   &  0.33   &  9.4 $\times10^5$ &  $\simeq 2 \times 10^3$  &11&  8  & 14 & 17 & 27\\%& -- &\\
B1+  &       &            &            &             &                   &&  8  & 15& 8  &  \\%& -- &\\
B1++  &       &            &            &             &                   &&  8  & 16& 4  &  \\%& -- &\\
\hline
C1    & 1.1 & 0.05    &  0.084 &  6.0 $\times10^7$ & $\simeq 10^5$ &22&  8  & 14&4 & 3 \\%& -- &\\
C1+     &       &           &          &               &                & & 8  & 15 & 2 &   \\%& -- &\\
C1--     &       &            &          &              &                &&  8  & 13 & 8 &    \\%& -- &\\
C1-- --  &       &            &          &              &              & & 8  & 12& 17  &   \\%& -- &\\
C1t15   & 1.5 &         &          &                    &        &30&  8  & 14  &4  & 2.2 \\%& -- &\\
C1t05   & 0.5 &           &          &                  &         &10&  8  & 14 &4  & 6.6 \\%& -- &\\
C1t03   & 0.3 &           &          &                  &          &6&  8  & 14&4  & 11 \\%& -- &\\
C1t01   & 0.1 &          &          &                   &         &2& 8  & 14 &4  & 13 \\%& -- &\\
\hline
D1    & 1.1 & 0.03   &  0.042 &  4.8 $\times10^8$ & $\simeq 8 \times 10^5$  &50&  7  & 13 & 4 & 1.3 \\% & -- &\\
\hline
\end{tabular}
\end{table*}

Since the smooth density profile and the seeded turbulent field are not fully self-consistent, 
we ran a relaxation phase before the actual simulation. 
For each run, the system was evolved without self-gravity during $0.3 ~ t_\mathrm{tc}$ at the lowest level of refinement ($2^8$) to prepare a more realistic density field with local fluctuations. 
The relaxation effects are discussed in Appendix \ref{appen_r}.

\subsection{Numerical setup}
The simulations were run with the adaptive mesh refinement (AMR) magnetohydrodynamics (MHD) code RAMSES \citep{Teyssier02,Fromang06}. 
The system was evolved following ideal MHD equations. 
Although these simulations were not magnetized, the MHD solver was employed to avoid numerical uncertainties when the magnetic field might be included in follow-up studies. 
All simulations were run on a base grid of $2^8$, corresponding to different physical resolutions for different density cases, except for the densest case D, which was run on $2^7$. 
After we switched on gravity, the refinement required that the local Jeans length always be resolved by ten cells. 
Canonical runs were refined to level 14,  
and we varied the resolution to check for numerical convergence over up to 4 AMR levels (see Table~\ref{table_compactness}). 
The lowest resolution in all runs was 38 AU, and the best resolution
reached 2 AU. 

\subsection{Sink particles}
Sink particles were used in our simulations as a subgrid model of stars \citep{Krumholz04,federrath2010}. 
The algorithm we used was developed by \citet{Bleuler14}, and it proceeds as follows: 
first, density peaks are identified. 
Over a given density threshold ($n_{\rm sink} = 10^{10}~\cc$), a sink particle is placed when the local gas properties satisfy the criteria of maximum refinement level, 
local virial boundedness, and flow convergence. 
The sink particle then interacts with the gas through gravity and continues to accrete mass from its surrounding. 
 A sink accretes 75 percent of the mass that exceeds $n_{\rm sink}$ from surrounding cells with the accretion radius, $r_\mathrm{acc}$, that is, four times the smallest cell size. 
 The angular momentum of the accreted gas is added to the sink. The effect of angular momentum in regulating mass accretion or fragmentation may 
 modify the resulting sink particle mass spectrum in the presence of a magnetic field, and it is currently ignored in the this simplified study as the magnetic field is not considered, and even if existed, it is thought to have no strong influence on the core at such small scales.  
The pre-stellar core (a few tens of AU) is resolved in the simulations, and this guarantees that the sink particles represent individual stars. 

\subsection{Missing physics}\label{missing}
We performed an ensemble of numerical experiments to quantify the influence of initial conditions
and spatial resolution, and the physics used here is greatly simplified.
Many other processes not included here are nonetheless important, 
and may even be dominant in some circumstances. For example, the role of the accretion luminosity 
that emanates from the protostars has been stressed by \citet{krumholz07}, \citet{Bate09b}, and \citet{commercon2011}. 
Accretion luminosity provides a substantial heating source to the gas, which leads 
to greater thermal support and may reduce or even suppress the fragmentation, that is, the formation of several
 objects. The magnetic field also likely plays 
an important role  \citep{hennebelle2011,peters2011,myers2013}. In particular, due to magnetic braking and magnetic support, it tends to reduce the fragmentation within clusters.  
This discussion may become even more crucial in the context of massive disks, which, as we show, form in some of the 
simulations presented below. A magnetic field may drastically modify the picture there 
\citep{hennebelle2008,machida2008}, even when non-ideal MHD processes are included \citep{masson2016,hennebelle2016,wurster2017}.
The same is true for the accretion luminosity, which can also efficiently heat the disk and stabilize it 
against fragmentation \citep{Offner09,commercon2010}, 
although \citet{stamatellos2012} argued that intermittent accretion may 
limit the effect of radiative heating.

\section{Overall evolution}

\subsection{Qualitative description}

\subsubsection{Large scales}
\setlength{\unitlength}{1cm}
\begin{figure*}[]
\begin{picture} (0,9.2)
\put(0,0){\includegraphics[width=6.5cm]{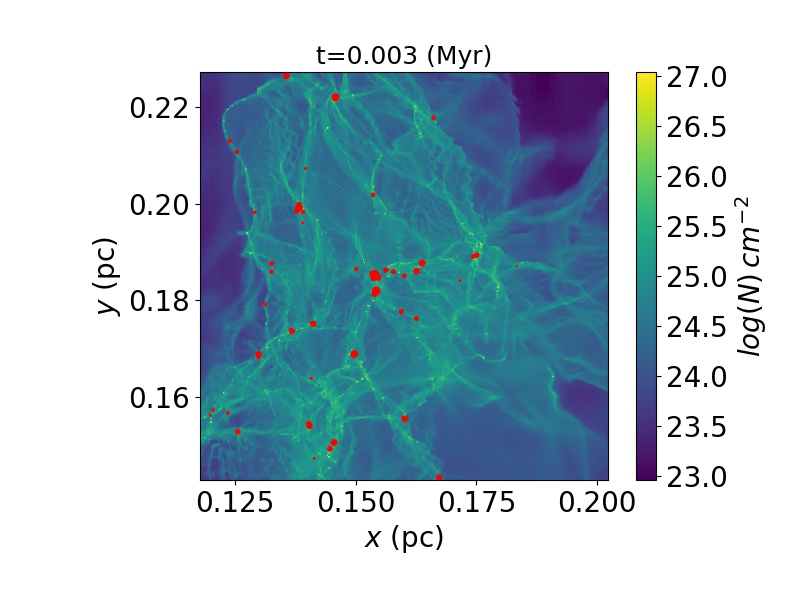}}
\put(6,0){\includegraphics[width=6.5cm]{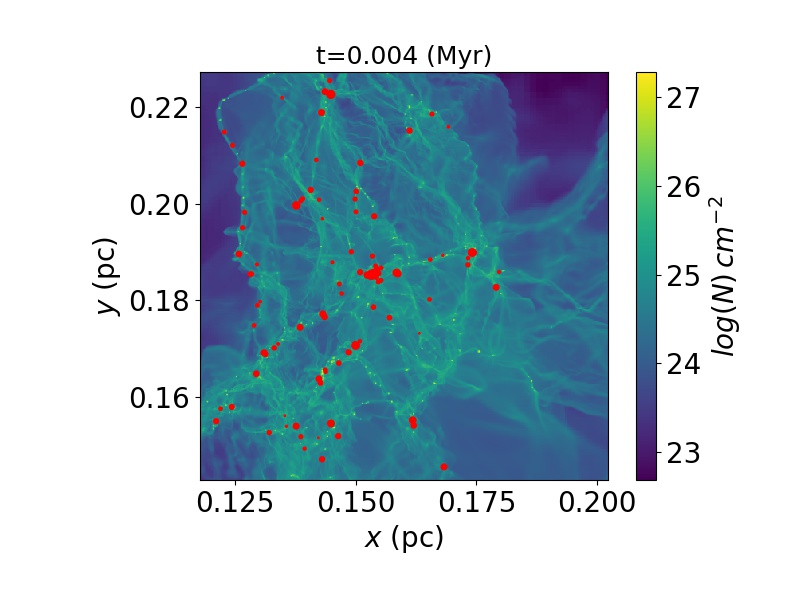}}
\put(12,0){\includegraphics[width=6.5cm]{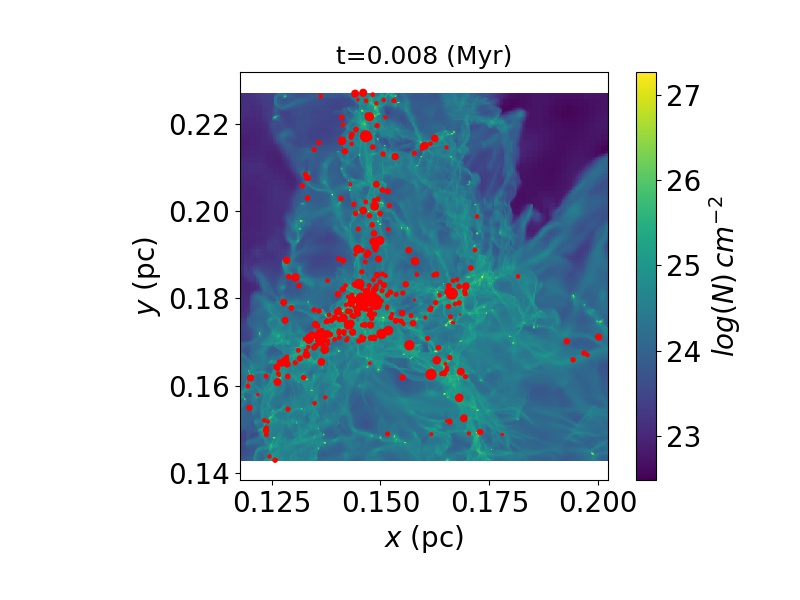}}
\put(0,4.6){\includegraphics[width=6.5cm]{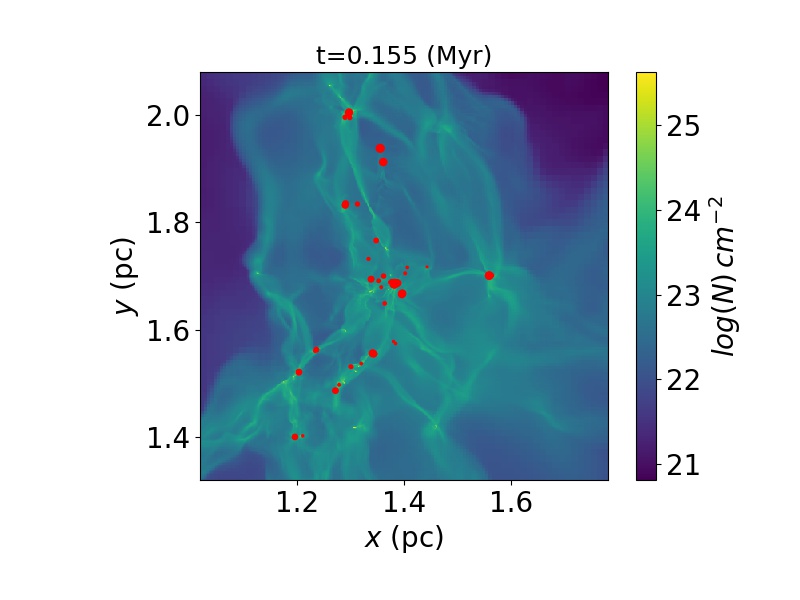}}
\put(6,4.6){\includegraphics[width=6.5cm]{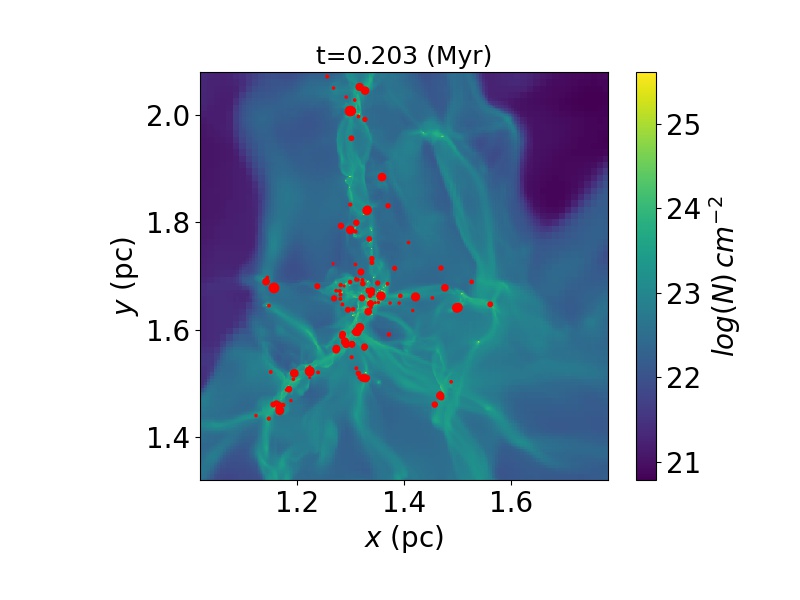}}
\put(12,4.6){\includegraphics[width=6.5cm]{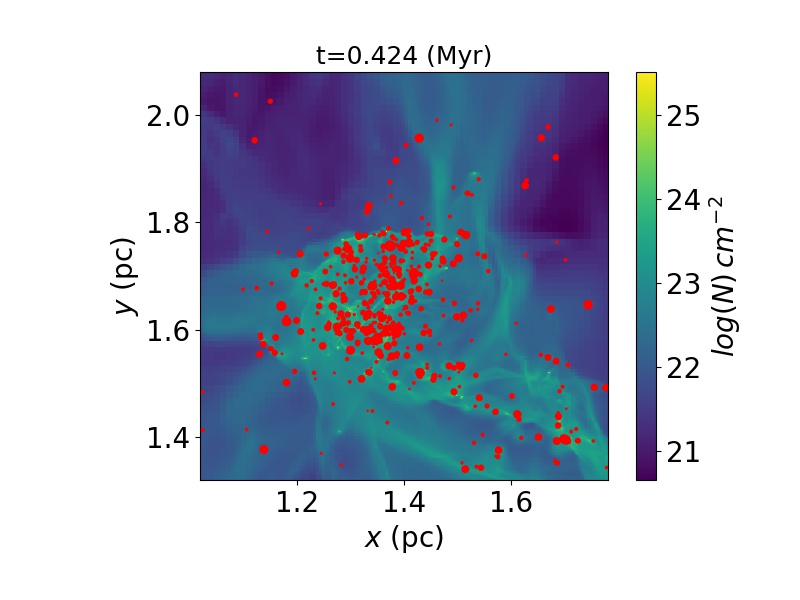}}
\end{picture}
\caption{Column density maps of runs A1++ at 0.155, 0.203, 0.424 Myr (\emph{top row from left to right}) and C1+ at 0.003, 0.004, 0.008 Myr (\emph{bottom row from left to right}). Run A1++ is more globally collapsing, and the sinks, initially forming inside the filamentary network, are more concentrated in a central cluster. Run C1+ is initially much denser and the evolution of the filamentary structure is more pronounced, with more widespread sink formation.}
\label{fig_coldens}
\end{figure*}

Figure~\ref{fig_coldens} shows the column density along the z-axis for three snapshots
of simulations A1++ (top panels) and C1+ (bottom panels).
The first snapshot corresponds to an early time where less than 20 $\Ms$ of gas 
has been accreted onto sink particles, while for the later times, 200-300 
 $\Ms$ of gas has been accreted. 

At time 0.155 Myr, a network of dense filaments is already present in run A1++. 
The sink particles, represented by red dots with sizes proportional to the particle masses, 
are located within these filaments and sometimes even at the intersection between filaments. 
This trend becomes clearer at 0.203 Myr, 
while the correlation appears less tight at 0.424 Myr. This is because the
sink particles have accreted a substantial fraction of the initial cloud (about one third of the cloud 
mass) 
and have also undergone N-body interactions.  

At time 0.003 Myr, run C1+ displays significantly more filaments than run A1++ because the 
thermal support is lower in run C1+. 
The sink particles are also located inside filaments, and they tend to be more broadly distributed as a result of the widespread filament formation. 
This is even more obvious at time 0.004 Myr. 

\subsubsection{Small scales}

\setlength{\unitlength}{1cm}
\begin{figure*}[]
\begin{picture} (0,9.2)
\put(0,0){\includegraphics[width=6.5cm]{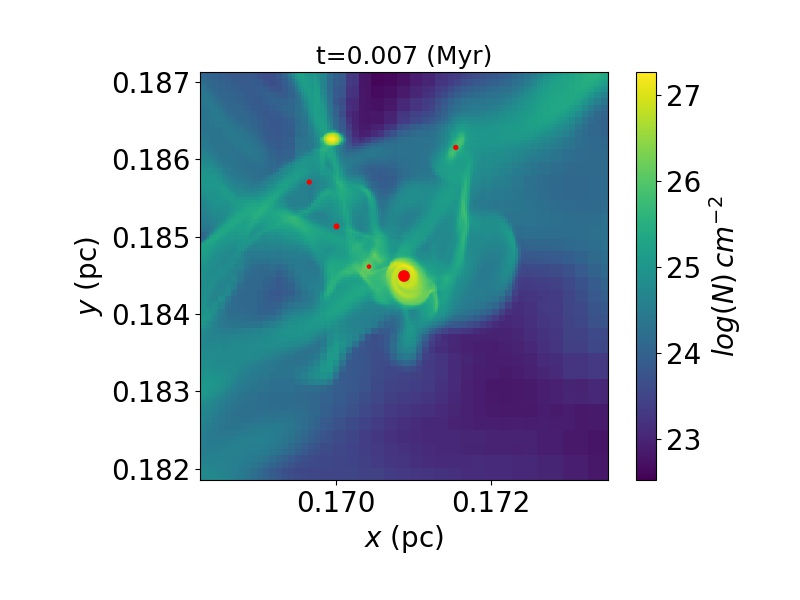}}
\put(6,0){\includegraphics[width=6.5cm]{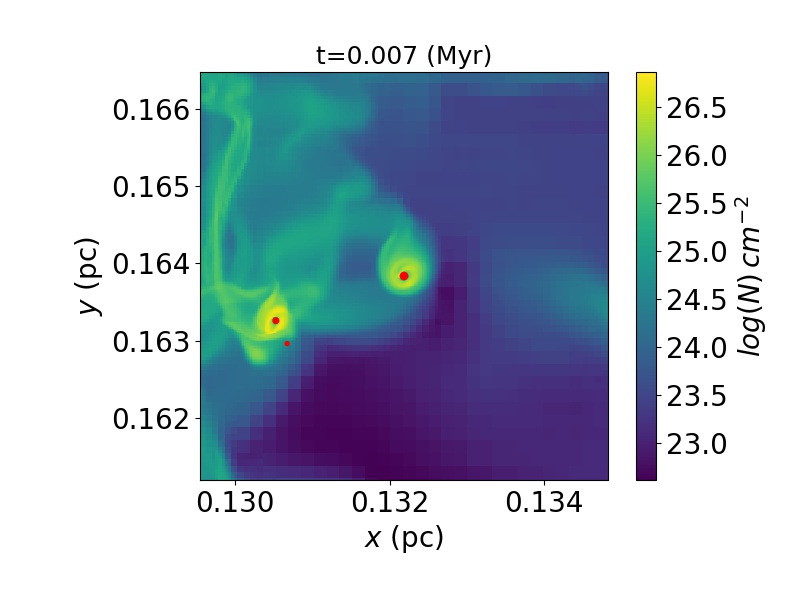}}
\put(12,0){\includegraphics[width=6.5cm]{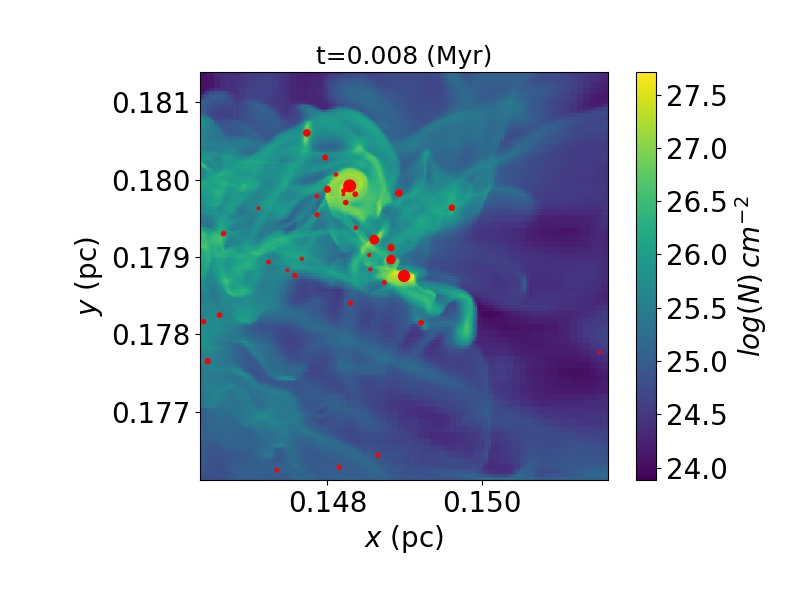}}
\put(0,4.6){\includegraphics[width=6.5cm]{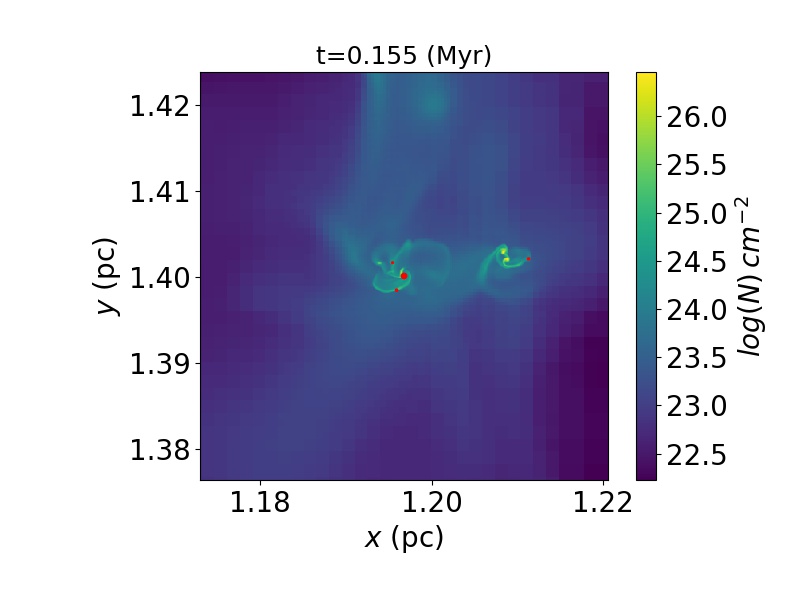}}
\put(6,4.6){\includegraphics[width=6.5cm]{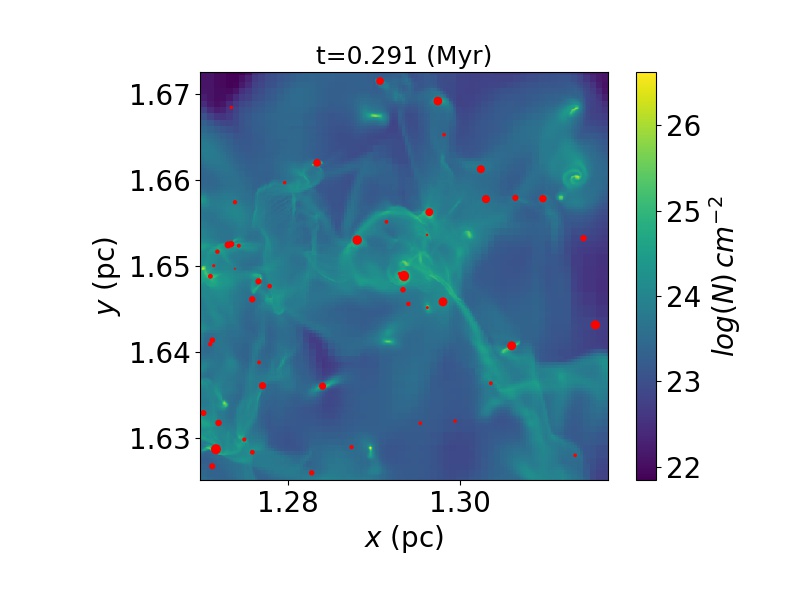}}
\put(12,4.6){\includegraphics[width=6.5cm]{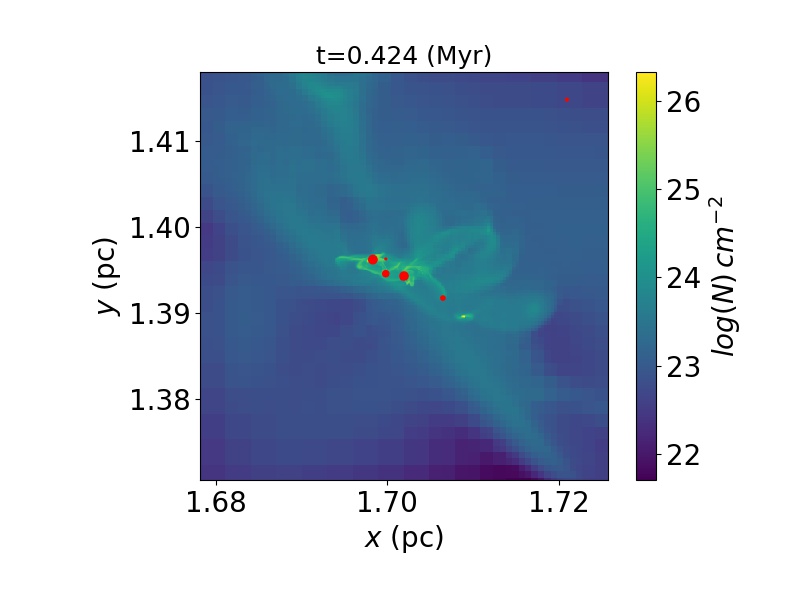}}
\end{picture}
\caption{Zoomed column density maps of runs A1++ (\emph{top row}) and C1+ (\emph{bottom row}). Red dots represent the sink particles, and their sizes correspond to the sink mass. Disk formation around sink particles is observed in both cases, while the disks in run A1++ are more extended, showing likely signs of rotation and fragmentation, and the disks in run C1+ are significantly smaller, with no sign of fragmentation.}
\label{fig_zoom_coldens}
\end{figure*}

Figure~\ref{fig_zoom_coldens}
shows three zooms for simulations A1++ (top panels) and C1+ (bottom panels) at various times. 

The zooms corresponding to simulation A++ reveal two massive fragmenting disks at time 0.155 Myr (top left panel, located at $x \simeq 1.195$, $x \simeq 1.21$ and $y \simeq 1.4~\pc$) seen face-on and 
one disk at time 0.424 Myr (top right panel, located at the center) seen edge-on. Their typical size is a 
few hundreds of AU, and they likely exhibit a rotating 
pattern to which several sinks are associated. 
In the two cases, the disks are relatively isolated without other objects in their immediate vicinity (see also Fig.~\ref{fig_coldens}). 
On the other hand, the zoom at time 0.291 Myr (top middle panel) shows the presence of a many sink particles, 
with no sign of rotating and fragmenting structures. Small disks can be seen around some 
sink particles. The trend observed here is typical of the situation in run A. 
Massive fragmenting disks are relatively common in regions that are not too crowded, but rare, if not absent, in regions where numerous sink particles are found. 
This suggests that in run A1++, disk fragmentation is a significant mode that leads to the formation of new objects
(however, see Sect.~\ref{missing}).

The zooms corresponding to run C1+ (bottom panels) also show disks 
(for example, three disks are located at the centers of each of the three panels). 
These disks, however, show no signs of fragmentation, even though the left and middle panels are relatively isolated. This trend appears to be generic for run C. 
Disk fragmentation does not seem to be an important mode. The reason is that
disks are much smaller, with a typical size on the order of a few tens of AU.  The absence of large disks
is most probably due to the high number density of sink particles, which produces dynamical interactions
that lead to efficient transport of angular momentum. Disk truncation by stellar encounter 
has indeed been recognized as an efficient process \citep{clarke1993,breslau2014,jilkova2016}.

\subsection{Density PDF}
Figure~\ref{fig_pdf} shows the probability distribution function (PDF) of the density field
 at four time steps and for six simulations, corresponding 
to four initial densities and three initial turbulence levels. The PDFs are 
very similar to one another. They present a peak that is roughly  $n_0/400$, with $n_0$ being the central density, 
and then a power law at high densities with a slope of about -1.5. 
In Table~\ref{table_compactness}, we list the position of the density PDF peak, $n_{\rm pdf peak}$, 
because it may be more representative than the initial peak density, $n_0$, 
in particular, because we let the cloud relax before running the simulation with self-gravity. 
For most cases, the PDF is almost invariant in time. 
A slight decrease in turbulence level does not make much difference (C1t03), while with significantly lowered turbulence (C1t01), 
the global collapse shifts the PDF toward high densities in time. 
This PDF shape is indeed expected and has been inferred in previous calculations \citep{kritsuk11,HF12,girichidis2014}.  
The $n^{-1.5}$ power law stems from the $n \propto r^{-2}$ density profile that develops during the collapse \citep[e.g.,][]{shu77}. At very high density, that is to say, 
$n > 10^{10} ~\cc$, the PDF becomes shallower. This is a consequence of the eos, which
 becomes adiabatic. At low density, that is, below the mean density, we see a behavior that 
broadly looks lognormal with a plateau near the peak and then a stiff decrease. 

\setlength{\unitlength}{1cm}
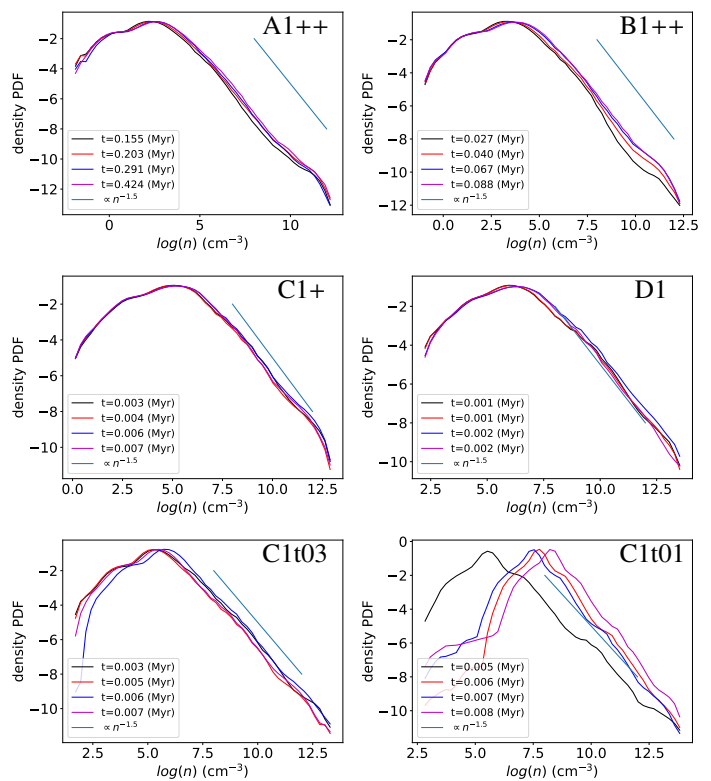
\begin{figure}[]
\begin{picture} (0,10.5)
\put(-0.1,0){\includegraphics[trim=20 20 20 20,clip,width=4.9cm]{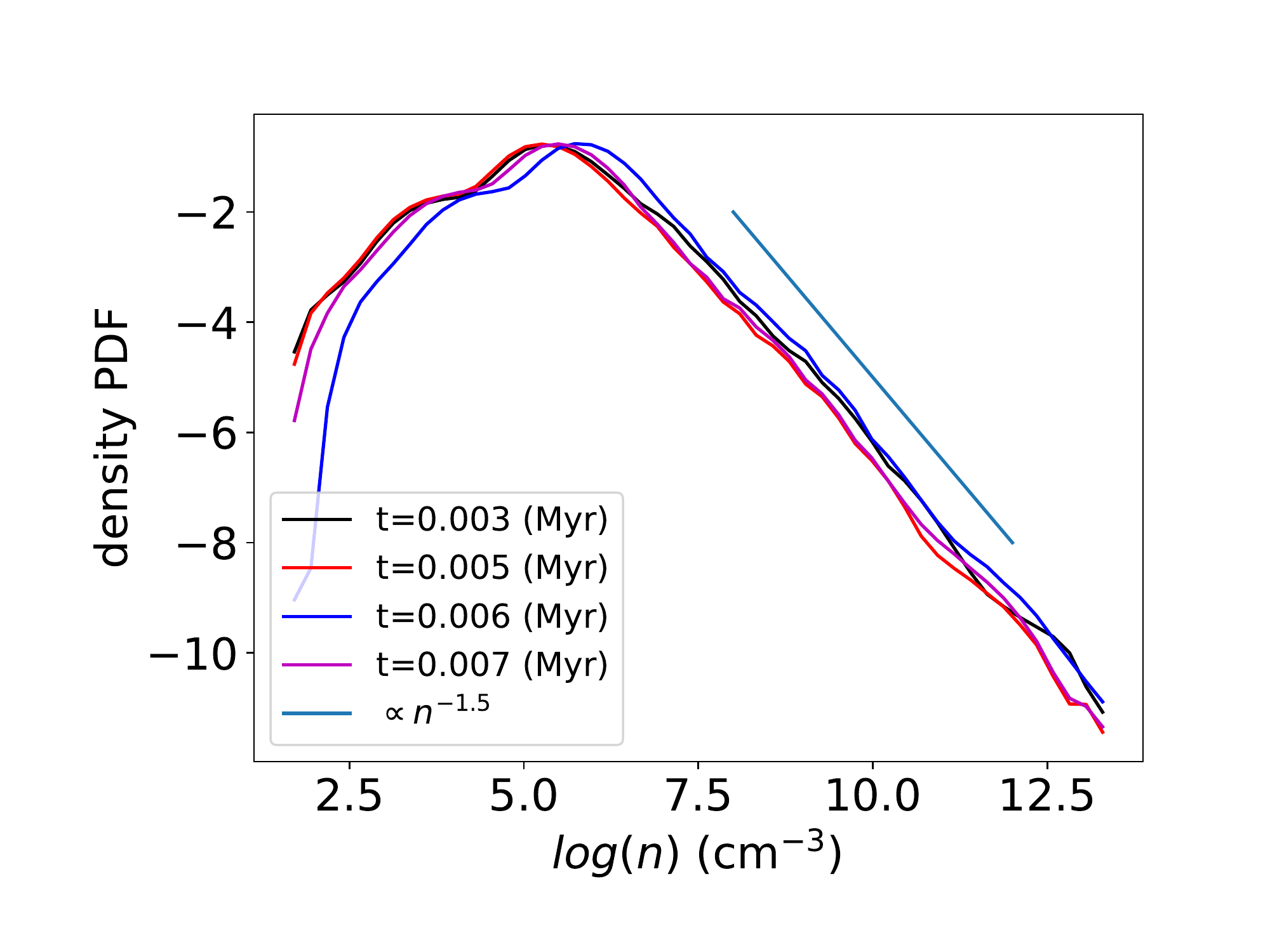}}
\put(3.4,3){C1t03}
\put(4.5,0){\includegraphics[trim=20 20 20 20,clip,width=4.9cm]{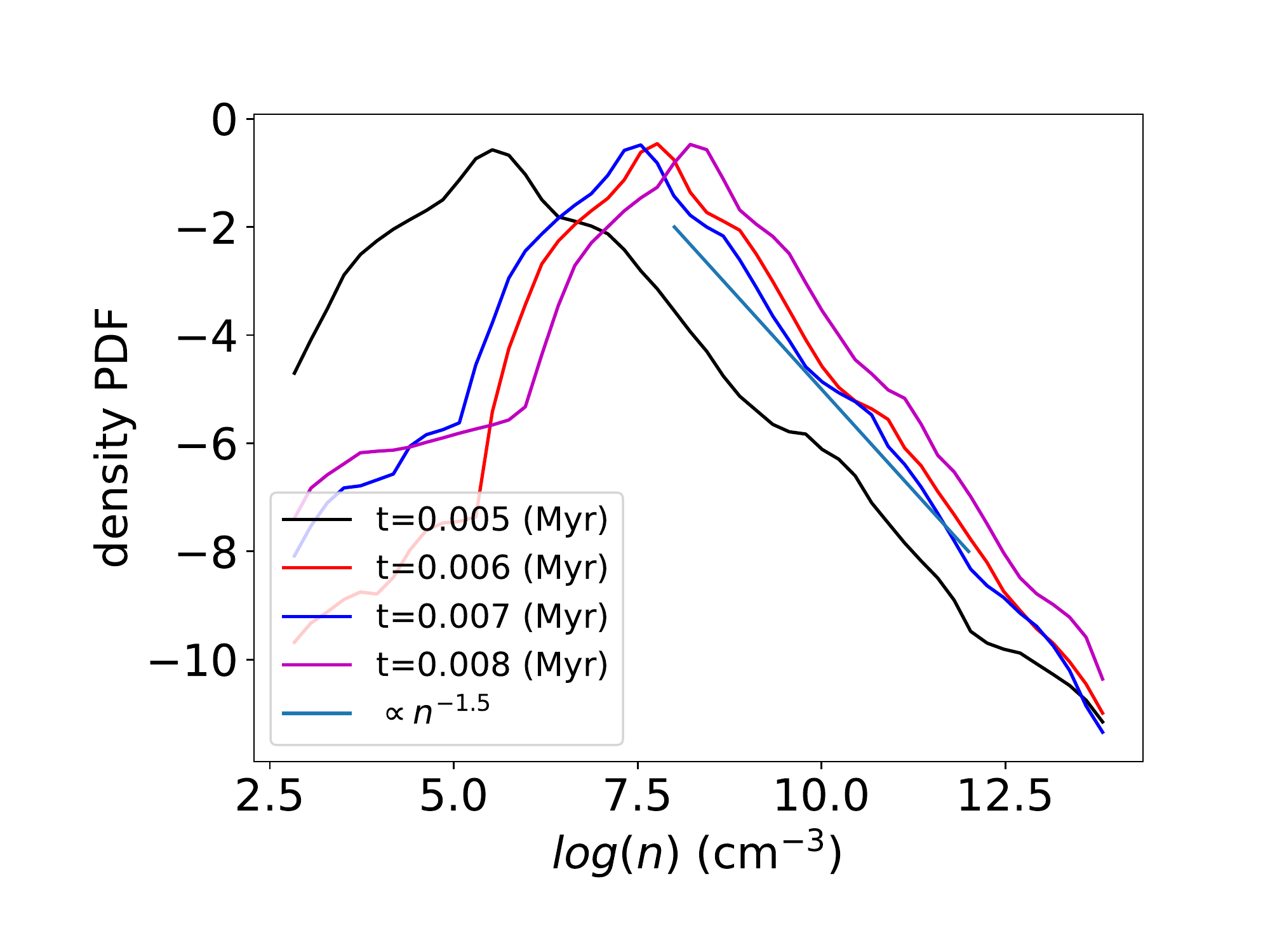}}
\put(8.1,3){C1t01}
\put(-0.1,3.5){\includegraphics[trim=20 20 20 20,clip,width=4.9cm]{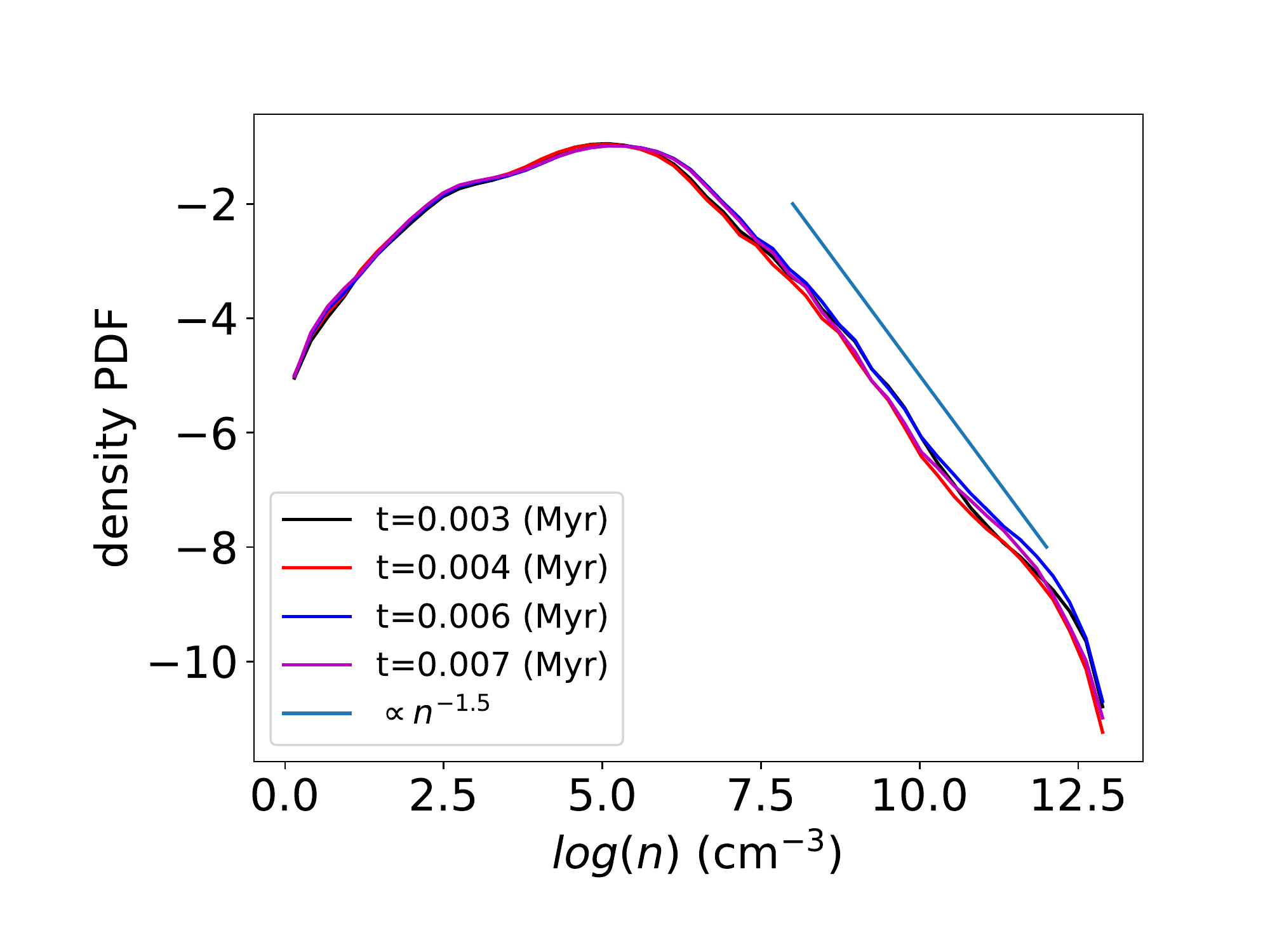}}
\put(3.6,6.5){C1+}
\put(4.5,3.5){\includegraphics[trim=20 20 20 20,clip,width=4.9cm]{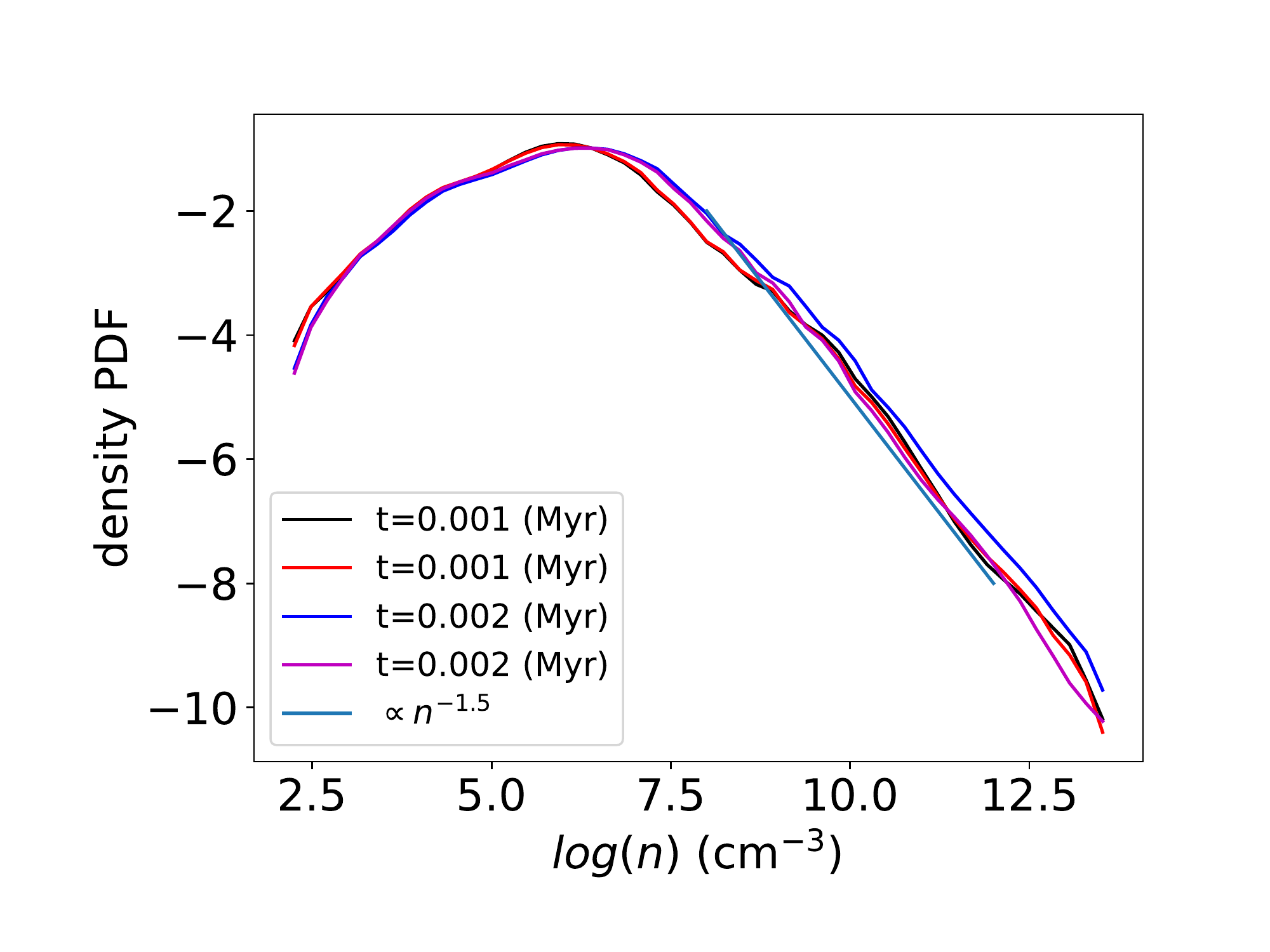}}
\put(8.3,6.5){D1}
\put(-0.1,7){\includegraphics[trim=20 20 20 20,clip,width=4.9cm]{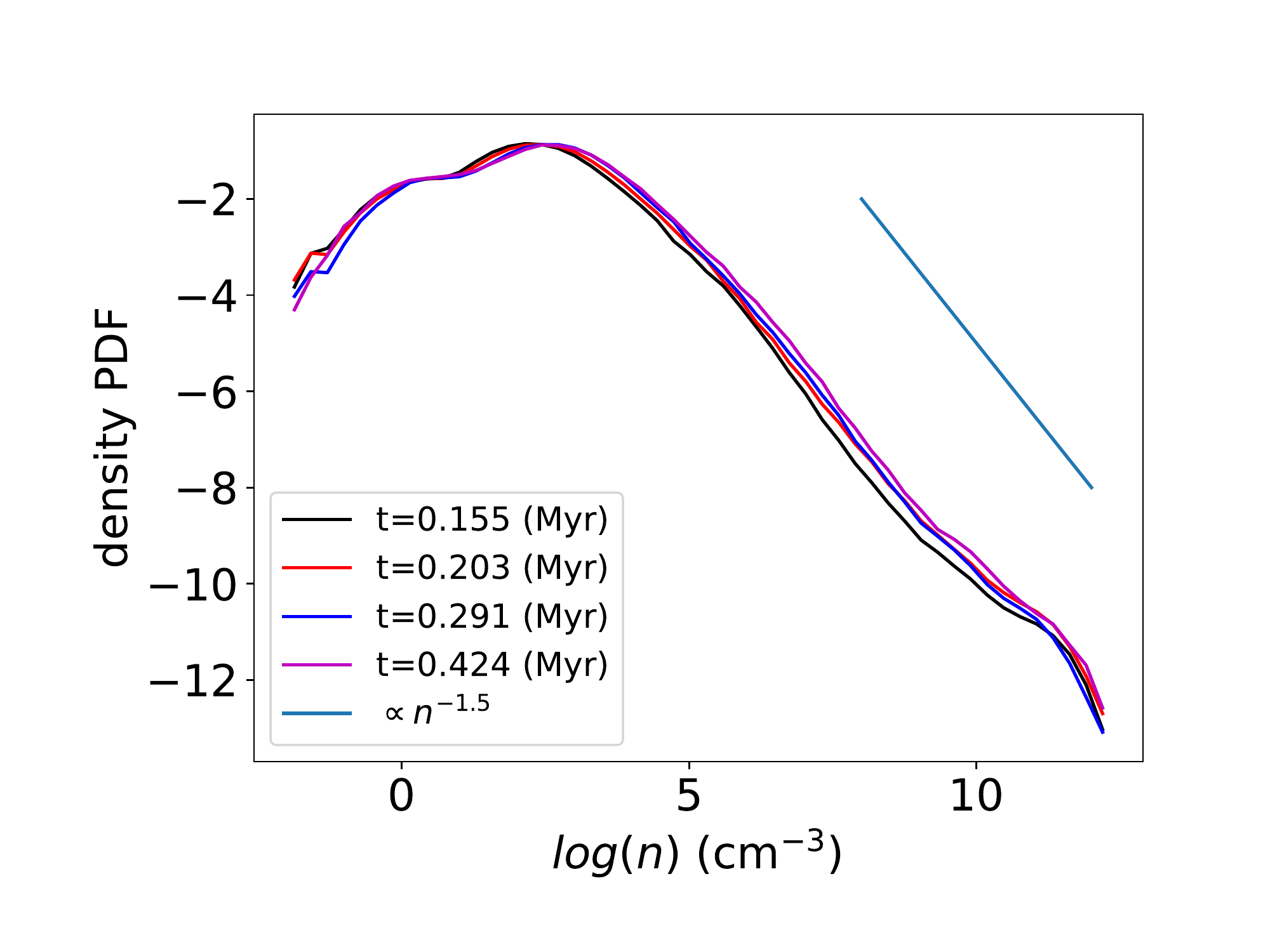}}
\put(3.4,10){A1++}
\put(4.5,7){\includegraphics[trim=20 20 20 20,clip,width=4.9cm]{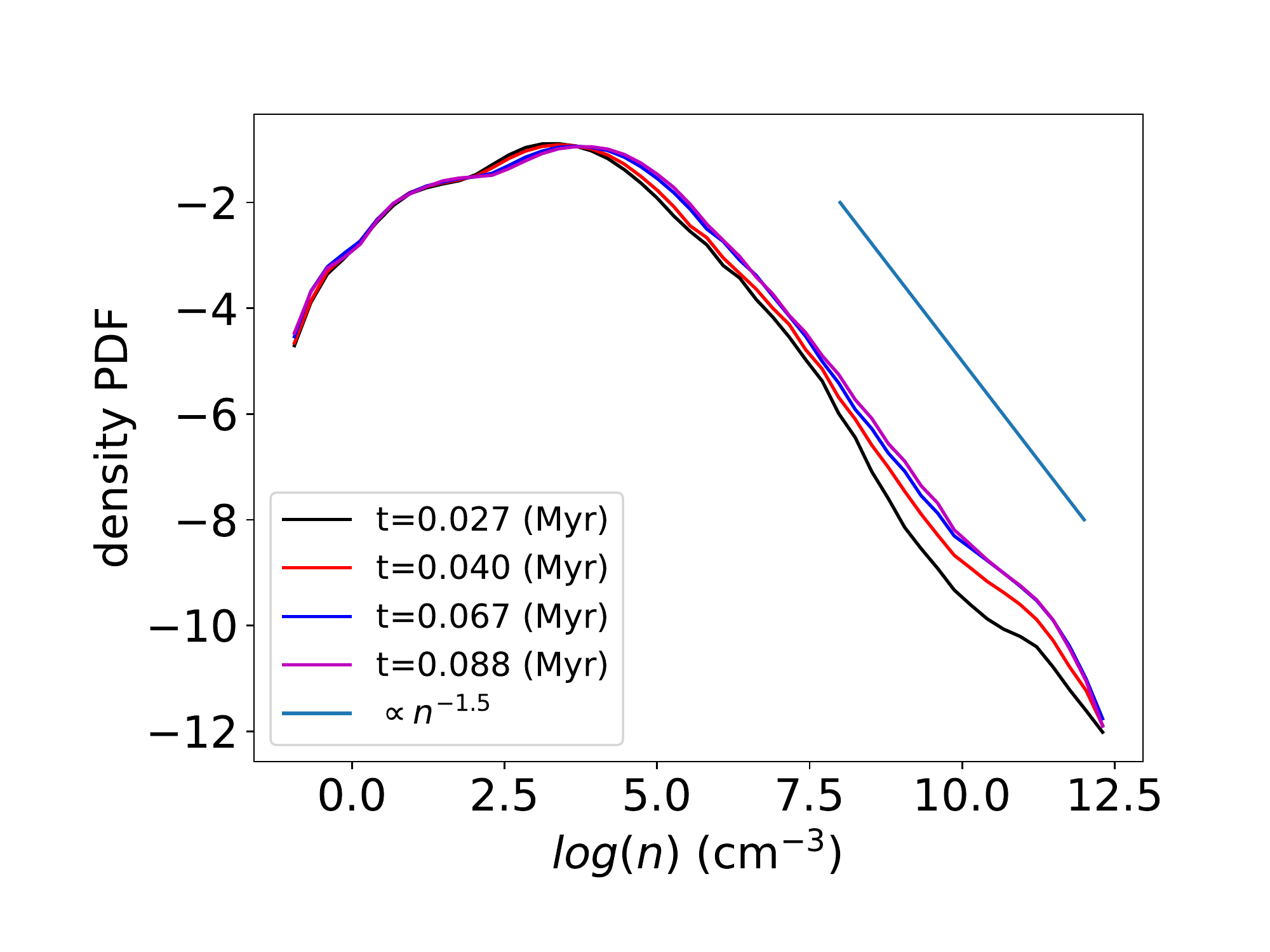}}
\put(8.1,10){B1++}
\end{picture}
\caption{Density PDF (normalized in random units) for runs A1++, B1++, C1+, D1, C1t03, and C1t01. At low densities, the shape of the PDF is
broadly lognormal, while at high density, it is $\propto n^{-1.5}$. The PDF is almost invariant in time given that the cloud is sufficiently supported, as it is in most cases. In run C1t01, the initial turbulent support is very low and thus the cloud global collapse is strong, giving a density PDF that shifts toward high densities in time.}
\label{fig_pdf}
\end{figure}

\subsection{Sink particle analysis}\label{sink}

\subsubsection{Density profiles around sink particles}
Figure~\ref{fig_dens_sink} shows the mean density profile around sink particles in 
simulation A1++ (top panels) and C1+ (bottom panels). 
The profile is averaged among all radial directions and among sinks, 
starting from the distance of twice the smallest cell size. 
When there exists a nearby sink, the density profile is considered only up to half of the distance in between. 
The solid lines represent the mean value 
while the shaded areas show the standard deviation. Three ranges of mass are 
considered, below $0.01 ~\Ms$, between $0.01 ~\Ms$ and $0.1 ~\Ms$, and 
between $0.1 ~\Ms$ and $1 ~\Ms$. 
For A1++, results are presented for sinks younger than 1000 yr (left), 5000 yr (middle), and all sinks (right), while for C1+, 300 yr (left) and 1000 yr (middle) were used for age filtering.

Several interesting trends can be inferred.  First of all, at younger ages
(i.e., left and middle panels), the density profiles are broadly $\propto r^{-2}$ and 
proportional, by a factor of a few, to the density profile of the singular isothermal sphere (SIS), $\rho_\mathrm{SIS} = c_\sound^2/(2\pi G r^2)$, where the isothermal sound speed $c_\sound = 200~ {\rm m~ s}^{-1}$ at 10 K 
\citep[e.g.,][$n_\mathrm{SIS} = \rho_\mathrm{SIS}/(\mu m_{\rm p})$ with the black solid line]{shu77}. Second, the density at early times (left panels) is significantly
above $n_\mathrm{SIS}$, particularly between 20-50 AU, where it can be as high 
as $\sim 20 ~n_\mathrm{SIS}$. This value will be used in the companion paper (hereafter paper II) to analyze the gas surrounding 
the first Larson core and estimate the expected peak of the stellar distribution. 
At later times (middle and right panels), the density drops to lower values that are comparable to or 
even lower than $n_\mathrm{SIS}$. 
There is large dispersion both due due the age dispersion of sinks and the asymmetric environment in which older sinks are found. 
This is particularly the case in the right panels that include sinks of all ages,  
since the sinks decorrelate from the local overdensity in which they form at later evolutionary stage.
The sink particle environment changes rapidly, that is to say, in a few kyr.

\setlength{\unitlength}{1cm}
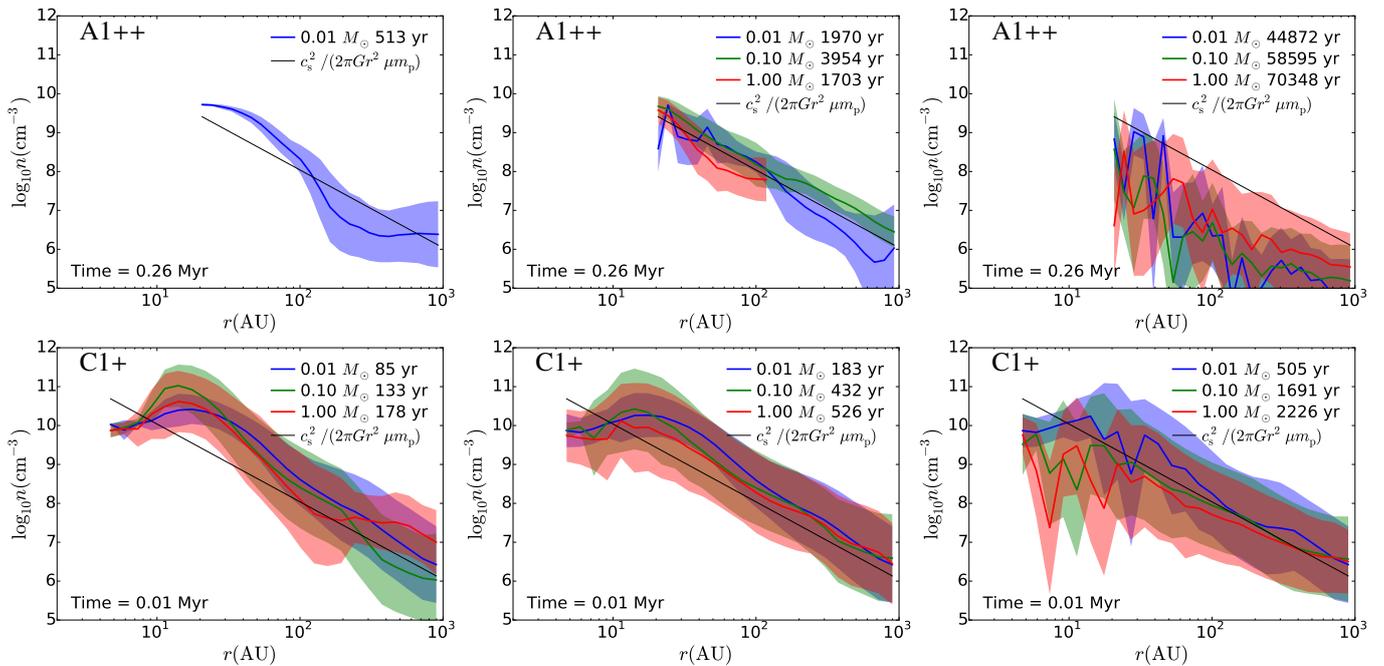
\begin{figure*}[]
\begin{picture} (0,8.8)
\put(0,4.4){\includegraphics[trim=10 20 10 12,clip,width=6cm]{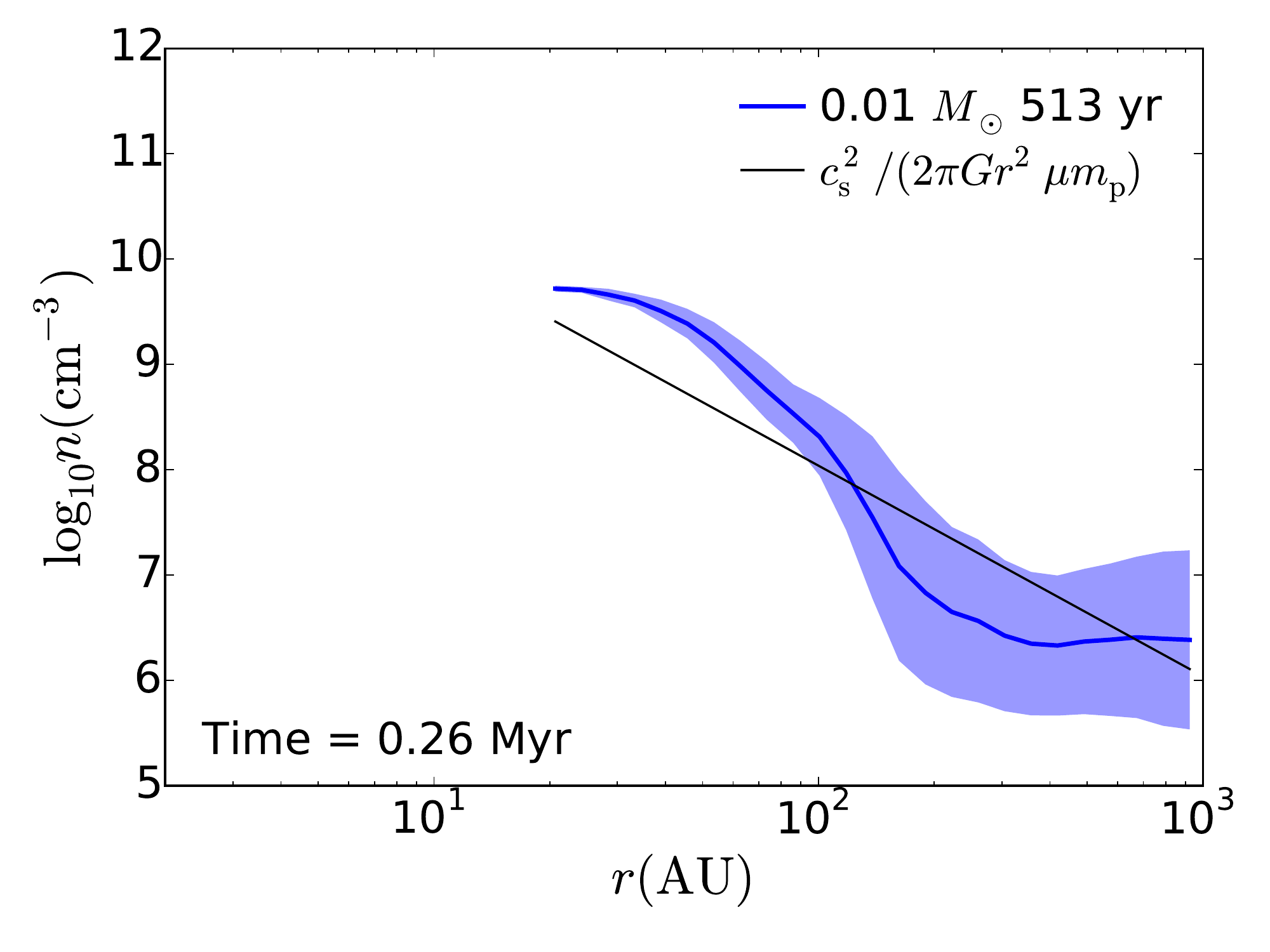}}
\put(1,8.3){A1++}
\put(6,4.4){\includegraphics[trim=10 20 10 12,clip,width=6cm]{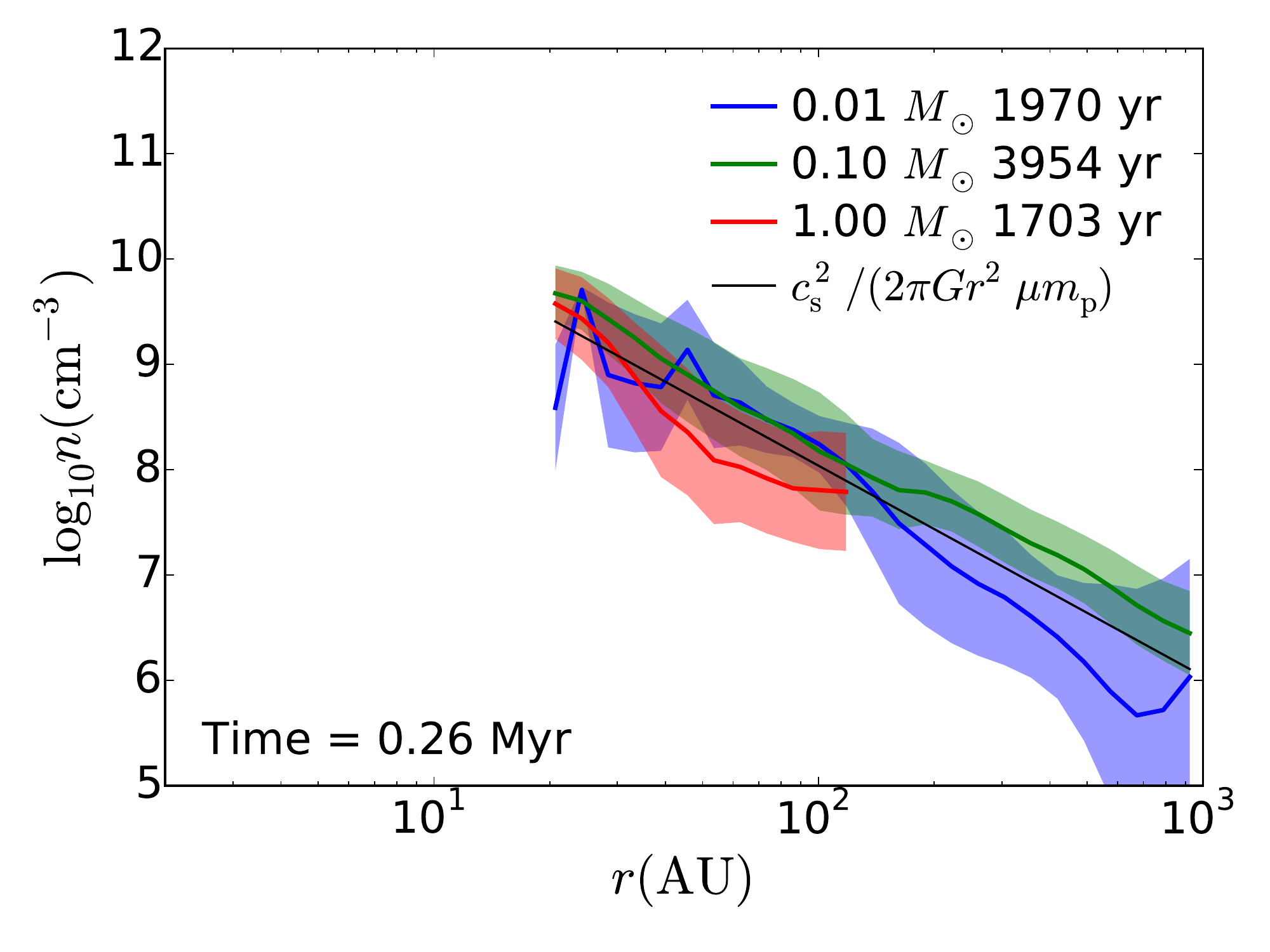}}
\put(7,8.3){A1++}
\put(12,4.4){\includegraphics[trim=10 20 10 12,clip,width=6cm]{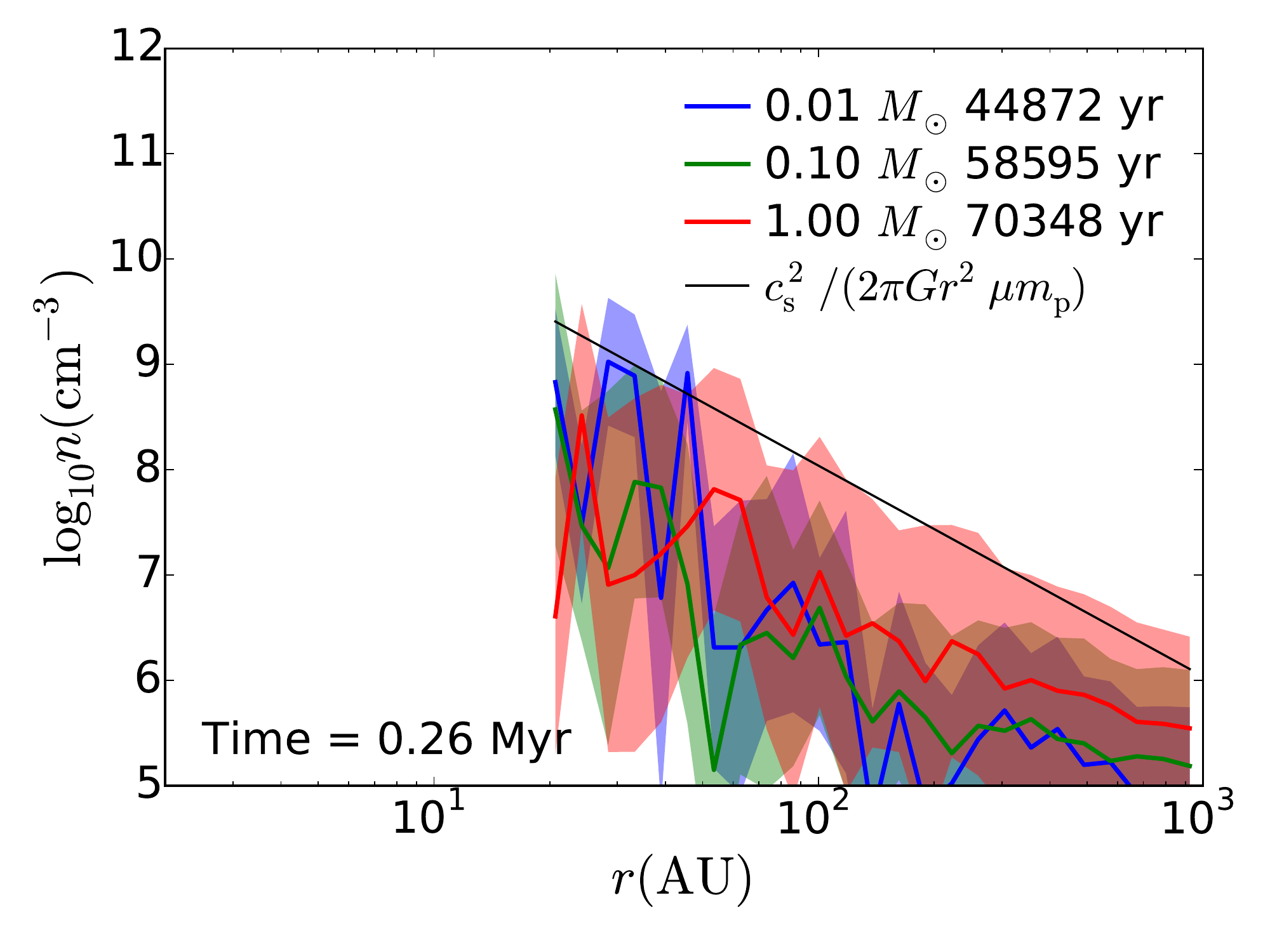}}
\put(13,8.3){A1++}
\put(0,0){\includegraphics[trim=10 20 10 12,clip,width=6cm]{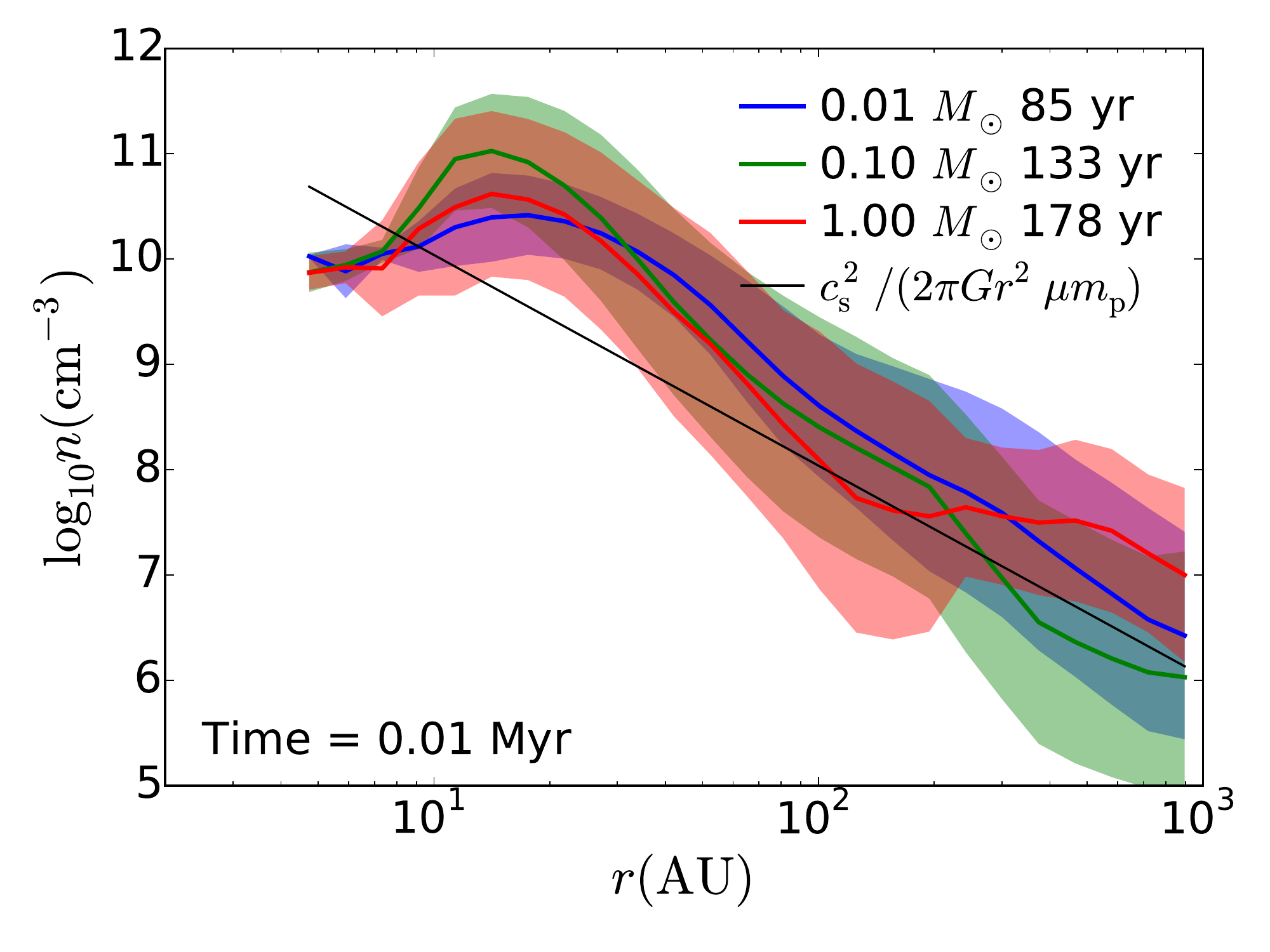}}
\put(1,3.9){C1+}
\put(6,0){\includegraphics[trim=10 20 10 12,clip,width=6cm]{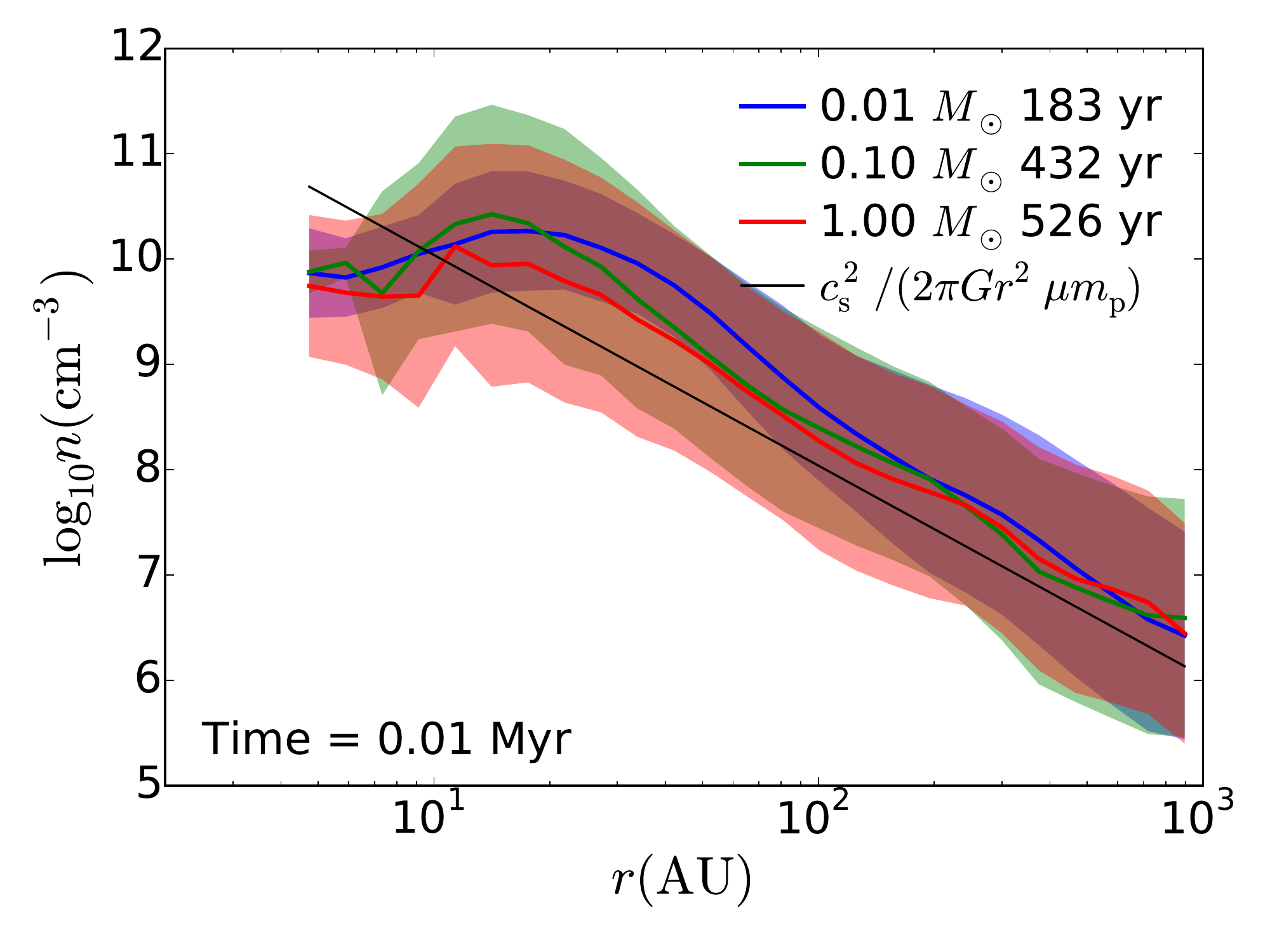}}
\put(7,3.9){C1+}
\put(12,0){\includegraphics[trim=10 20 10 12,clip,width=6cm]{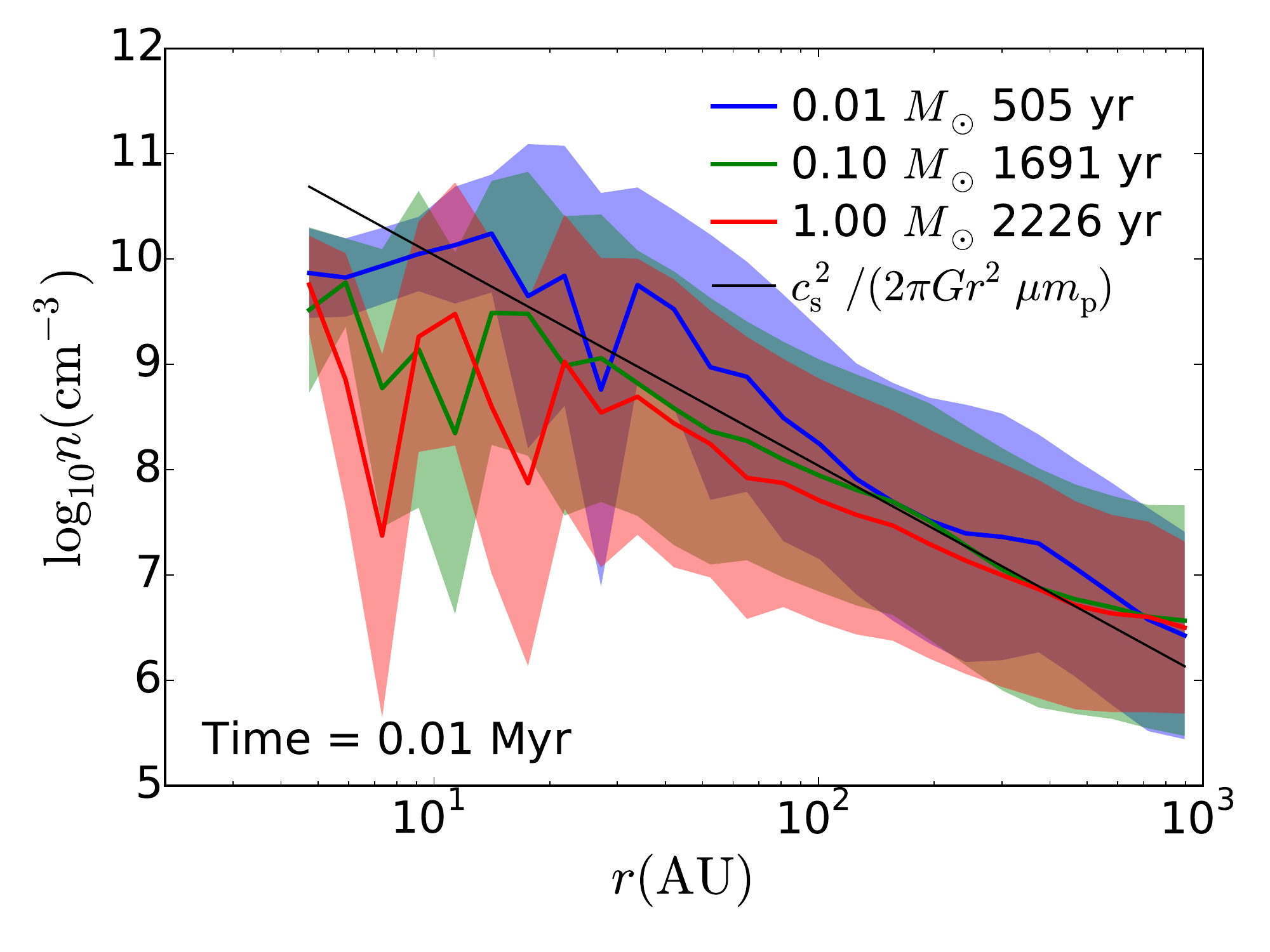}}
\put(13,3.9){C1+}
\end{picture}
\caption{Density profiles around sink particles  for runs A1++ at 0.26 Myr ({\it top panels}) and C1+ at 0.008 Myr ({\it bottom panels}). 
From left to right, sinks aged under 1000 yr, 5000 yr, or all sinks are used to produce the average profiles for run A1++, while 300 yr and 1000 yr are used as thresholds for run C1+. 
Sink particles are separated into three mass bins $<0.01~\Ms$ ({\it blue}), $0.01-0.1~\Ms$ ({\it green}), and $0.1-1~\Ms$ ({\it red}), average age of selected sinks is shown in the legend. Shaded regions represent the standard deviation of the density profiles for all sinks and all radial directions. 
The SIS density profile, $n_\mathrm{SIS} = c_\sound^2/(2\pi G r^2 \mu m_{\rm p})$, is plotted in black for reference. 
The density profile at younger ages is always broadly $\propto r^{-2}$, with a factor $\sim 10$ higher than the SIS, while sinks of older ages have probably accreted all mass from the core in which it is embedded or have been ejected, thus decorrelating from the  $r^{-2}$ profile. This is reflected by the large dispersion that is particularly evident with more massive sinks. }
\label{fig_dens_sink}
\end{figure*}

\subsubsection{Time-dependence of accretion} \label{sec_time}

\setlength{\unitlength}{1cm}
\begin{figure*}[]
\begin{picture} (0,8.6)
\put(0.5,4.3){\includegraphics[trim=15 10 15 10,clip,width=8cm]{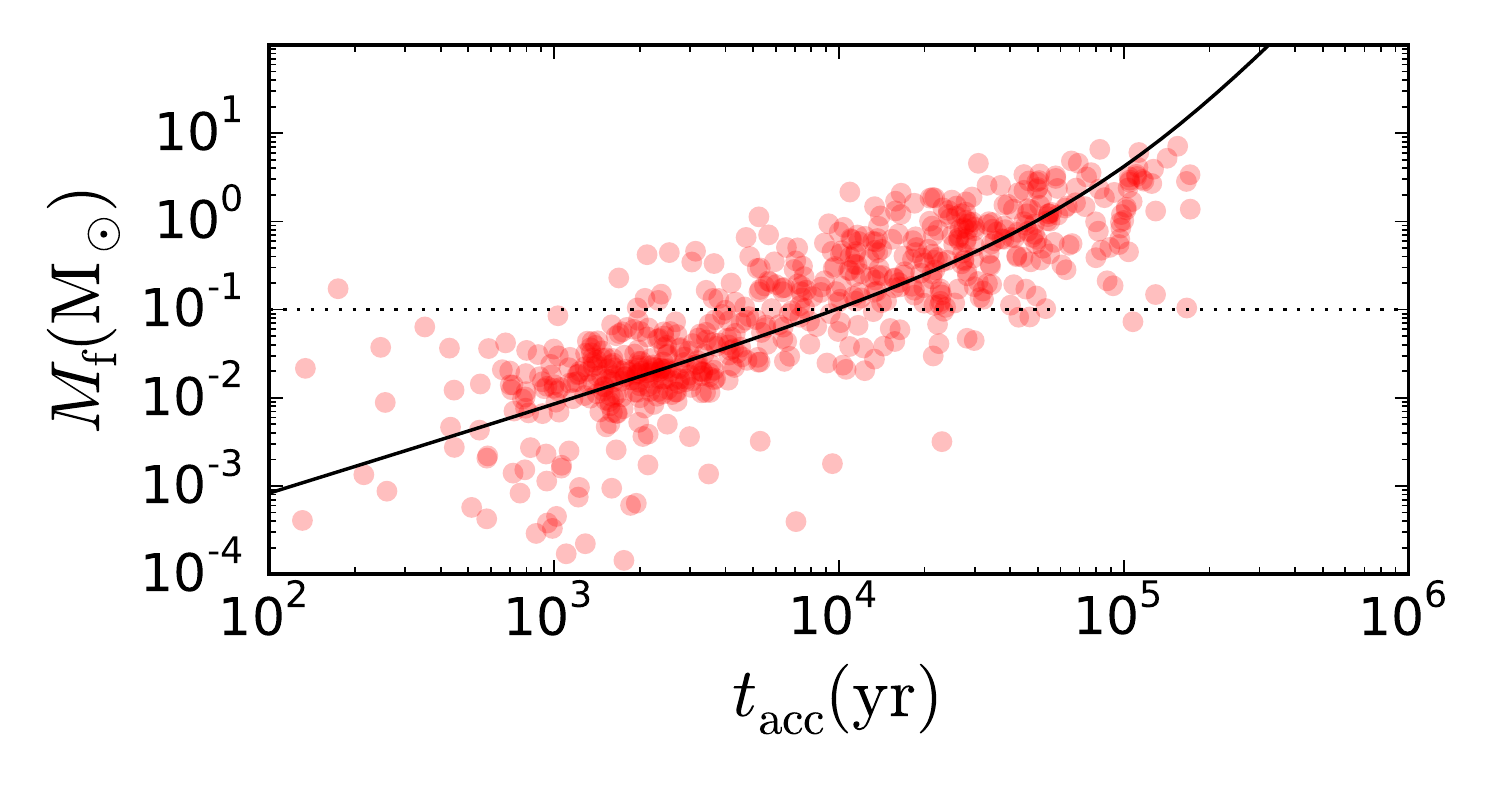}}
\put(6.9,5.8){A1++}
\put(9.,4.3){\includegraphics[trim=15 10 15 10,clip,width=8cm]{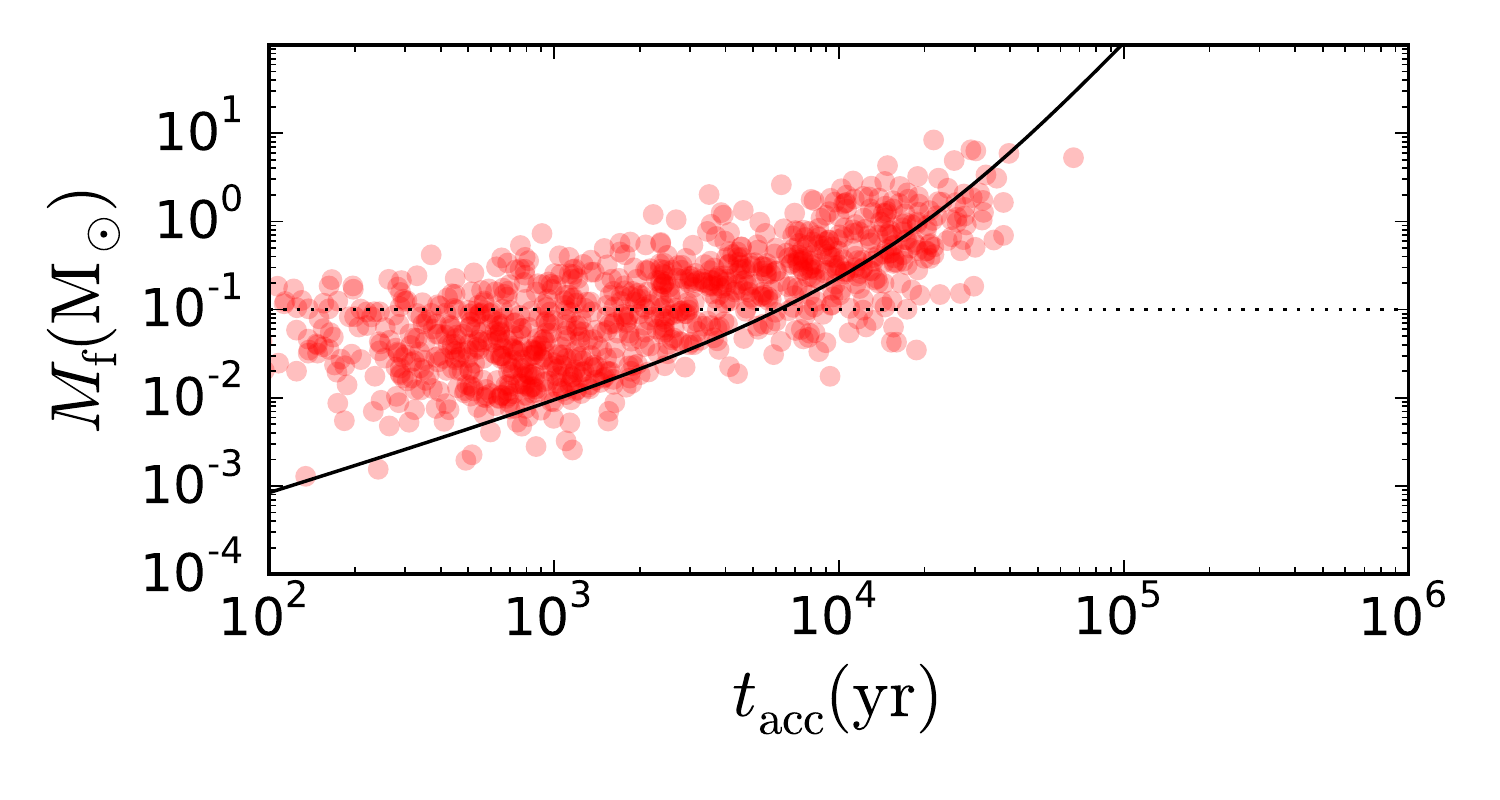}}
\put(15.4,5.8){B1++}
\put(0.5,0){\includegraphics[trim=15 10 15 10,clip,width=8cm]{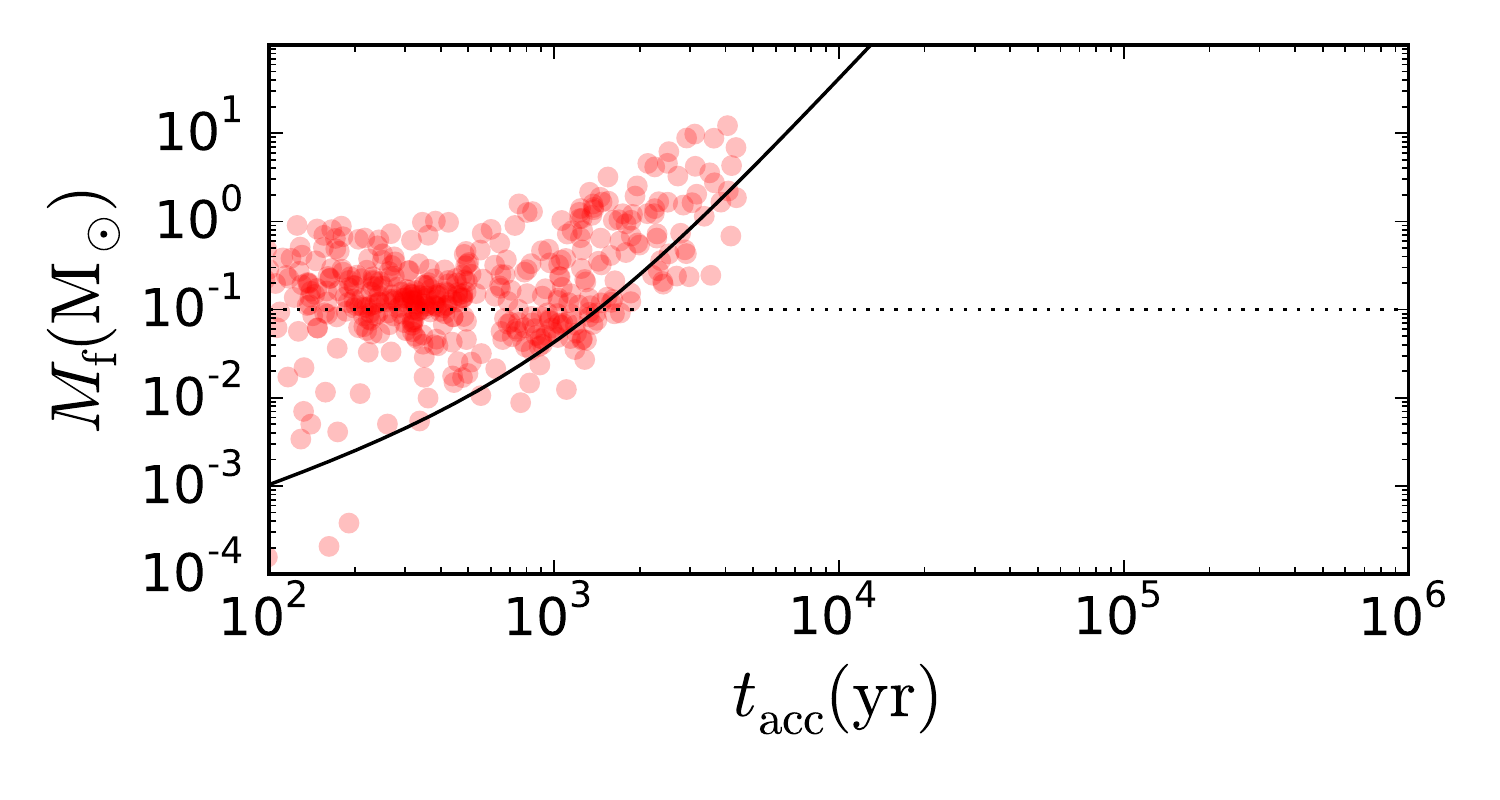}}
\put(7,1.5){C1+}
\put(9.,0){\includegraphics[trim=15 10 15 10,clip,width=8cm]{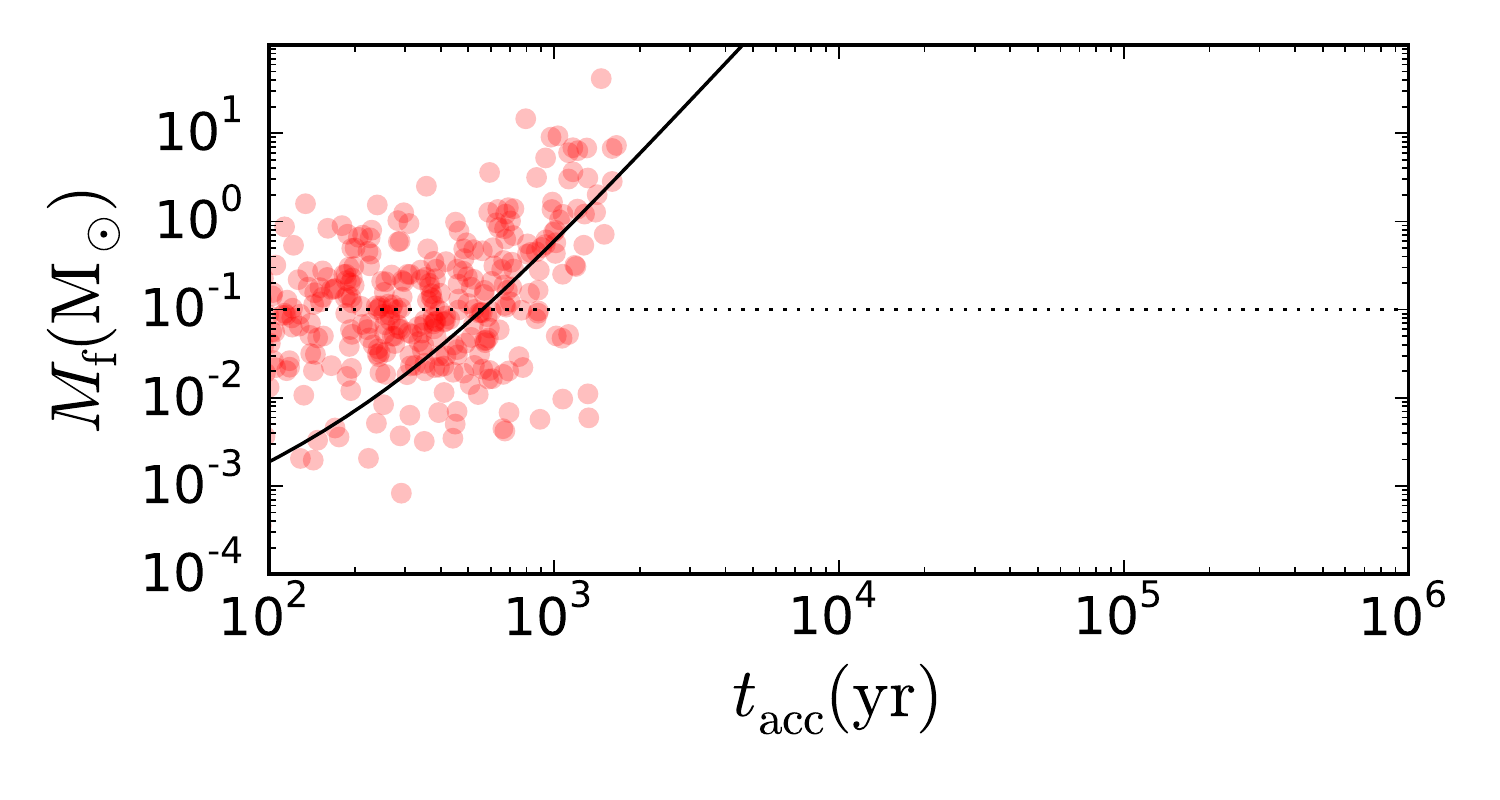}}
\put(15.5,1.5){D1}
\end{picture}
\caption{
Final sink mass against accretion time ($t_\mathrm{acc,60}$ is used) for runs A1++, B1++, C1+, and D1. 
 The solid black lines represent the free-fall time of the mass reservoir derived from the analytical model (see Sect.~\ref{accret_time}). 
The behavior of the distribution and the analytical model are very similar
except for short accretion times (corresponding to mass $\lesssim 0.1~\Ms$), where the physics to be considered is different. This strongly suggests that the mass that built the stars essentially comes from a preexisting deterministic reservoir.}
\label{fig_timescale}
\end{figure*}

Figure \ref{fig_timescale} shows sink mass against accretion time for runs A1++, B1++, C1+, and D1. 
For the four runs, 
the final sink mass, $M_\mathrm{f}$, and the accretion timescale, $t_\mathrm{acc,60}$, 
taken as the time that a sink needs to accrete $60$ percent of its mass, although with a large dispersion, 
are correlated, and $t_\mathrm{acc,60}$ increases with $M_\mathrm{f}$ on average. 

In the case of run A1++, $M_\mathrm{f}$ and $t_\mathrm{acc,60}$ are nearly proportional.
 Interestingly, there is an excess of points 
at $M_\mathrm{f} \simeq 10^{-2} \, \Ms$ and  $t_\mathrm{acc,60} \simeq 3 \times 10^3$ yr.
The abundance of low-mass sinks is a possible  signature of the disk fragmentation discussed above. 
For run B1++, there is no clear sign of bi-modality. The relation between $M_\mathrm{f}$
and $t_\mathrm{acc,60}$ is still broadly linear, but a deviation seems to occur at high mass. 
Finally, runs C1+ and D1 present similar behaviors, while the latter exhibits a slightly larger dispersion. 
There is a flat dependence of $M_\mathrm{f}$ on $t_\mathrm{acc,60}$ for $M < 0.1~\Ms$ 
and a much stiffer one for larger $M_\mathrm{f}$ with $M_\mathrm{f} \propto t_\mathrm{acc,60}^{3-4}$.
This behavior is quantitatively interpreted in Sect. \ref{accret_time}.

%------------------SMF COMPARISON------------------------
%------------------COMPACTNESS AND TURBULENCE STUDIES------------------------
\section{Mass spectra of sink particles}
Figures~\ref{fig_SMF_density} and \ref{fig_SMF_turbulence} show the mass functions of sink particles at total accreted mass of 
$20,~50,~100,~150,~200,~\text{and}~300~\Ms$, where applicable, since some runs are less evolved. 
To assess the numerical convergence of the runs, we performed runs at several spatial resolutions for most cases,  
and they are presented in the same row. 
This allows us to compare the various runs at more or less the same physical resolution, therefore asserting that the 
differences are not mere consequences of different resolutions. 
The stellar distribution from the simulations were fitted with lognormal distributions, and their characteristic mass and width of distribution are 
listed in Appendix \ref{appen_g}. The purpose was to measure the position of the peak in a systematic way. 
We also drew power-law distributions, which constitute good fits above $\sim 0.1 ~\Ms$.

As a general remark, the shape of the mass spectrum is determined rather early for all runs, 
that is to say, even when only $20-50 ~\Ms$ have been accreted. 
This is compatible with the analysis in Sect.~\ref{sink}, where it has been found that the accretion time is short ($<10^5$ yr for model A and $<500$ yr for model C in Fig. \ref{fig_timescale}) and that once the 
sinks have decorrelated from their environment (sinks in the
low-density environment in Fig. \ref{fig_dens_sink}), they no longer accrete significantly. 

\setlength{\unitlength}{1cm}
\begin{figure*}[]
\begin{picture} (0,18)
\put(12,13.5){\includegraphics[width=6cm]{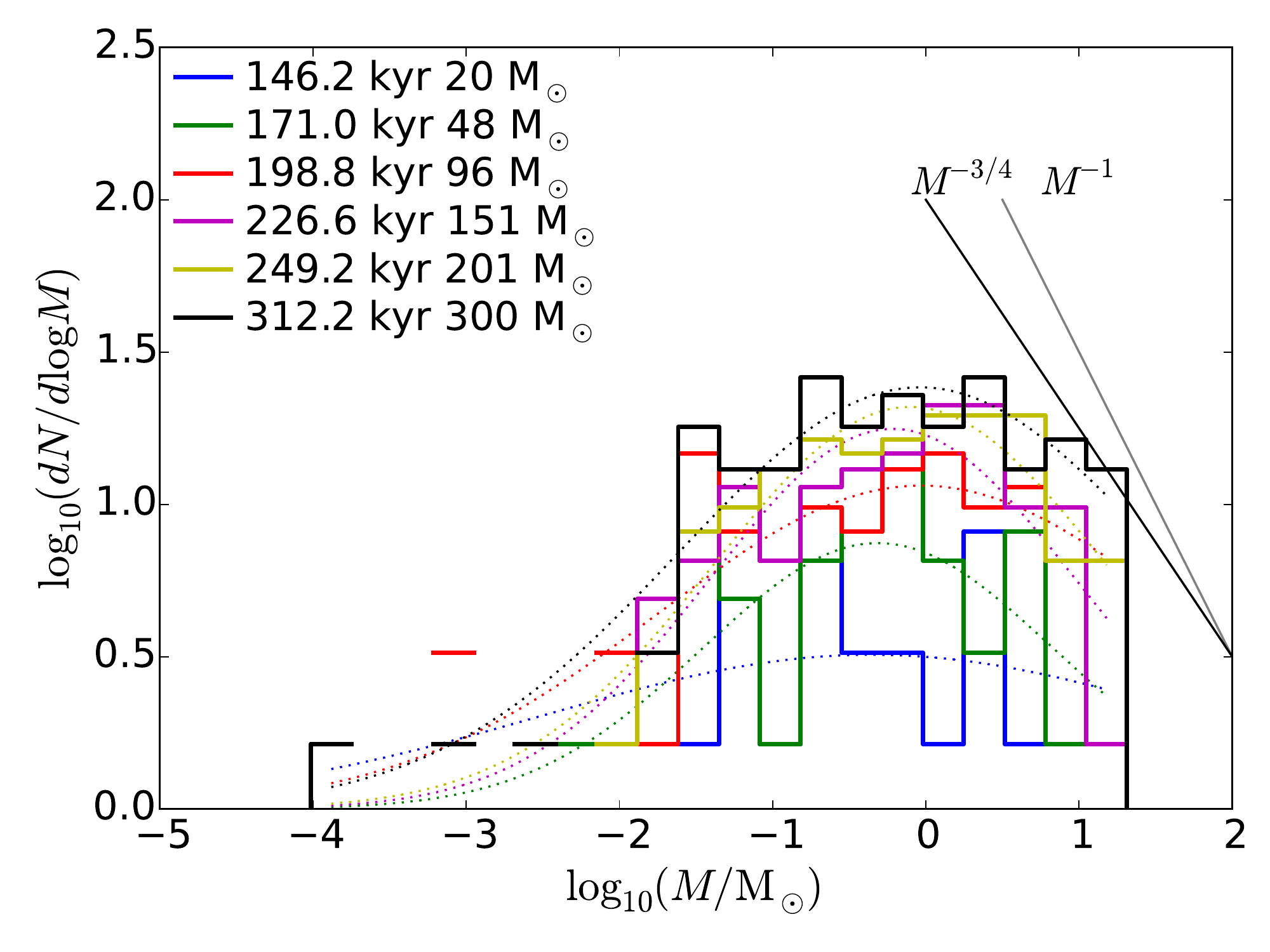}}
\put(6,13.5){\includegraphics[width=6cm]{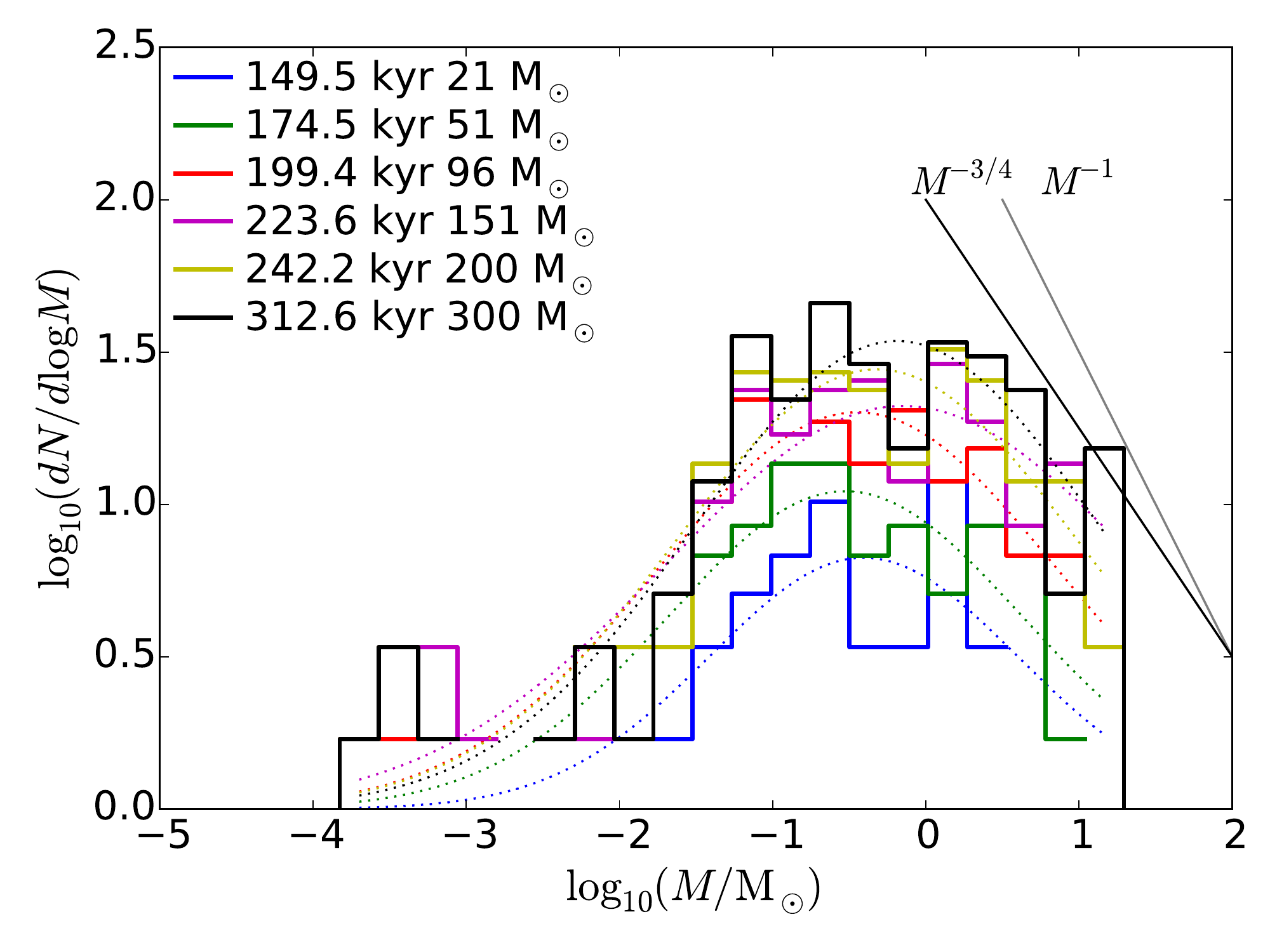}}
\put(0,13.5){\includegraphics[width=6cm]{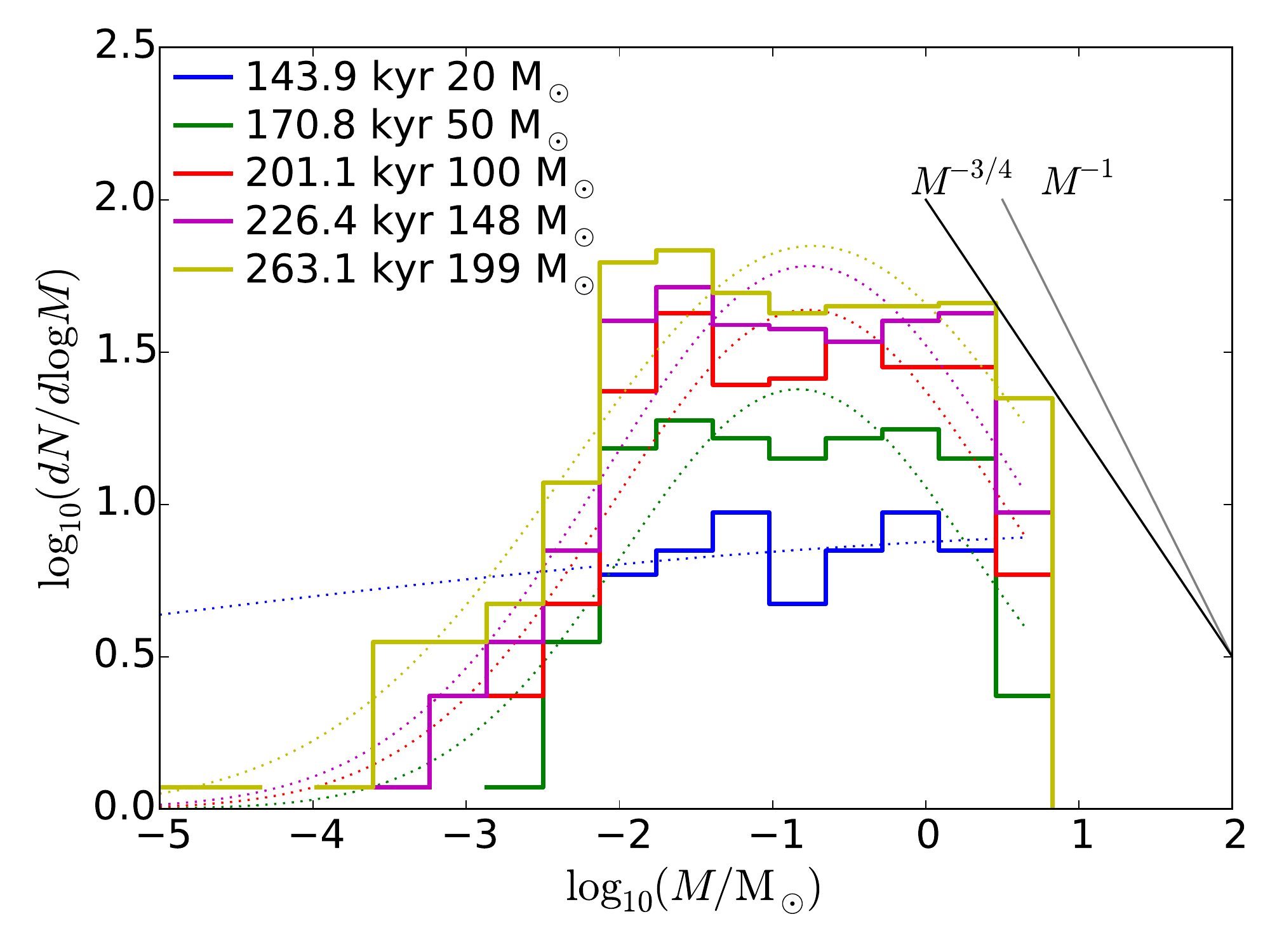}}
\put(4,17.4){A1++, 9 AU}
\put(10,17.4){A1+, 19 AU}
\put(16.2,17.4){A1, 38 AU}

\put(12,9){\includegraphics[width=6cm]{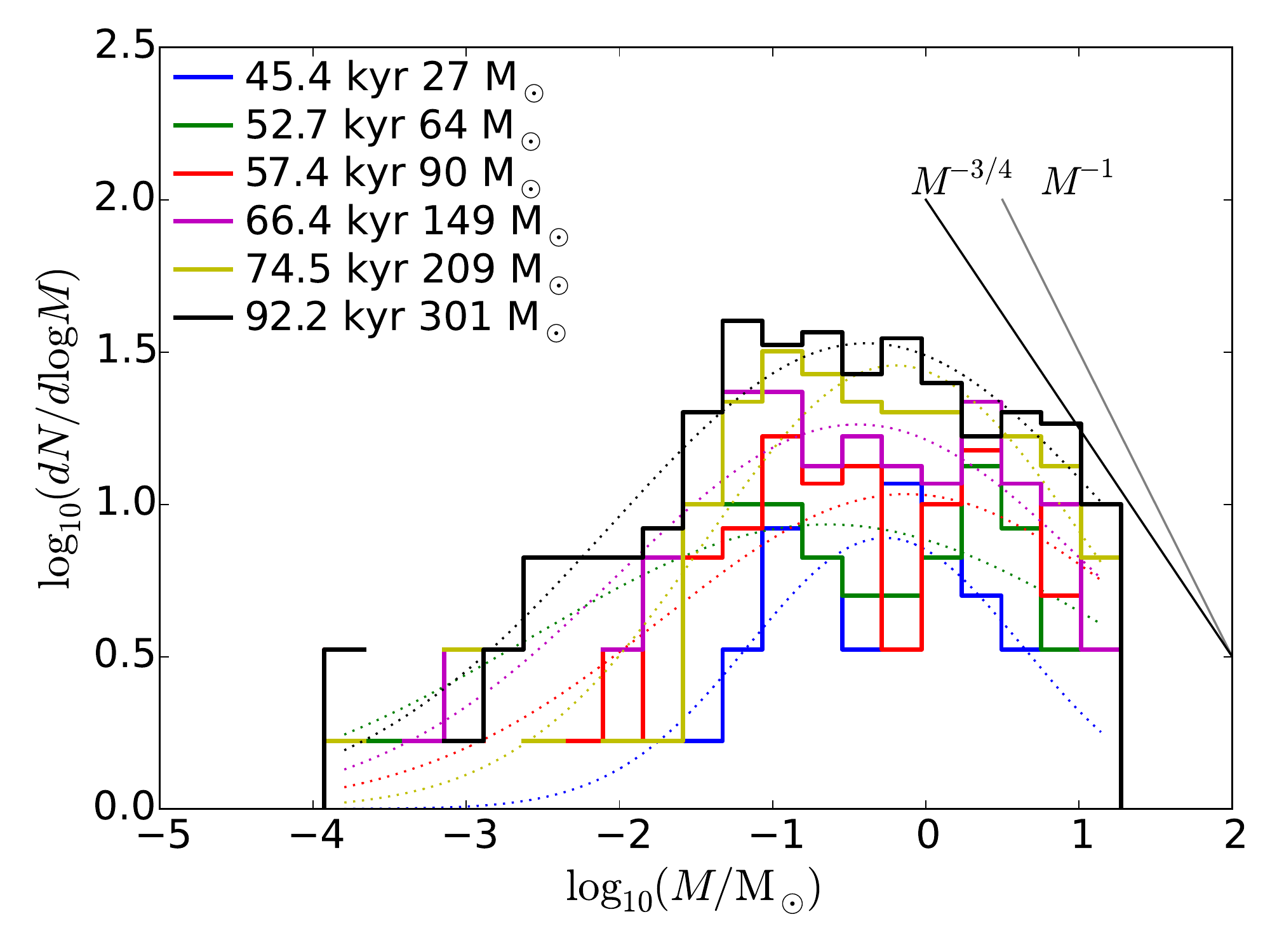}}
\put(6,9){\includegraphics[width=6cm]{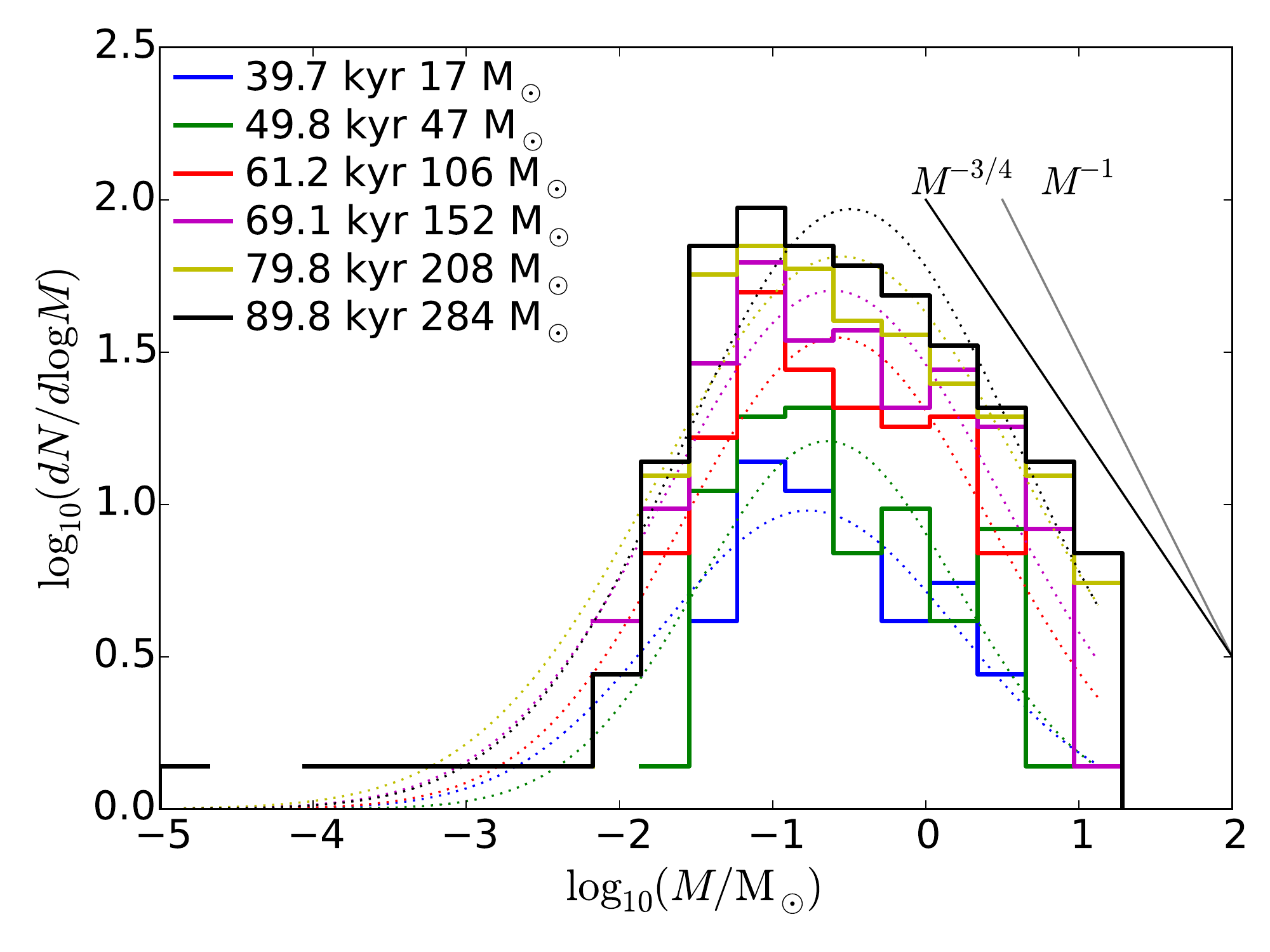}}
\put(0,9){\includegraphics[width=6cm]{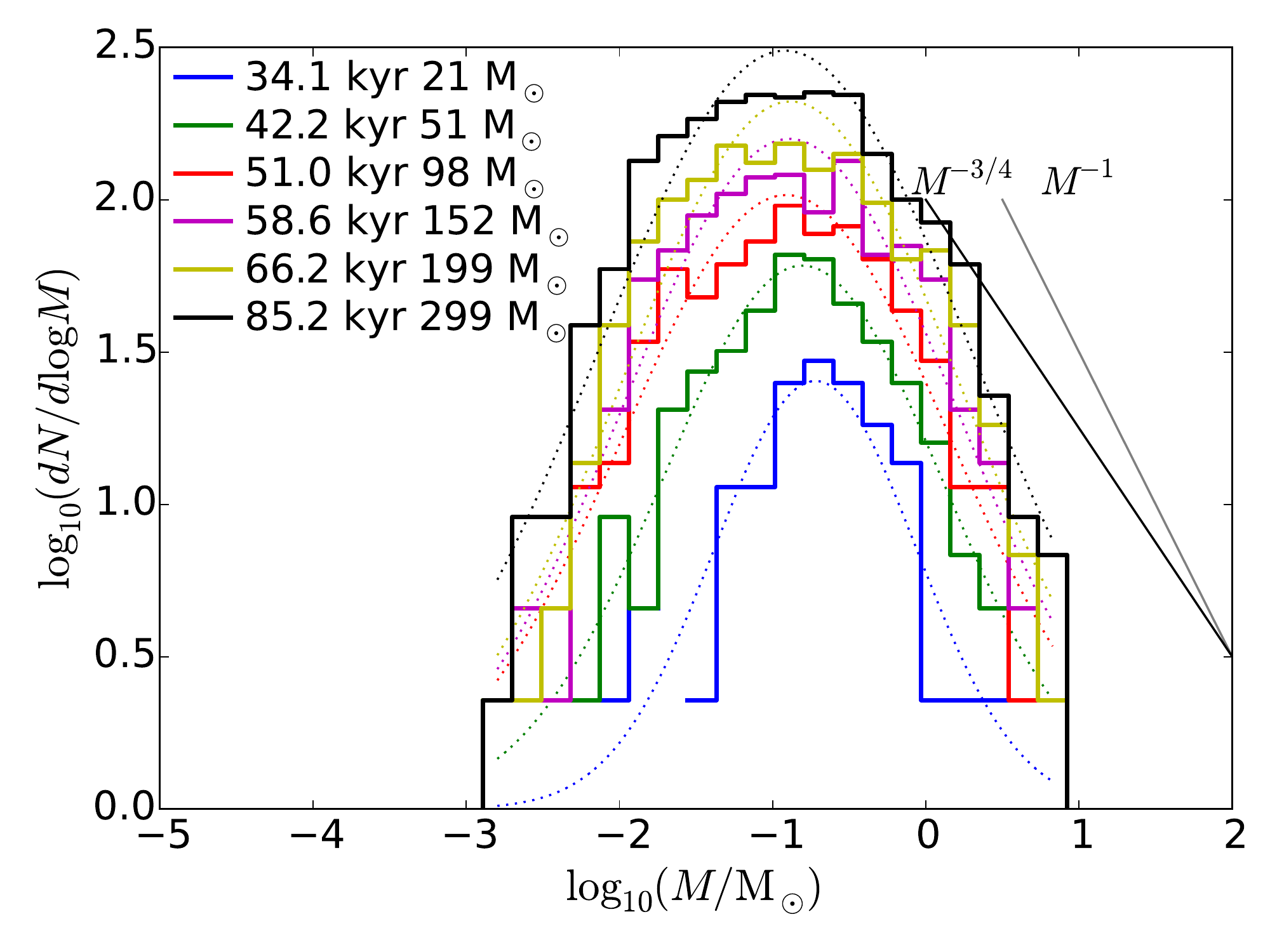}}
\put(4.06,12.9){B1++, 4 AU}
\put(10.1,12.9){B1+, 8 AU}
\put(16.2,12.9){B1, 17 AU}

\put(12,4.5){\includegraphics[width=6cm]{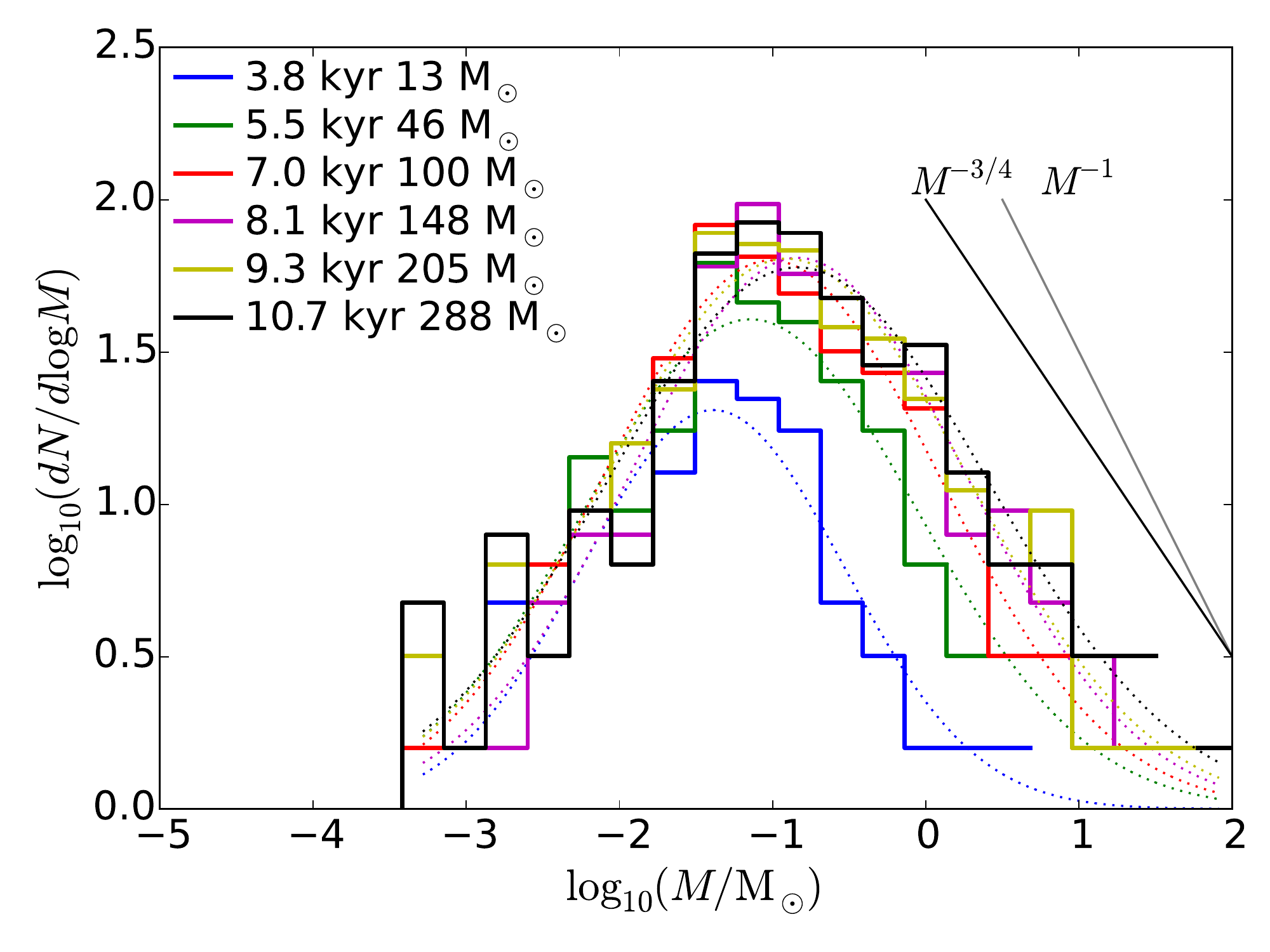}}
\put(6,4.5){\includegraphics[width=6cm]{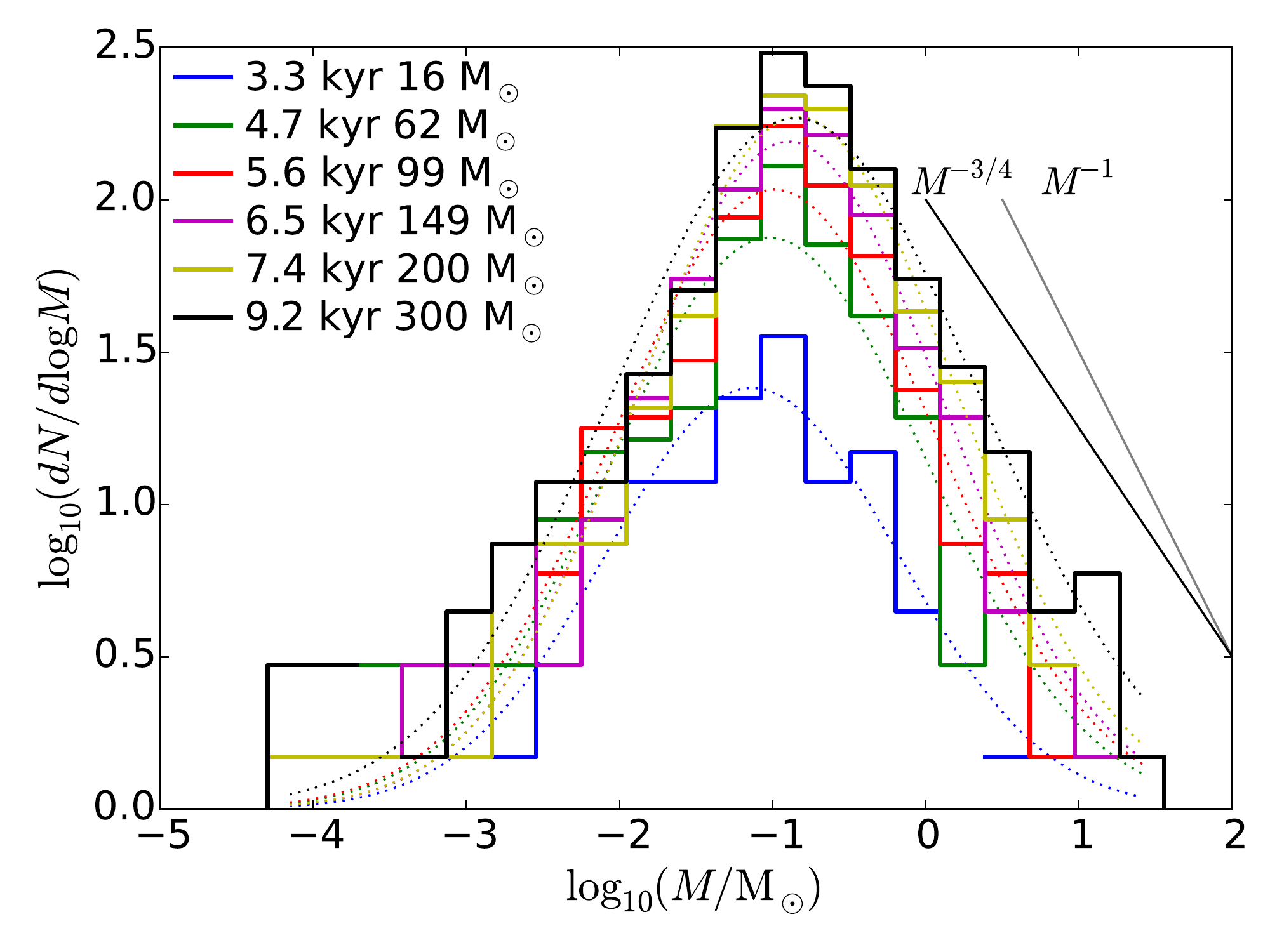}}
\put(0,4.5){\includegraphics[width=6cm]{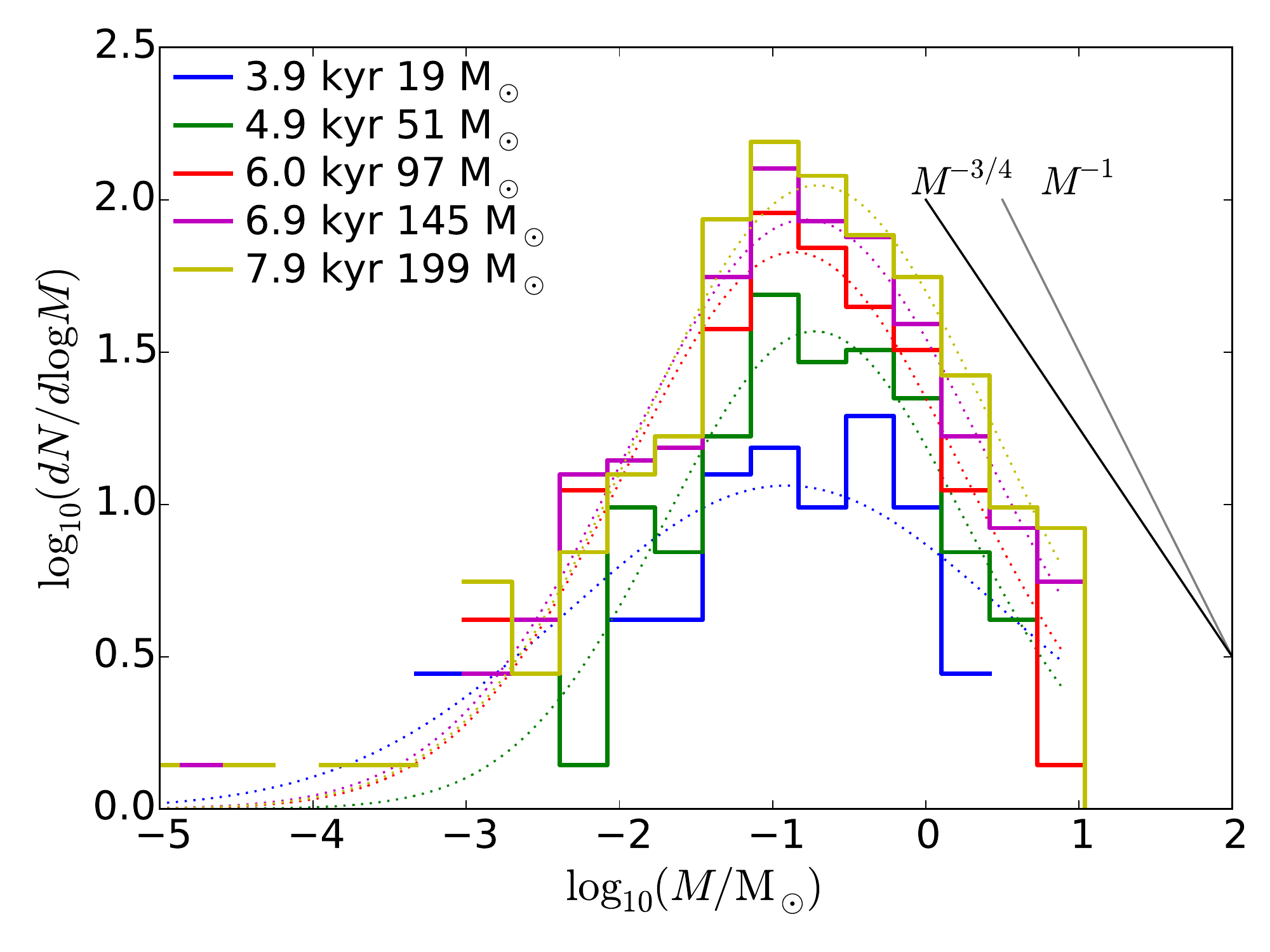}}
\put(4,8.4){C1+, 2 AU}
\put(10.2,8.4){C1, 4 AU}
\put(16.1,8.4){C1--, 8 AU}

\put(12,0){\includegraphics[width=6cm]{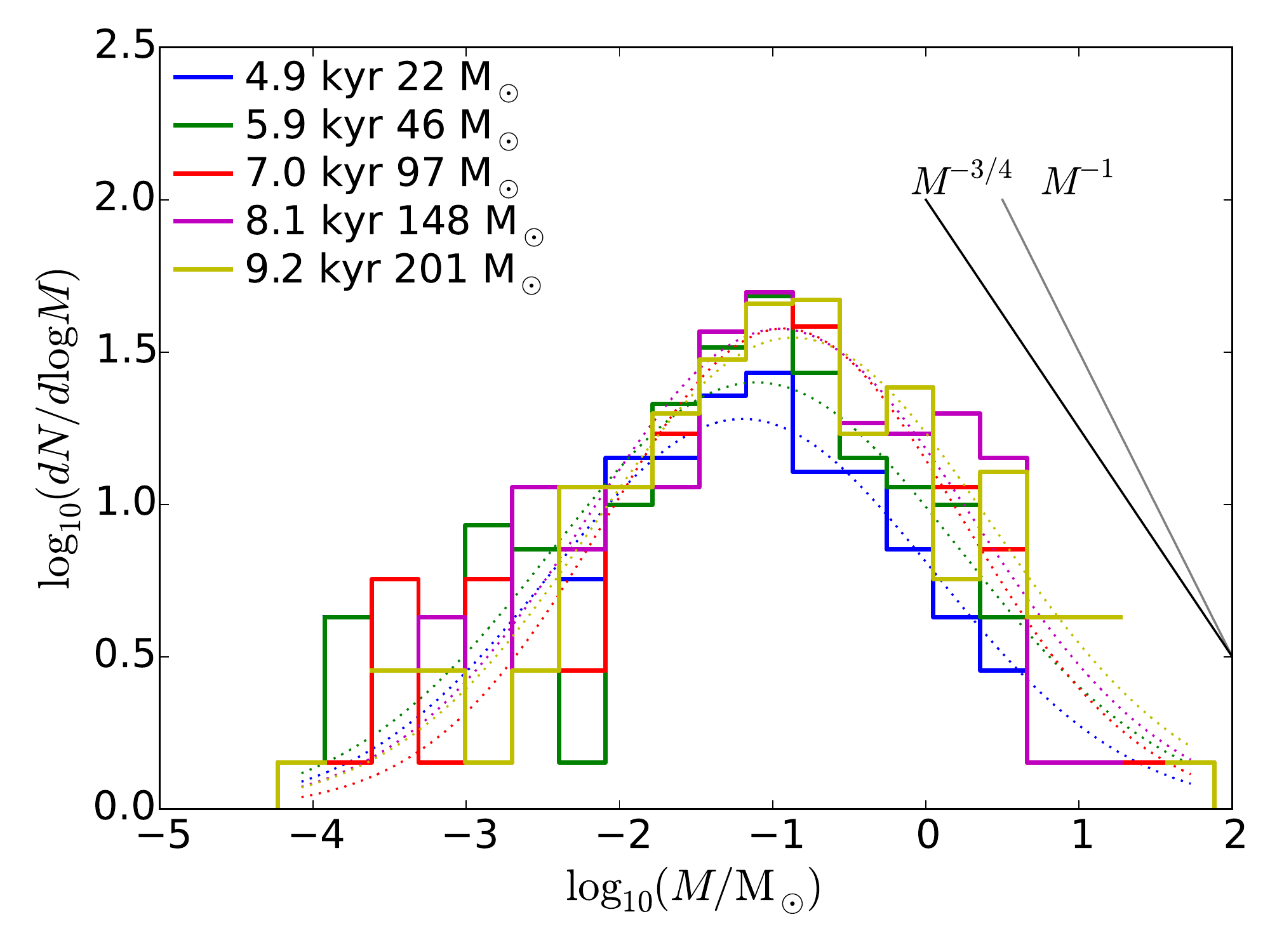}}
\put(15.8,3.9){C1-- --, 16 AU}

\put(0,0){\includegraphics[width=6cm]{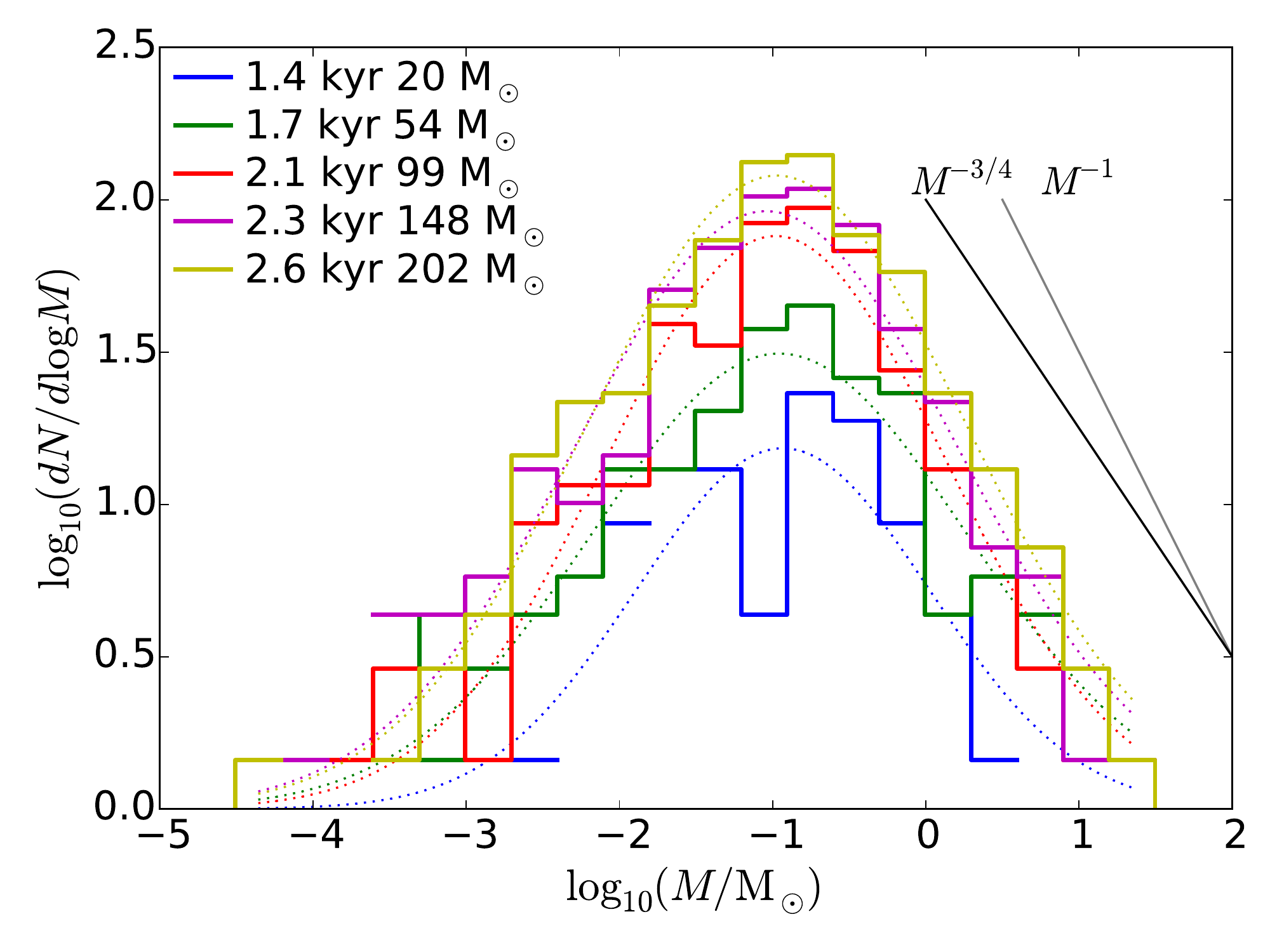}}
\put(4.2,3.9){D1, 2 AU}
\end{picture}
\caption{Sink mass spectra for models A1, B1, C1, and D1 at various time steps, 
corresponding to a comparable amount of accreted mass. Various spatial 
resolutions are explored for the purpose of verifying numerical convergence and direct comparison 
between models. For models B1, C1, and D1, a peak at $\simeq 0.1~\Ms$ is obtained independently of the numerical resolution. While model A1 presents a flat mass spectrum, models 
C1 and D1 present at high mass a power-law behavior that is compatible with $M^{-3/4}$ (see also Fig.~\ref{model}).}
\label{fig_D1}\label{fig_SMF_density}
\end{figure*}

\setlength{\unitlength}{1cm}
\begin{figure*}[]
\begin{picture} (0,4.5)
\put(0,0){\includegraphics[width=6cm]{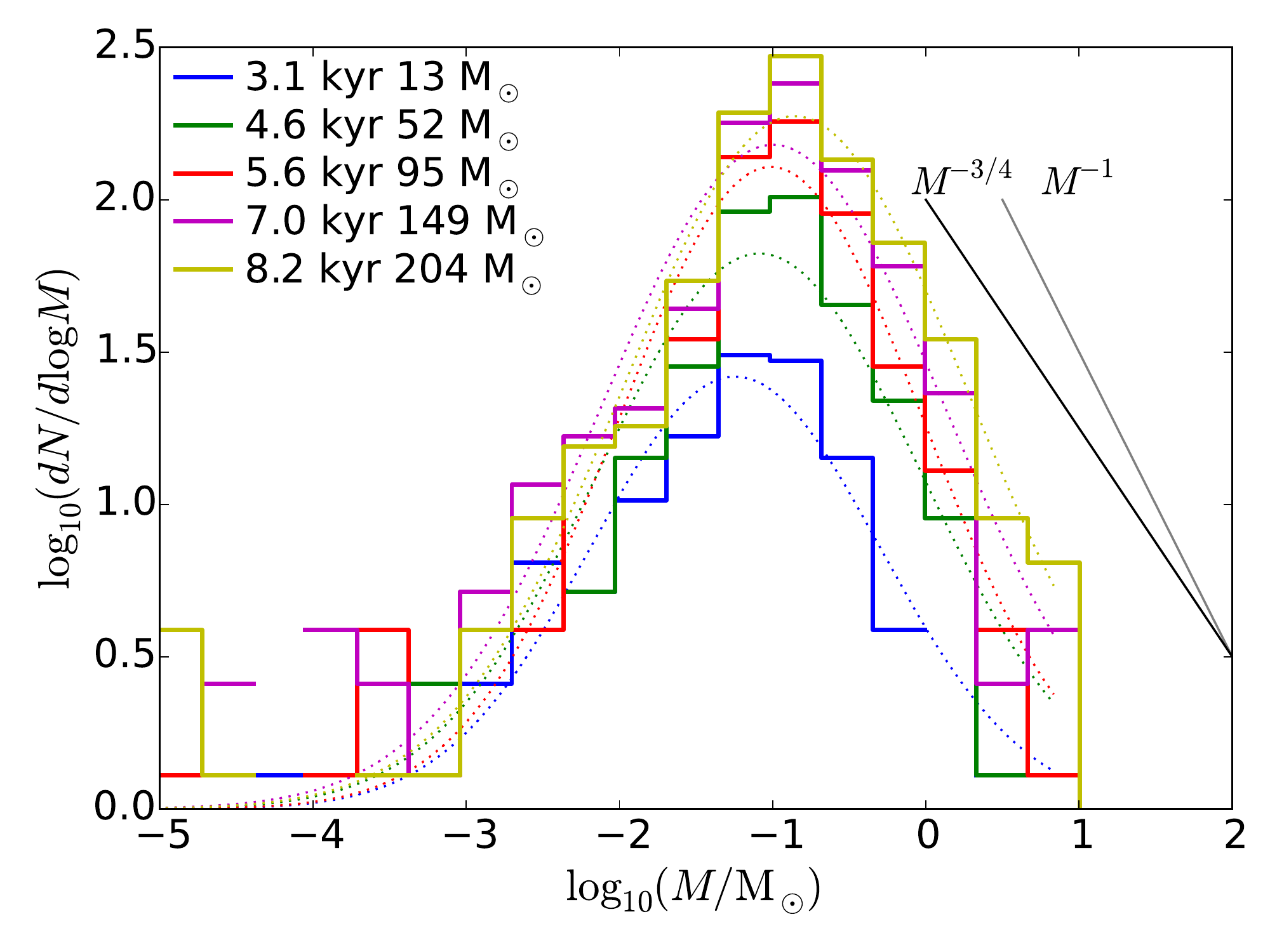}}
\put(4,3.9){C1t15, 4 AU}
\put(6,0){\includegraphics[width=6cm]{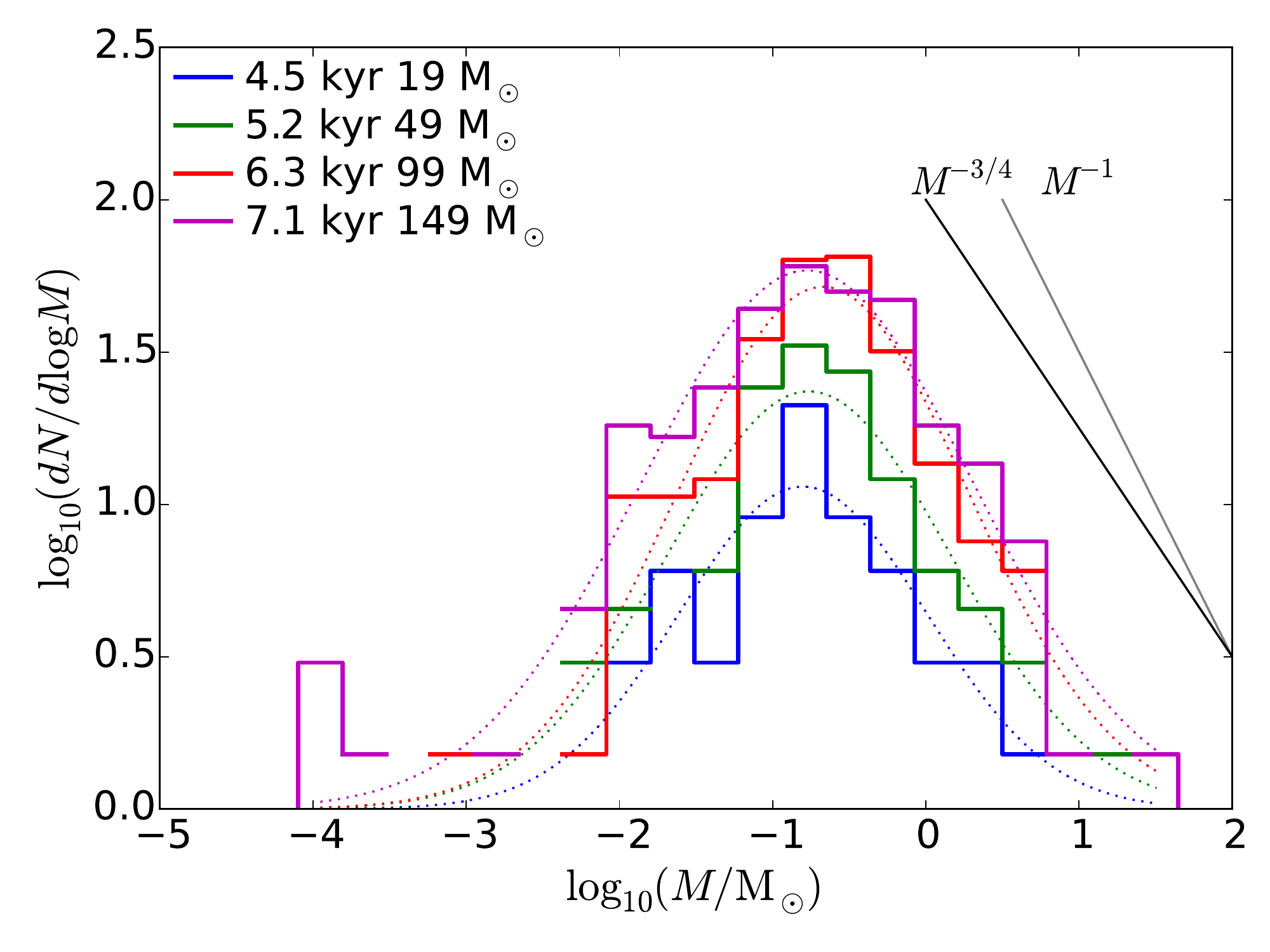}}
\put(10,3.9){C1t03, 4 AU}
\put(12,0){\includegraphics[width=6cm]{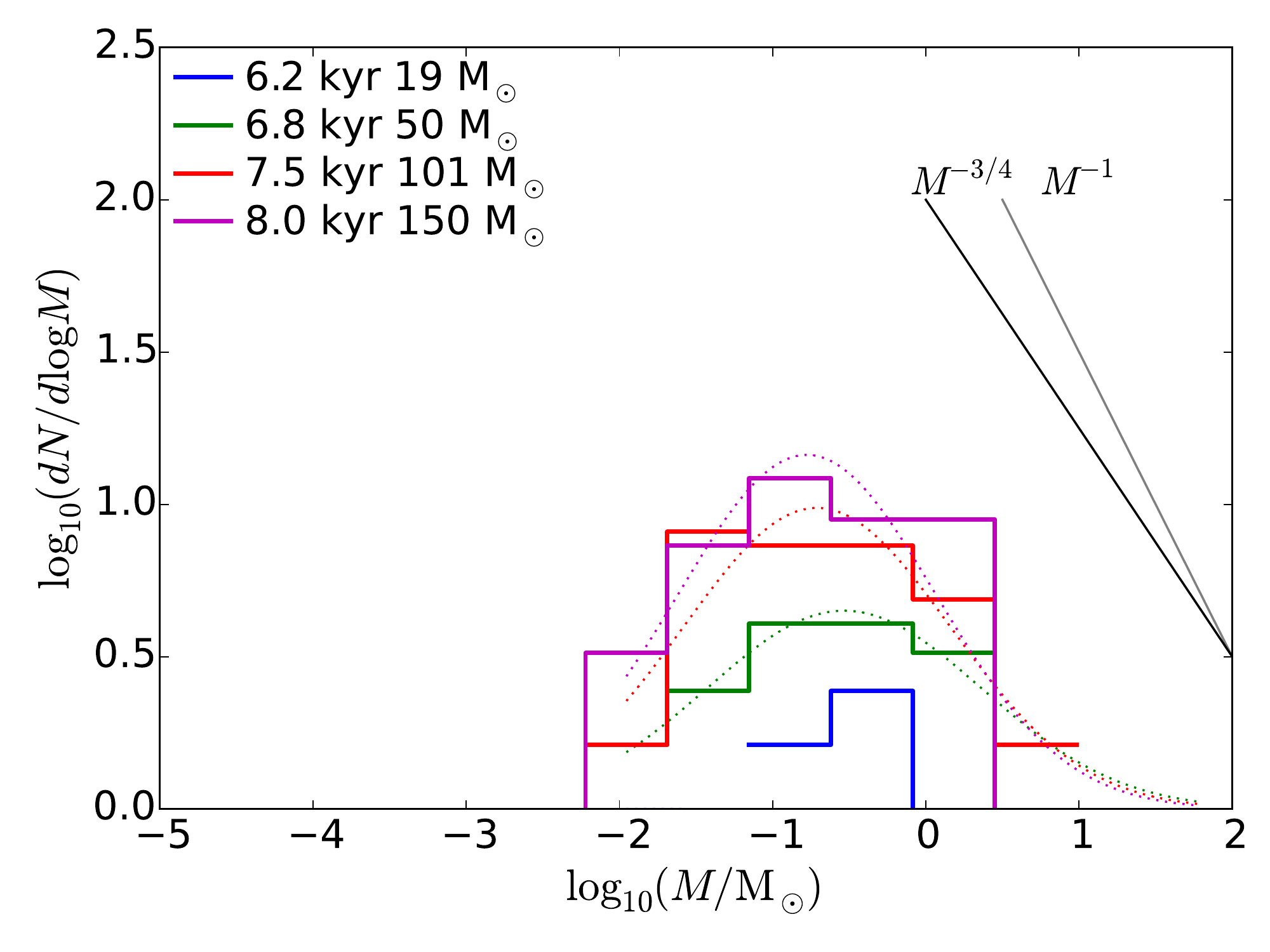}}
\put(16,3.9){C1t01, 4 AU}
\end{picture}
\caption{Mass spectrum evolution for models C1t15, C1t03, and C1t01 corresponding to 
three levels of initial turbulence. Except for C1t01, which initially
has a very low 
turbulent energy, the mass spectra are not significantly different.}
\label{fig_SMF_turbulence}
\end{figure*}

\subsection{Dependence on density}
The first row of Fig.~\ref{fig_SMF_density} shows results for runs A1. 
In run A1++, several features appear. 
First of all, the mass spectrum is essentially flat between 0.1 and 3 $\Ms$. 
For masses $> 3 ~\Ms$, there is a stiff decrease. 
At masses $\simeq 0.01 ~\Ms$, we
see an excess of objects, as we noted in Sect.~\ref{sink}, that is likely 
a consequence of disk fragmentation and is possibly a spurious effect in the simulation that may disappear with more complete physics. 
At lower resolutions, runs A1+ and A1 show mass spectra that are compatible with the 
spectrum of run A1++, although they are noisier.  

The second row of Fig.~\ref{fig_SMF_density} shows results for runs B1.
In run B1++, we can distinguish a flat mass spectra for $0.05~ \Ms < M < 0.3 ~\Ms$
and then a power-law type distribution at higher masses with an exponent $\simeq -0.5$ to -1. At masses 
$M > 2 ~\Ms$, there might even be another regime, described by a stiffer power law
(this remains to be confirmed), as shown by the yellow and black histograms. The shapes of 
the mass spectra of run B1+ are broadly compatible with those of B1++, but the flat part is absent, and instead 
a peak appears at 0.1 ~$\Ms$. The mass spectra of run B1 are significantly flatter
than those of B1+ and B1++, illustrating the need for resolution and careful 
numerical convergence tests.  

The third row of Fig.~\ref{fig_SMF_density} shows results for runs C1 (see also the right panel of the bottom row). 
In run C1+, a peak appears at $0.1~\Ms$, while at high mass, 
a power law with an exponent equal to about -0.75 is inferred. 
Run C1 exhibits very similar mass spectra, 
with the peak and the power law almost identical. Even the total 
number of objects is comparable (when comparing at 200 accreted $\Ms$). 
Thus it seems reasonable to claim that numerical convergence has been reached for runs C1+ and C1.
Runs C1-- and C1-- -- also show a peak at 0.1 $\Ms$, 
while the power-law behavior is noisier. 
The slope may be slightly flatter for C1-- --.

The bottom left panel of Fig.~\ref{fig_SMF_density} shows results for run D1. 
The mass spectra are very similar to those of run C1 with a peak at 
$0.1~\Ms$ and a power law with exponent about -0.75 to -1. There are 
possibly more low-mass objects ($< 0.1~\Ms$) than in run C1+.

\subsection{Dependence on turbulence}
The runs with various levels of turbulence are presented in Fig.~\ref{fig_SMF_turbulence}.
Run C1t15 has twice more kinetic energy than run C1, while runs C1t03 and C1t01 have  
10 and 100 times less kinetic energy, respectively. 
Increasing the turbulence level to slightly super-virial (C1t15) does not make any visible difference the mass spectra,
which present the same peak and power-law behavior as run C1. 

For run C1t03, the peak of the mass spectra seems to be shifted to slightly higher masses, 
say, tentatively $\simeq 0.2 ~\Ms$ and there are slightly fewer objects. The exponent of the power law is unchanged. 
When the turbulence is lowered even more (C1t01),
one dominating heavy star is formed at the center.  This means
that less mass is available to form 
other objects, and their number is significantly reduced. The mass spectra 
are now very flat over almost two orders of magnitude in mass.

\section{Analytical modeling and physical interpretation}

\subsection{General formalism}
The theory developed by \citet{HC08,HC09,HC13} consists of identifying the 
mass that at scale $R$ is gravitationally unstable, that is, the mass 
contained within the clumps in which gravity dominates
other supports. This is achieved by writing that at all scales the mass at a
density that exceeds a scale-dependent threshold (obtained through 
the virial theorem) is equal to the mass contained in the structures
of mass lower than or equal to the (turbulent) Jeans mass.

As discussed in \citet{HC13}, a difficult question is to what extent 
the distribution should take into account the crossing timescale 
of the density perturbations. Small perturbations can indeed
be 
rejuvenated several times while the large perturbations are still collapsing.
In the case of cold dark matter halos, there is no such contribution 
because the fluctuations are imprinted at the beginning of the Universe and
are then amplified by gravity. In the case of a turbulent molecular cloud, 
where a turbulent cascade proceeds, it seems that the small scales are 
continuously reprocessed.

In the present case, the clouds we considered collapse in about one 
free-fall time. Moreover, it is unclear whether a turbulent cascade truly develops 
because the crossing time is typically a few free-fall times. 
In this context, it therefore seems reasonable not to weight the 
mass spectra by local free-fall time. As we show below, this 
is entirely consistent with the numerical results.

As in \citet{HC08}, we write that 
\begin{eqnarray}\label{bal_mass}
{ M_{\rm tot}(R) \over V_\mathrm{c}} =
 \int\limits^{\infty}_{\delta_R^\mathrm{c}} \overline{\rho} \exp(\delta)   {\cal P}_R(\delta)  d\delta
= \int\limits_0^{M_R^\mathrm{c}} M'  {\cal N}\! (M')  P(R,\!M')\, dM', 
\end{eqnarray}
where $V_\mathrm{c}$ is the cloud volume, $\overline{\rho}$ is
the mean density, ${\cal P}_R$ is the density PDF, and the local fluctuation is $\delta=\log(\rho/\overline{\rho})$. 
This equation is the same as the 
one inferred in \cite{HC08}. 
The first expression for $M_{\rm tot}(R)$ represents 
 the mass contained within structures of mass $M \le M_R^\mathrm{c}$.
It is equal to the mass of the gas, which, smoothed at scale $R$, has a logarithmic
density higher than a critical threshold $\delta_R^\mathrm{c}$.
The second expression signifies that 
the number of structures of mass $M \le M_R^c$  is $ {\cal N}(M') P(R,M')  dM'$,
where ${\cal N}(M') dM'$ is the number density of structures of mass
between $M'$  and $M'+dM'$ and $P(R,M')$ is the probability to find a gravitationally unstable cloud 
of mass $M'$ embedded inside a cloud of gas, which at scale $R$ has a logarithmic density
higher than $\delta_R^\mathrm{c}$. 
It has been demonstrated by \citet{HC08} that $P(R,M')=1$ is a reasonable assumption. 

Taking the derivative of Eq.~(\ref{bal_mass}) with respect to $R$, we obtain
\begin{eqnarray}\label{spec_mass1}
 {\cal N} (M_R^c)  = 
  { \overline{\rho} \over M_R^c} 
{dR \over dM_R^c} \,
\left( -{d \delta_R ^c \over dR} \exp(\delta_R^c) {\cal P}_R( \delta_R^c) \right).
\end{eqnarray}
We note that we have ignored here any dependence of ${\cal P}_R$ on $R$, which would 
introduce a second term that does not play a significant role in most circumstances. 

When the expression of the critical density threshold, $\delta_R^c$, is specified, 
the mass spectrum of the self-gravitating pieces of fluid can be inferred 
using the geometrical expression 
\begin{eqnarray}
M=C_m \rho R^3, ~\text{where typically $C_m = 4 \pi/3$.}
\label{masse}
\end{eqnarray}

\subsection{Density probability function}
As suggested by Eq.~(\ref{spec_mass1}), the density PDF plays a major role, and
it is necessary to use the right PDF to obtain reliable results. 
In particular, most analytical models have been using a lognormal PDF, 
which is not a valid approximation for collapsing clouds. 

\subsubsection{Lognormal distribution}
Numerical simulations of turbulent isothermal clouds have revealed
 \citep{vazquez94,padoan1997,Kritsuk07,Federrath08} that 
a reasonable approximation of ${\cal P}$ is the lognormal distribution 
\begin{eqnarray}\label{Pr0}
{\cal P}(\delta) = {1 \over \sqrt{2 \pi \sigma_0^2}} 
\exp\left(- { (\delta - \overline{\delta})^2 \over 2 \sigma_0 ^2} \right),\\
\text{where} \; \overline{\delta}=-\sigma_0^2/2 \;
 \; \text{and}  \; \; \sigma_0^2=\ln (1 + b^2 {\cal M}^2). \nonumber
\end{eqnarray}
In this expression, ${\cal M}$ is the Mach number and $b$ a non-dimensional coefficient 
that depends on the forcing, which typically 
varies between 0.25 when the forcing is purely solenoidal and almost 1 when 
the forcing is applied on compressible modes \citep{Federrath10}. 
For simplicity and because it is usually not so important, we neglect here the 
scale dependence of $\sigma_0$ \citep[see, e.g.,][for a discussion]{HC13}.

\subsubsection{Power-law distribution}
When the collapse proceeds, the density PDF that develops
in the inner part of the clump is simply \citep{kritsuk11,HF12,girichidis2014}
\begin{eqnarray}
{\cal P}(\rho) =  {\cal P}_0 \left( { \rho \over \rho_0 } \right)^{-1.5},
\label{PL}
\end{eqnarray}
where ${\cal P}_0$ and $\rho_0$ are constants of normalization.

\subsection{Instability criterion}
The last step is to specify the density threshold by writing the virial theorem.
As shown in \citet{HC08,HC13},
the condition for a density fluctuation at scale, $R$, to be unstable is
\begin{eqnarray}
M > M_J = a_J { \Bigl[ (c_\sound)^2 + (V_0^2 / 3) (R /   1 {\rm pc} )^{2 \eta} \Bigr]^{3 \over 2}   \over \sqrt{G^3 \overline{\rho} \exp(\delta) }  },
\label{cond_tot}
\end{eqnarray}
with $a_J$ being a dimensionless geometrical factor of order unity. 
Taking for example the standard definition of the Jeans mass, 
the mass enclosed in a sphere of diameter equal to the Jeans length, we obtain $a_J=\pi^{5/2}/6$.
With Eq.~(\ref{masse}), this implies
\begin{eqnarray}\label{crit_Mtot}
M > M_R^c = a_J^{2 \over 3} C_m^{1 \over 3}
\left( {  (c_\sound)^2 \over G    } R + {V_0^2  \over 3\, G  } \left({R \over 1 {\rm pc}}\right)^{2\eta} R \right),  
\end{eqnarray}  
where $M_R^c$ is the critical mass at scale $R$. 
In Eq.~(\ref{cond_tot}), $\eta$ is the exponent that enters the velocity dispersion, and its value is typically expected to be 0.4-0.5, 
and $V_0$ is the velocity dispersion at 1 pc (and not at the cloud scale). 

After normalization, Eq.~(\ref{crit_Mtot}) becomes
\begin{eqnarray}
\widetilde{M}_R^c =  M / M_J^0 = \widetilde{R}\,
(1+ {\cal M}^2_* \widetilde{R}^{2 \eta}),
\label{mass_rad}
\end{eqnarray}
where $\widetilde{R} = R/\lambda_J^0$. 
The Jeans mass $M_J^0$, Jeans length $\lambda_J^0$, and Mach number at the cloud scale ${\cal M}_*$ are given by
\begin{eqnarray}
M_J^0&=& a_J\,{ c_\sound^3 \over \sqrt{G^3 \overline{\rho}}}, \label{mjeans} \nonumber
\end{eqnarray}
\begin{eqnarray}
\lambda_J^0&=& \left( {a_J \over C_m} \right)^{1\over 3}{c_\sound\over \sqrt {G{\overline \rho}}}, ~~\text{and} \label{ljeans}\nonumber 
\end{eqnarray}
\begin{eqnarray}
{\cal M}_* &=& { 1  \over \sqrt{3} } { V_0  \over c_\sound}\left({\lambda_J^0 \over  1 {\rm pc} }\right) ^{ \eta}.\label{mass_star}
\end{eqnarray}

\subsection{Asymptotic analysis}

\subsubsection{Expected behavior during collapse: two support
regimes}
In the case of a the power-law density PDF, 
we obtain the mass spectrum as a function of mass with Eqs.~(\ref{spec_mass1}), (\ref{PL}), and (\ref{mass_rad}).
To gain further insight, we performed an asymptotic analysis by writing 
\begin{eqnarray}
M \propto R^{1+2 \eta},
\end{eqnarray}
which leads to 
\begin{eqnarray}
\rho \propto M^{(2\eta -2) / (1+2 \eta)}.  
\label{rho_M}
\end{eqnarray} 
If the thermal support is dominant, then $\eta=0$, 
while $\eta \simeq 0.5$ if the turbulent dispersion is dominant.
From Eq.~(\ref{spec_mass1}), we obtain
\begin{eqnarray}
{\cal N}(M) \propto  { \sqrt{\rho}^{-1} \over M^2 } \propto {1 \over M^2 } \exp \left( -{\delta \over 2} \right) \propto M ^{- (1 + 5 \eta) / (1 + 2 \eta)}. 
\label{asymp_pl}
\end{eqnarray}
Therefore, two asymptotic behaviors are expected. 
If the thermal support is dominant, we expect ${\cal N}(M) \propto M^{-1}$, while if the turbulent support 
is dominant, then ${\cal N}(M) \propto M^{-7/4}$ (assuming $\eta=0.5$).  This leads to 
$dN / d \log M \propto M ^0$ and $dN / d \log M \propto M ^{-3/4}$, respectively.

This qualitatively matches the numerical results very well since simulations A1, which have the 
lowest initial density and therefore the highest thermal support, present a flat mass spectrum, that is, $dN / d \log M \propto M^0$, while runs C1 and D1, which have the densest initial conditions and therefore the lowest thermal support, present mass spectra 
entirely compatible with $dN / d \log M \propto M ^{-0.75}$ for $M > 0.1 ~\Ms$. 

If the weighting by the free-fall time were taken into account, 
that is to say, if we were to account for the possibility that several generations of low-mass stars might form while massive stars are forming, 
the mass spectrum would need to be  multiplied by $\tau_\mathrm{ff}^{-1} \propto \rho^{1/2} \propto \exp(\delta/2)$. 
Then we would infer $dN / d \log M \propto M ^{-1}$  , regardless
of the value of $\eta$ and of the relative importance between thermal and turbulent support. 
From our results, this does not seem to be the case.

\subsubsection{Lognormal case}
For comparison, it is worth recalling the asymptotic behaviors that have been inferred for the case of lognormal PDF.
\citet{HC08} inferred (in the region where the PDF is relatively flat, that is to say, where $\delta^2$ 
inside the exponential of Eq.~(\ref{Pr0}) does not dominate) that 
\begin{eqnarray}
{\cal N}(M)  \propto  {\sqrt{\rho} \over M^2}  \propto  M ^{- 3 (1 +  \eta) / (1 + 2 \eta)}, 
\label{asymp_ln}
\end{eqnarray}
which leads to $dN / d \log M \propto M ^{-2}$ if the thermal support dominates the turbulent dispersion and 
$dN / d \log M \propto M ^{-1.25}$ otherwise. 
Taking into account the weighting by the free-fall time would 
modify these conclusions. \citet{HC13} inferred that 
\begin{eqnarray}
{\cal N}(M)  \propto  { \rho \over M^2}  \propto  M ^{- (4 +  2 \eta) / (1 + 2 \eta)}, 
\label{asymp_ln2}
\end{eqnarray}
leading to to $dN / d \log M \propto M^{-3}$ if the thermal support dominates and 
$dN / d \log M \propto M^{-1.5}$ if the turbulent support dominates. \\ 

It might be surprising at first  that  for a lognormal PDF, thermal support 
leads to spectra that are stiffer than that with turbulent support,
while the reverse is true for the power-law PDF $\propto \rho^{-1.5}$. 
The reason lies in the dependence of ${\cal N}$ on $\rho$ (see Eqs.~(\ref{asymp_pl}),~(\ref{asymp_ln}), and ~(\ref{asymp_ln2}))
and the density-mass relation inferred from the virial theorem (Eq.~(\ref{rho_M})). 
If the PDF decreases too stiffly
with density, the dense material is rare, and since for a specific clump mass, thermal support leads to 
clumps more diffuse than turbulent dispersion, thermal support tends to make larger clumps. 
In this respect, the critical exponent 
for the PDF is -1 if free-fall time is not taken into account and -1.5 otherwise.

\subsection{Comparison with the simulation results}
\setlength{\unitlength}{1cm}
\begin{figure*}[]
\begin{picture} (0,6.7)
\put(-0.2,0){\includegraphics[width=4.8cm]{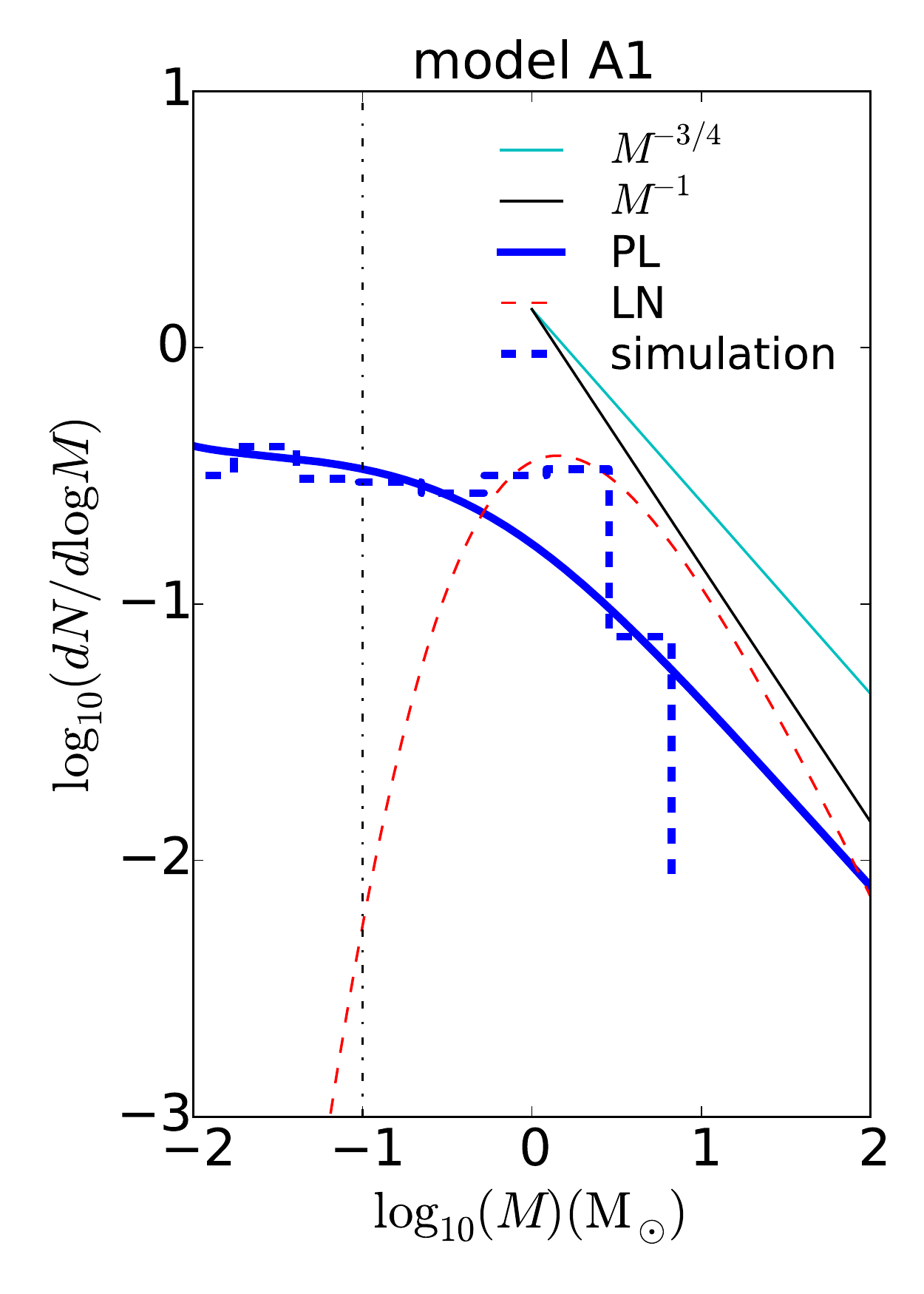}}
\put(4.4,0){\includegraphics[width=4.8cm]{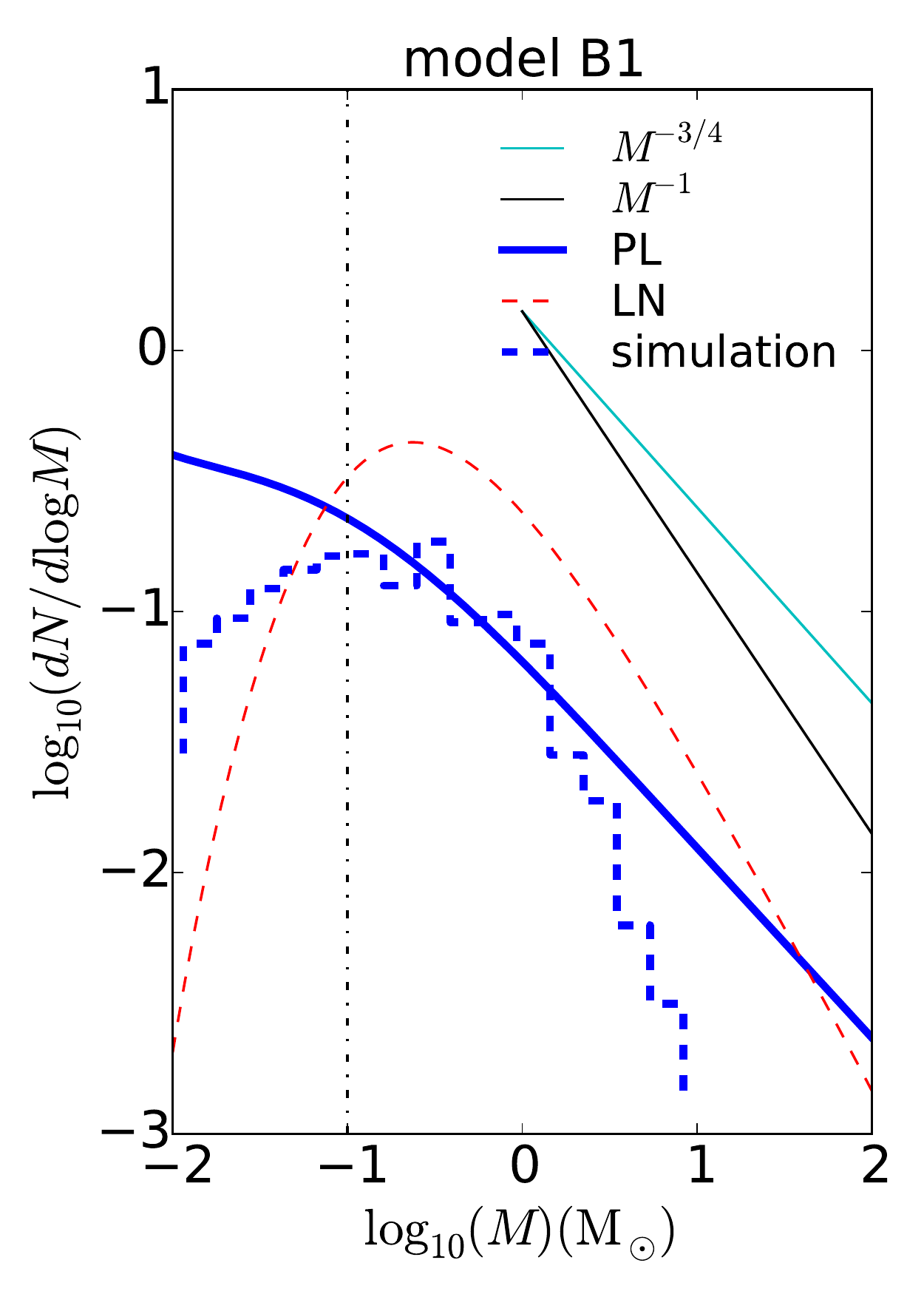}}
\put(9.,0){\includegraphics[width=4.8cm]{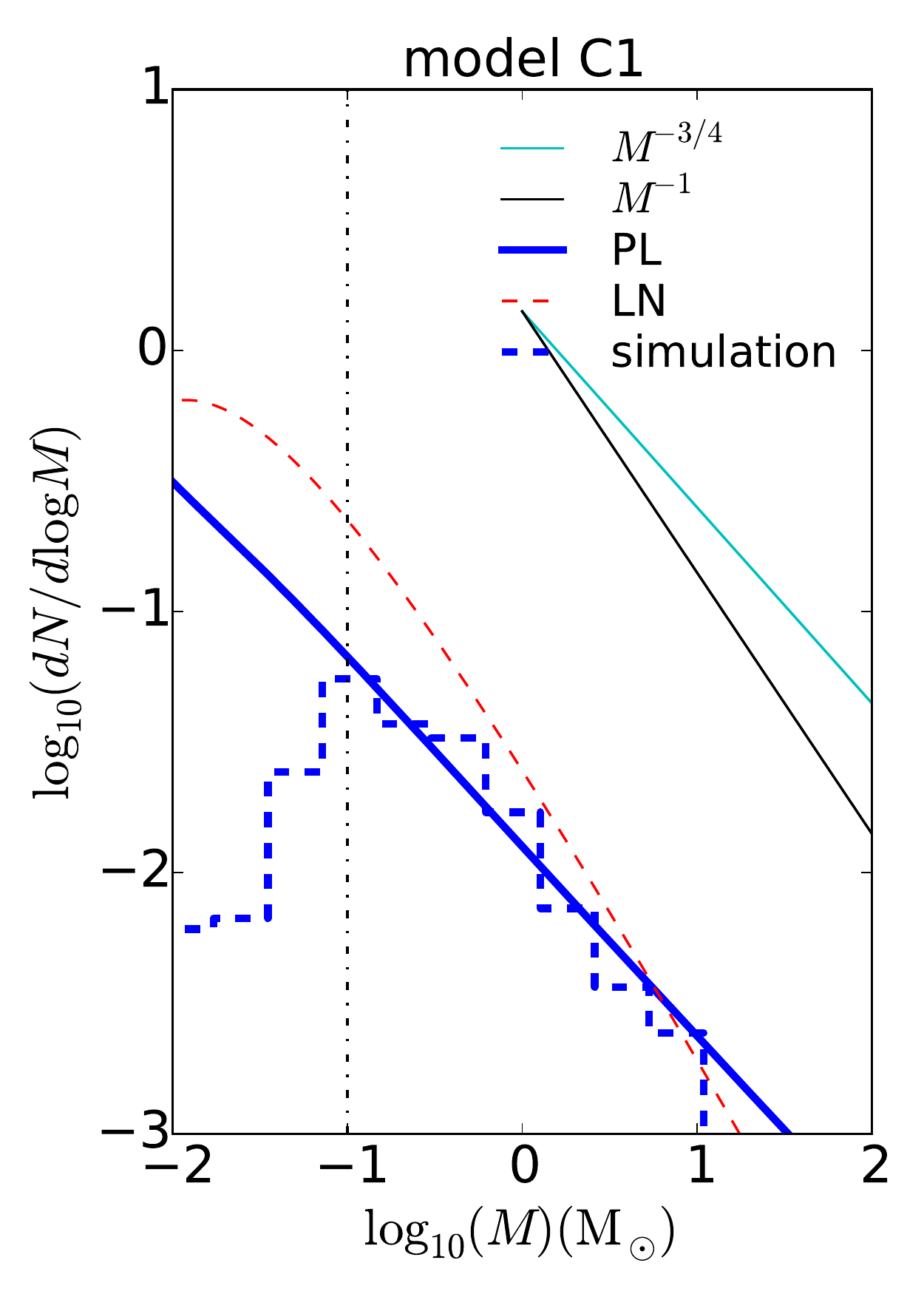}}
\put(13.6,0){\includegraphics[width=4.8cm]{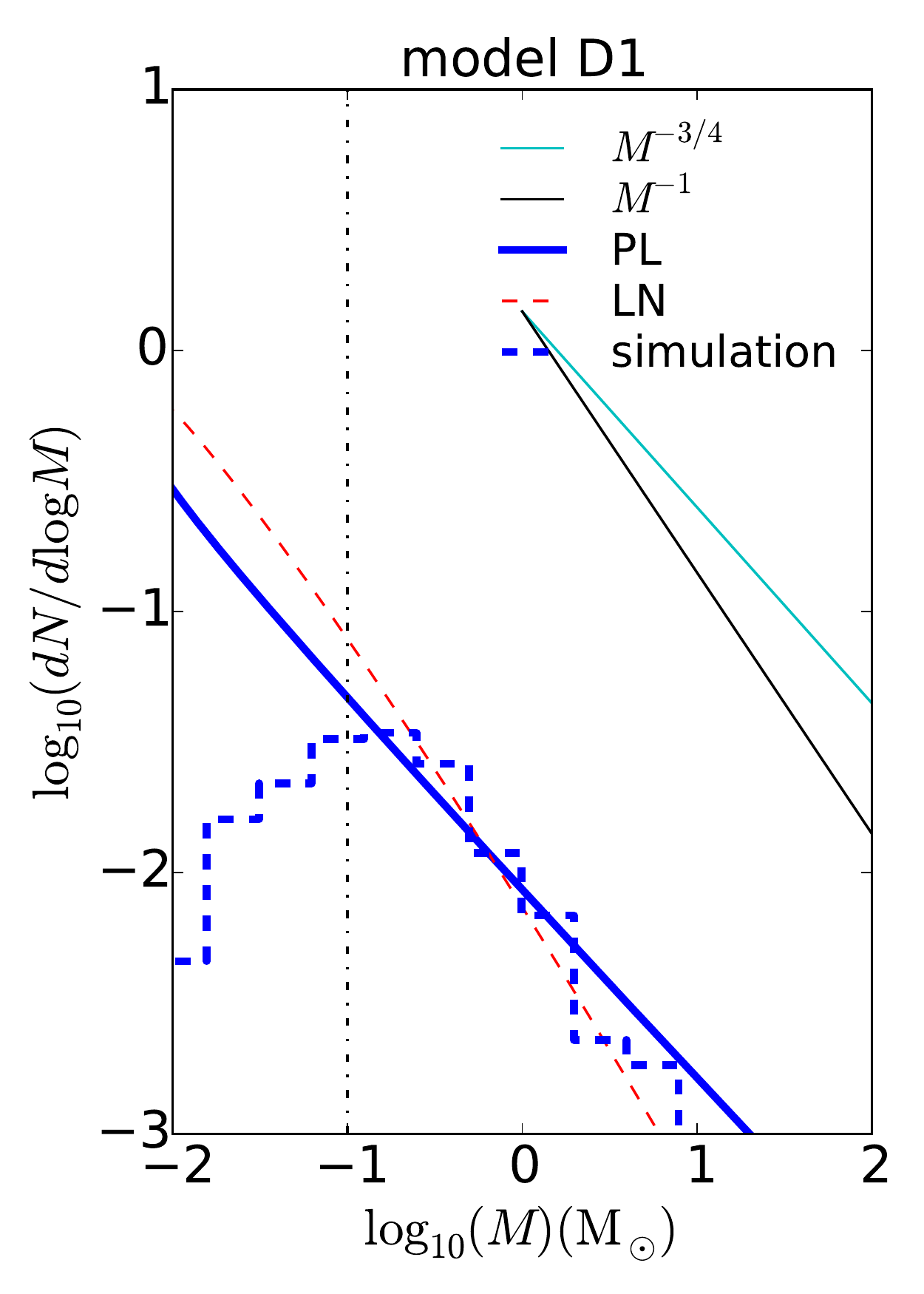}}
\end{picture}
\caption{Mass spectra as predicted from analytical model for four sets of 
parameters corresponding to runs A1++, B1++, C1+, and D1. Simulation results at $150 ~\Ms $ accreted (blue dashed, same as the magenta histogram in Fig. \ref{fig_SMF_density} with absolute value shifted to compare to models), model with power-law (blue solid) and lognormal (red dashed) density PDF are plotted, as well as the power-law relation (-3/4 in cyan and -1 in black) for reference. The vertical dot-dashed line indicates a mass lower limit above which the model is applicable (see paper II).}
\label{model}
\end{figure*}

To compare the simulation results with the analytical predictions, we need to specify a mean density 
and a Mach number. For the first, we use a radius, $R_{\rm c}$, 30 percent larger than the radius 
indicated in Table~\ref{table_compactness} since this latter value corresponds to the radius before the 
relaxation phase. A velocity dispersion also has to be specified in order to derive the Mach number.
A difficulty arises because a substantial part of the motions is due to the collapse itself,
and it is not straightforward to distinguish infall and {\it \textup{turbulence}}. For the cloud velocity dispersion, 
we take $\sigma _\mathrm{c} = \sqrt{G M / (2R_\mathrm{c})} $. For a virialized cloud, we may have chosen 
$\sqrt{G M / R_\mathrm{c}} $, while infall should be excluded for our present purpose of estimating the turbulent dispersion. 
This leads to the normalized velocity dispersion at 1 pc
\begin{eqnarray}
V_0= \sigma_\mathrm{c} ( R_\mathrm{c} / 1 \, {\rm pc} )^{-\eta}.
\label{V_disp}
\end{eqnarray}

\subsubsection{Comparison of the accretion times}
\label{accret_time}
We first compare the accretion time-scale displayed in Fig.~\ref{fig_timescale} and the prediction of the formalism. 
For this purpose we use Eqs.~(\ref{masse}) and (\ref{mass_rad}). 
Since the free-fall time is given by $\tau_\mathrm{ff} = \sqrt{3 \pi / (32 \rho G)}$, we have a link between the 
accretion time, assumed to be the free-fall time, and the mass. Before a quantitative comparison, it is enlightening 
to consider the asymptotic behavior. Writing again $M \propto R^{1+2 \eta}$, we obtain 
\begin{eqnarray}
\tau_\mathrm{ff} \propto M^{1 - \eta \over 1 + 2 \eta}. 
\end{eqnarray}
Two regimes appear: if thermal support dominates turbulent dispersion, we obtain
$M \propto \tau_\mathrm{ff}$, while $M \propto \tau_\mathrm{ff}^4$ is obtained with $\eta=0.5$ in the opposite case (${\cal M}_* \widetilde{R}^{1+2 \eta} > 1 $). This qualitatively matches 
 the behaviors observed in Fig.~\ref{fig_timescale} very well
since for run A1++, $M \propto t_\mathrm{acc,60}$
, while for run C1+ and D1, a much stiffer relation is obtained. 

Using the complete Eq.~(\ref{mass_rad}), while applying $1.3~R_\mathrm{c}$ and $V_0$ from Eq. (\ref{V_disp}), we plot $M$ against $0.6~\tau_\mathrm{ff}$ for the 
four runs in Fig.~\ref{fig_timescale} (solid black lines). 
It is justified with a $\rho \propto r^{-2}$ density profile that the fraction of accreted mass is proportional to the collapse time. 
Sinks of mass close or below 0.1 $\Ms$ are strongly affected by the adiabatic equation 
of state, as described in paper II, and the physics that governs their formation and timescale is 
not the one described by the present analytical model. Therefore, the present comparison is meaningful 
only for masses $\gtrsim 0.1~\Ms$. With this in mind, the model provides a very reasonable 
fit for the four runs at masses $\gtrsim 0.1 \Ms$.

The stiff dependence of the mass on the accretion time for run C1+ and D1 is well 
described by the analytical model. This is particularly interesting because this behavior 
is directly linked to the turbulent dispersion (which in a loosely sense can be called turbulent \textup{{\it \textup{support}})}. 
This implies that the origin of the massive stars is indeed a consequence of the coherence induced by the support, which can be either 
thermal or turbulent,  and is not due to purely competitive accretion.

\subsubsection{Comparison of the mass spectra}
The analytical prediction is obtained by combining Eqs.~(\ref{spec_mass1}),~(\ref{PL}) and~(\ref{mass_rad}) with 
$V_0$ given by Eq.~(\ref{V_disp}). For the sake of comparison, we also computed the predicted mass spectra employing a
 lognormal density PDF. 
Figure~\ref{model} shows the results for four runs. Mass spectra at the time when the total sink mass equals to $150~\Ms$ were used for comparison. 
Since the shape of the mass spectrum varies little with time, we normalized 
the number of objects from the simulations to match the number predicted by the analytical model. 
The analytical model with the power-law density PDF 
stated by Eq.~(\ref{PL}) and the mass spectra inferred from simulations, solid and dashed blue lines, respectively, 
generally agree pretty well for $M > 0.1~\Ms$, where the analytical models are applicable 
(see paper II), except for models B1, C1, and D1 at high mass, where the lognormal PDF seems to provide better fits. 
The changing slope of the mass spectrum around $\simeq 1~\Ms$ 
 may be a consequence of the PDF, which  shows a transition between a lognormal behavior at low density
and a power law at high density (see Fig.~\ref{fig_pdf}). 

To conclude, simulations and analytical models both predict the power-law change from $d N/ d\log M \propto M^0$ in the regime dominated by thermal energy (model A1) to $d N/ d\log M \propto M^{-3/4}$ in the turbulence-dominated regime (model C1 and D1), without any adjustment of free parameters.

%------------------DISCUSSIONS------------------------
\section{Discussions}
\begin{table}[t!]
\caption{Simulation parameters from the literature: molecular cloud mass, particle number density, and Mach number.}
\label{table_literature}
\centering
\begin{tabular}{l r r c}
\hline\hline
Label   & $M_\mathrm{tot}$ ($\Ms$) & $n$ ($\cc$) & $\mathcal{M}$ \\%& $\Gamma$ & thermodynamics \\
\hline
%\citet{Bonnell03} &$10^3$  & $3.4 \times 10^4$ & 10 \\%& -0.75 & isothermal \\
%\citet{Jappsen05} &120 & $8.4 \times 10^4$ & 3.2 \\%& & polytrope \\
%\citet{Bonnell08,Bonnell11} &$10^4$ & $3.3 \times 10^3$ & 23 \\%& &
%\citet{Bate09a} & 500 & $3.0 \times 10^4$ & 13.7 \\%& -0.75 & polytrope \\
%\citet{Krumholz11} &$10^3$ & $2.4 \times 10^5$ & 15 \\
%\citet{Girichidis11} &100 & $4.6 \times 10^5$ & 3.5 \\
%\citet{Bate12} & 500 & $3.0 \times 10^4$ & 13.7\\% & -0.75 & radiation \\
%\citet{BallesterosParedes15} & $10^3$ & $1.7 \times 10^4$  & 4, 8, 16\\
%\citet{bertelli2016} & 5750/516  & 100/330 & 1-20 \\
%\citet{liptai2017} & 50 & $6.1 \times 10^4$ & 6.4 \\
\!\!\citet{Bonnell03} &$1000$  & $34000$ & 10 \\%& -0.75 & isothermal \\
\!\!\citet{Jappsen05} &120 & $84000$ & 3.2 \\%& & polytrope \\
\!\!\citet{Bonnell08,Bonnell11} &$10000$ & $3300$ & 23 \\%& &
\!\!\citet{Bate09a} & 500 & $30000$ & 13.7 \\%& -0.75 & polytrope \\
\!\!\citet{Krumholz11} &$1000$ & $240000$ & 15 \\
\!\!\citet{Girichidis11} &100 & $460000$ & 3.5 \\
\!\!\citet{Bate12} & 500 & $30000$ & 13.7\\% & -0.75 & radiation \\
\!\!\citet{BallesterosParedes15} \!\!\!\!\!\!\!\!\! & $1000$ & $17000$  & \!4/8/16\!\\
\!\!\citet{bertelli2016} & \!\!\!5750/516  & 100/330 & 1-20 \\
\!\!\citet{liptai2017} & 50 & $61000$ & 6.4 \\
\hline
\end{tabular}
\end{table}

\subsection{Comparison with previous works}
Through the series of simulations A1 through D1 with increasing initial density,  
we have demonstrated that there are indeed two regimes of stellar distribution formation in massive collapsing clouds. 
First, the stellar distribution is flat in the low-density regime, where thermal energy is the dominant support against self-gravity. 
Second, $dN /d\log M \propto M^{-3/4}$ in the high-density regime, where turbulence is dominant. 
Alternatively, varying the turbulence strength in runs C1 at fixed density also shows this effect: 
the stellar distribution becomes flat as the initial turbulent energy is lowered. 
Table \ref{table_literature} lists a summary of simulation parameters from the literature and the regimes investigated in these works. 

Recently, two studies by \citet{bertelli2016} and \citet{liptai2017} concluded 
that in their numerical experiments, the mass spectrum 
 is not influenced by the initial turbulence, which is expected to have an effect on the IMF, according 
to the reservoir-based theories proposed by \citet{HC08}, \citet{schmidt2010} and \citet{hopkins2012}. 
Our results are partly compatible with their conclusion because we see from runs C1, C1t15, and C1t03 that varying the turbulence by a factor 5 does not 
lead to very significant change in mass spectrum. 
The reason is that self-gravity triggers the development of a power-law tail
that is not considered by \citet{HC08} and \citet{hopkins2012}. When the density PDF relevant for gravitational collapse
is considered, the comparison between the reservoir-based theory and the simulation results becomes very good for 
$M > 0.1~\Ms$ (near and below this value, the physics to be considered is different, as described in paper II), 
as shown in Fig.~\ref{model}. However, the mass spectra produced in our numerical 
simulations tend to be {\it \textup{too shallow}} compared to the Salpeter exponent. 
Therefore, while turbulence could possibly be less important 
in setting the low-mass part of the IMF than originally proposed by \citet{padoan1997}, \citet{HC08}, and  \citet{hopkins2012}, it is likely playing a determining role for the massive stars.

A numerical experiment comparable to ours has been conducted by  \citet{BallesterosParedes15}, 
who simulated a $1000~\Ms$ cloud at density $1.7 \times 10^4~\cc$ with Mach number varying between 4 and 16. 
The sink mass spectrum from their study is rather flat at the beginning, 
while a power law of $\Gamma \sim -1$ develops very quickly and there is no obvious difference between different initial conditions. 
Their conditions are broadly compatible with our runs B1 (or probably would stand between our runs B1 and C1). Given that the slope that they inferred is not very accurately determined, $\Gamma \simeq -1$ is close enough to the value $-3/4$ we that obtain, 
especially because for a high mass, we obtain some slightly steeper mass spectra (see Fig.~\ref{model}). 

Another similar set of parameters has been used by \citet{Bate09a}, which again falls between runs B1 and C1. 
From Fig.~3 of \citet{Bate09a}, the mass spectrum between 0.1 and 1 $\Ms$ seems to be 
shallower than the Salpeter exponent of $-1.35$ and compatible with $\Gamma=-0.75$. 
For higher masses, the statistics are less significant, but broadly compatible with $\Gamma \simeq -1$.

Furthermore, \citet{Girichidis11} also find mass spectra that are compatible with $\Gamma$
larger than or near -1 (particularly run TH-m-1 in their Fig.~8). 
Since they consider $100~ \Ms$, it is not easy to compare their
results with our runs, however. 

Finally, one important question is why the regime $\Gamma \simeq 0$ has not been reported in the works
listed in Table~\ref{table_literature}. The conditions for $\Gamma \simeq 0$ are not typical 
of present-day molecular clouds. However, this regime may have been observed in the context of primordial 
stars. The collapse of minihaloes, thought to be the progenitors of Pop III stellar systems, has
been studied and flat mass spectra have been inferred with $\Gamma \simeq -0.2$ \citep{stacy2013,stacy2016}.
Since the temperature of the gas in these minihaloes is on the order of 1000 K, the thermal support is 
much higher.

\subsection{Astrophysical interpretation: competitive accretion and mass reservoir}
A long-standing debate is concerned with the way stars acquire their mass, and  based 
on collapse calculations similar to those performed in this work, the role 
played by accretion has been emphasized, leading to the model named 
{\it \textup{competitive accretion}} \citep[e.g.,][]{bonnell2001,bonnell2004}. The underlying idea
is that massive stars tend to attract gas through their gravity and might accrete as $dM/dt \propto M^2$. 

Another type of models has been emphasizing the role of the mass reservoir
\citep{padoan1997,HC08,hopkins2012}. The fundamental difference with the 
competitive accretion scenario, which is stochastic in nature, is that the 
mass of the reservoir largely determines that of the star. 
Therefore, the mass of the stars is determined before their formation. The analytical model 
developed in this paper, identical  to the model proposed by \citet{HC08}
except for the density PDF, belongs to this category. The mass reservoir that predetermines
the mass of the star is a direct consequence of the thermal support and turbulent dispersion. 
The comparisons between our simulations and the analytical model (Figs.~\ref{fig_timescale} and~\ref{model})
show very good agreement (we recall that no free parameter had to be adjusted). 
This reinforces the role of the mass reservoir and suggests that it may be dominant. 

This does not mean that the 
conversion from the gas reservoir to the star is entirely deterministic and that there is no 
competition at all. The timescales displayed in Fig.~\ref{fig_timescale}
present significant dispersions as a consequence of all fluctuations in the 
system, which certainly include some kind of competitive accretion. The mean values and the 
general trends, however, can be well reproduced by a deterministic model that emphasizes
the role of a preexisting coherent reservoir of gas. Another important aspect is the 
choice of which particular star will accrete most of the mass of the reservoir in which it is 
embedded. It is likely the case that inside a turbulence-dominated reservoir, several thermally dominated 
reservoirs can be identified, leading to several low-mass stars that may eventually accrete the mass of the larger 
turbulence-dominated reservoir and become massive stars. 

It is worth stressing that massive stars form in the same way as 
low-mass stars. They simply result from larger reservoirs. It is true, however, that 
in many circumstances, the reservoir of the massive stars will be predominantly determined
by the turbulent dispersion (the velocity dispersion appearing in the virial criterion expressed by 
Eq.~(\ref{cond_tot})) instead of by the thermal support. Since turbulence presents many fluctuations, 
this reservoir is less clearly determined than the reservoir dominated by the thermal 
support.

\subsection{Different regimes leading to the IMF}
The results presented in this paper and its companion suggest that the IMF 
is a consequence of several different physical regimes. Here we try to clarify the 
situation by introducing some terminology and classification. 
Unfortunately, a difficulty arises because two independent types of processes play a
role and complicate the situation. 

First of all, we have to distinguish between the 
regimes for which the thermal support is dominant and the regime for which the turbulent dispersion dominates. Mathematically speaking, this is the respective amplitude of the two terms 
appearing in the right-hand side of Eq.~(\ref{crit_Mtot}). 
Second, 
there are also three {\it \textup{density regimes}} that need to be distinguished. They  
 broadly correspond to 
three ranges of mass, the most massive stars, intermediate-mass
stars, and
low-mass stars, or density regimes I, II, and III. Regime I corresponds
to the lognormal part of the density distribution, regime II the power-law density distribution, 
and regime III to the density at which the gas  is  adiabatic (formation of the first Larson core). 
Therefore,  density regimes I and  II
must be subdivided into thermally and turbulence-dominated regimes, or regimes Ia, IIa, and Ib, IIb. 
Below, {\it \textup{the term density regime}} refers to regimes
I, II, or III, while the term {\it \textup{regime}} refers to 
regimes Ia, Ib, IIa, IIb, or III (i.e., a combination of the density PDF and support against gravity).

Density regimes I and II 
 have in common that they present a preexisting deterministic mass reservoir that is due  either to thermal support or to turbulence dispersion and 
 gravity. Since  the density PDF of  density regime I is set by turbulence, while the density PDF of  density regime II  is due to collapse and  gravity,
 they are turbulently and  gravitationally driven, respectively. 
Regime III is  different and is described in paper II. It relies on the physics 
of the so-called first Larson core, which is related to the transition from isothermal
to adiabatic phase and to the density and temperature at which the second collapse
 occurs. In general, there is no clear relation to a pre-existing coherent mass reservoir in density regime III.
We stress that simulations A1 essentially present regimes IIa and III. 
Simulations C1 and D1 are essentially IIb and III with a possible transition 
to regime Ib at high masses. Simulation B1 is likely to represent regimes IIb, IIa, and III. 

The existence of these different density regimes may be a strong clue to understanding the apparent universality 
of the IMF \citep{Bastian10}. Regime III is largely universal, except in terms of metallicity, which may affect the 
properties of the first Larson core \citep{Masunaga99}. Regime Ib, in which the lognormal density PDF is 
set by turbulence and turbulent dispersion dominates thermal support, is likely 
the relevant regime (at least in the present-day Universe, and it could be different for Pop III, see \citet{stacy2013}) 
in a wide range of conditions and is also expected 
to be universal because gravity and turbulence are universal processes. The transition between density regimes I and II and  the existence of regimes Ia and IIa is not universal, however. 
Depending on the local density and Mach numbers, 
the masses for which the reservoir is built from a lognormal density PDF and those for which it is built from a power-law density 
PDF may vary. However, since the difference between the mass spectra expected in regimes Ib and IIb is small, they may not 
be easy to  distinguish observationally, especially because the statistics on a particular cluster usually present some noise. 

Finally, we stress that the existence of these various regimes is compatible with observations. For example, 
 in Fig.~11 of \citet{alves2013}, the mass spectrum is compatible with a cut at a mass lower than 
$0.1 ~\Ms$ (regime III), a power law with $\Gamma \simeq -3/4$ between 0.1 an 1 $\Ms$ (regime IIb), and then a power law with a Salpeter 
exponent at higher mass (regime Ib).

%------------------CONCLUSIONS------------------------
\section{Conclusions}
We investigated the fragmentation of a collapsing 1000 $\Ms$ clump 
and the resulting stellar mass distribution. 
In this paper we focused on the power-law part at the high-mass end of the distribution, which is related to 
the gas isothermal phase, while in paper II, we investigated the origin
of the peak of the distribution, which is is a consequence of the adiabatic eos at high density. 
We performed a series of numerical simulations in which we varied the initial density by four orders of magnitude and 
the turbulence energy by two orders of magnitude. We verified numerical convergence by 
systematically varying the resolution.  
To assess the physical processes at play in the simulations, 
we developed an analytical model that we compared to the simulation results. The model 
was adapted from \cite{HC08} and considered a power-law density PDF that was appropriated for a collapsing 
clump instead of a lognormal one, valid for a non-collapsing turbulent medium. 

We found two different regimes. When thermal support is dominant, 
it is inferred that $d N / d\log M \propto M ^ 0$, 
while when the local turbulent dispersion  dominates,
we obtain $d N / d\log M \propto M^{-3/4}$. As explained in paper II,
this is valid for masses higher than about ten times the mass of the first Larson core, $M_\mathrm{L}$, around which 
the thermodynamics and the tidal forces due to the first Larson core are essential. 
For masses lower than a few $M_\mathrm{L}$ in our simulations, the mass spectrum starts to decrease with the mass, 
except for the most diffuse clouds, where disk fragmentation leads to the formation 
of objects down to a mass of $M_\mathrm{L}$, that is, a few $10^{-2} ~\Ms$. 
We stress, however, that the physics included in the present simulations is too simplistic regarding disk formation and fragmentation. 

While the mass spectra of the  densest clouds qualitatively resemble the 
observed IMF, for masses higher than 0.1 $\Ms$ they
exhibit an exponent that is shallower
than the Salpeter exponent of $-1.35$. Nonetheless, we observe a possible transition 
toward a slightly steeper value that is broadly compatible with the 
Salpeter exponent for masses above a few solar masses. This change in behavior is associated 
with the change in density PDF, which switches from a power-law distribution to
a lognormal. Therefore, our results suggest that while gravitationally induced fragmentation may be  important 
 for low masses,  turbulently induced fragmentation is likely responsible for setting up the 
IMF above a few solar masses.

\begin{acknowledgements}
The authors thank the anonymous referee for a report that improved the manuscript.
We thank Gilles Chabrier for a critical reading of the manuscript
that improved the clarity of the paper significantly.
This work was granted access to HPC
   resources of CINES under the allocation x2014047023 made by GENCI (Grand
   Equipement National de Calcul Intensif). 
   This research has received funding from the European Research Council under
   the European Community's Seventh Framework Programme (FP7/2007-2013 Grant  Agreement no. 306483).
\end{acknowledgements}

\appendix
%------------------RELAXATION RUNS------------------------
\section{Influence of the initial fluctuations}
\label{appen_r}

\setlength{\unitlength}{1cm}
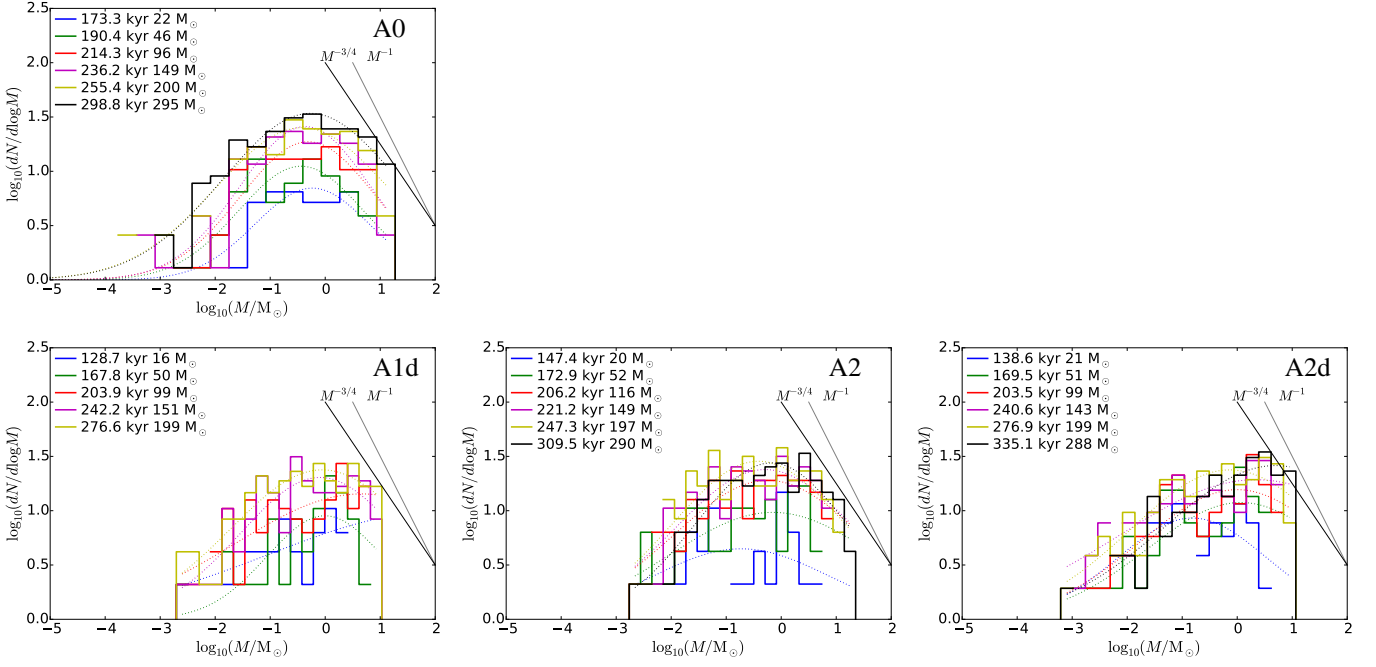
\begin{figure*}[]
\begin{picture} (0,9)
\put(0,4.5){\includegraphics[width=6cm]{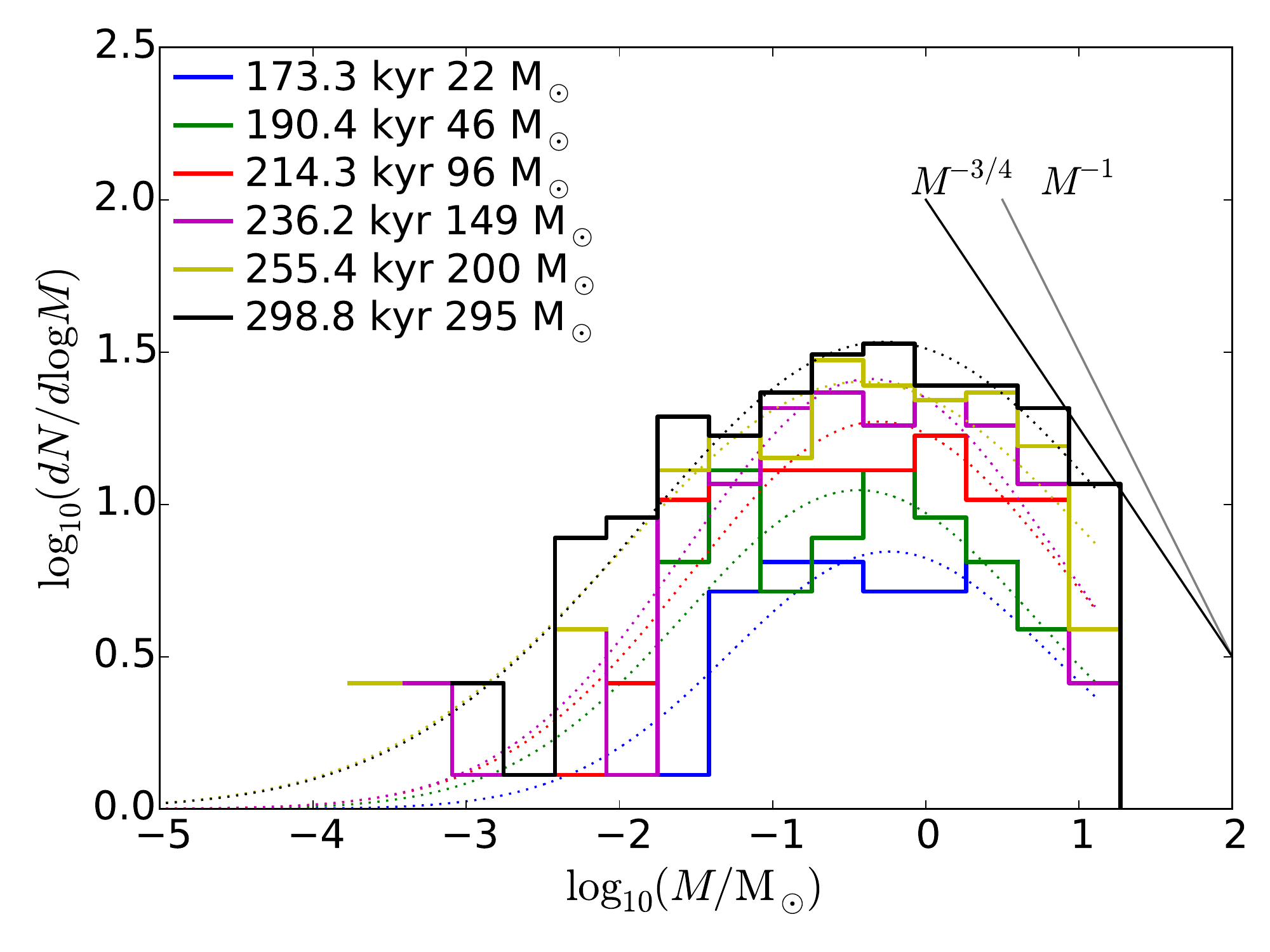}}
\put(0,0){\includegraphics[width=6cm]{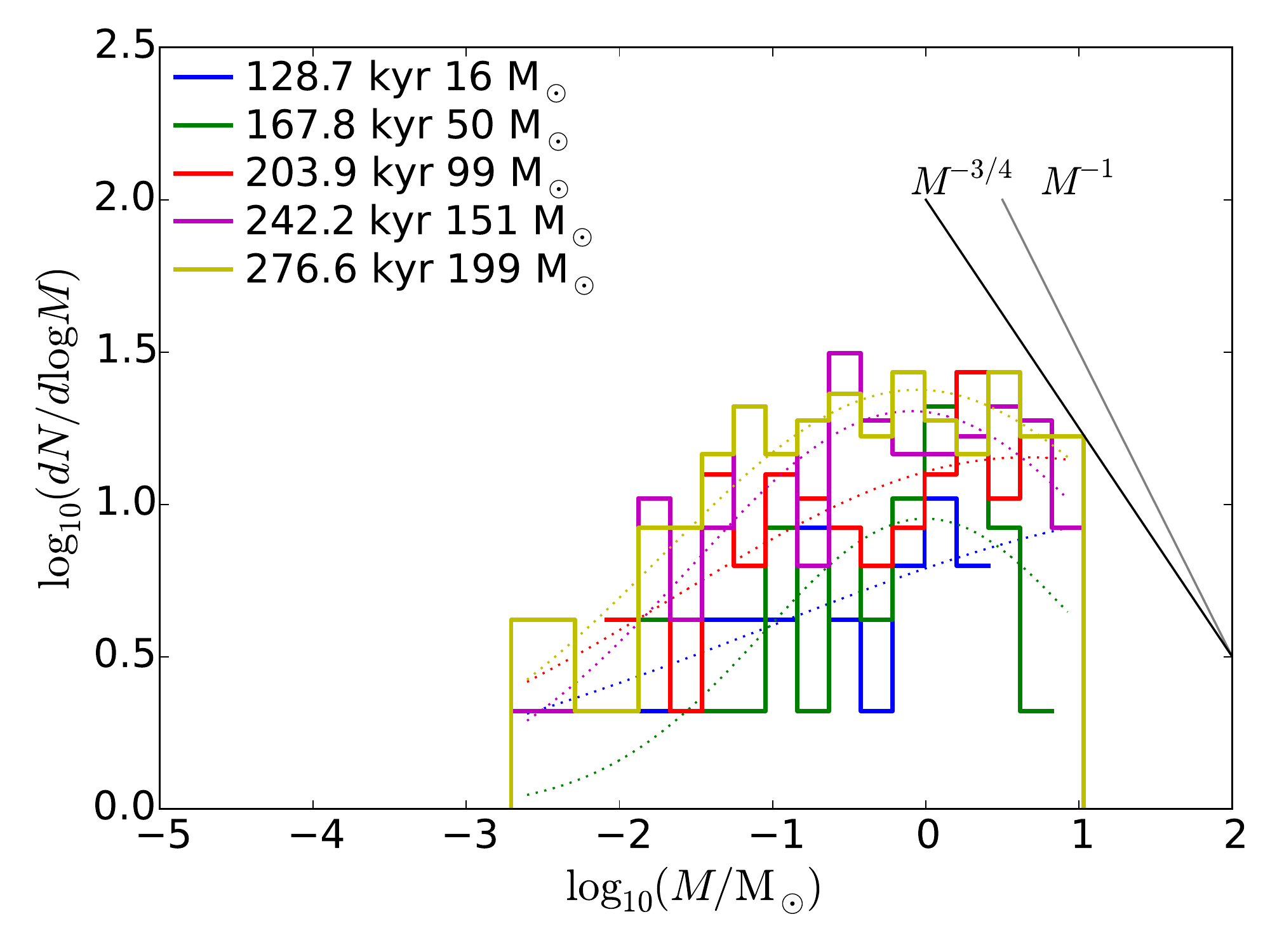}}
\put(6,0){\includegraphics[width=6cm]{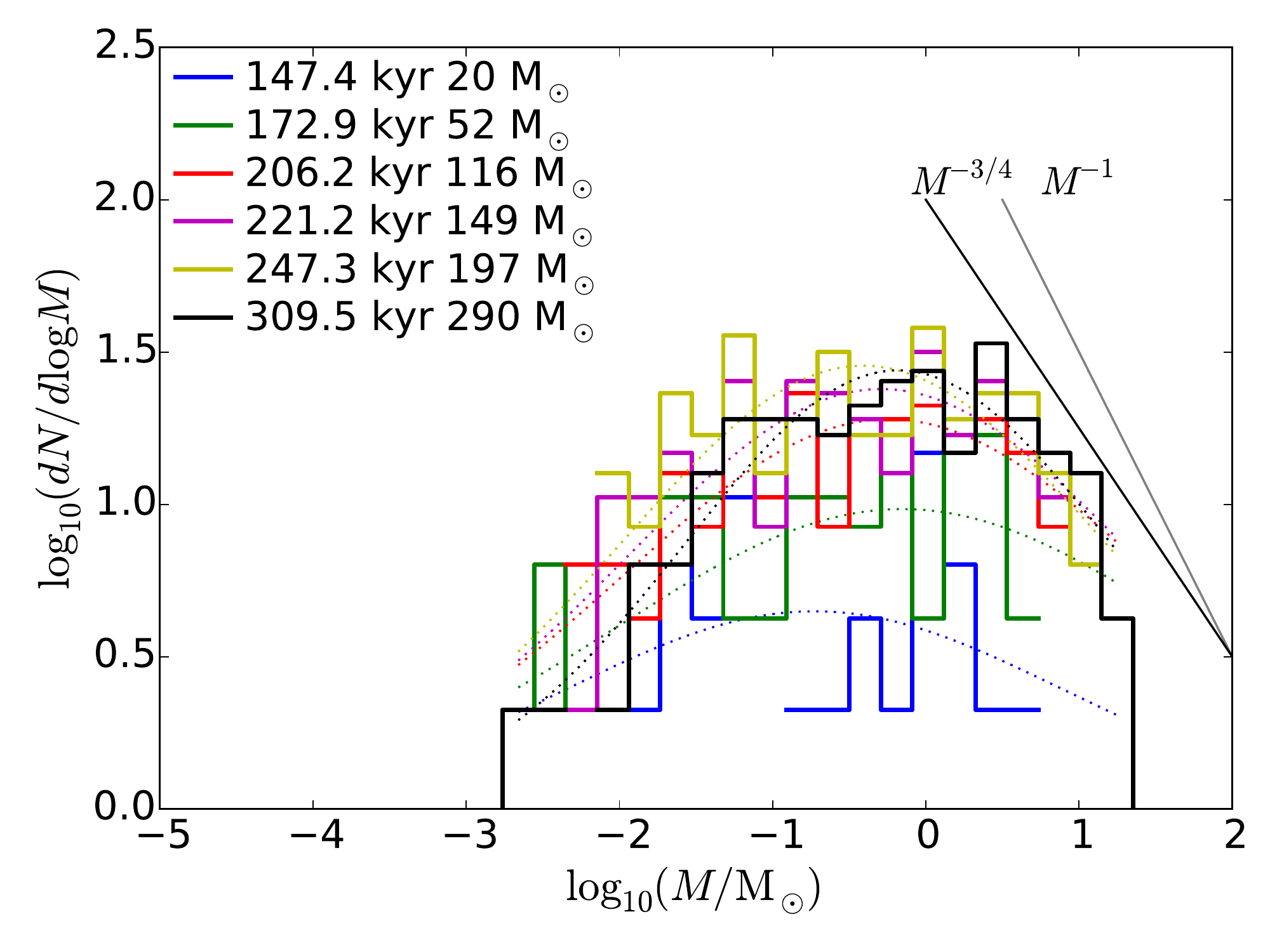}}
\put(12,0){\includegraphics[width=6cm]{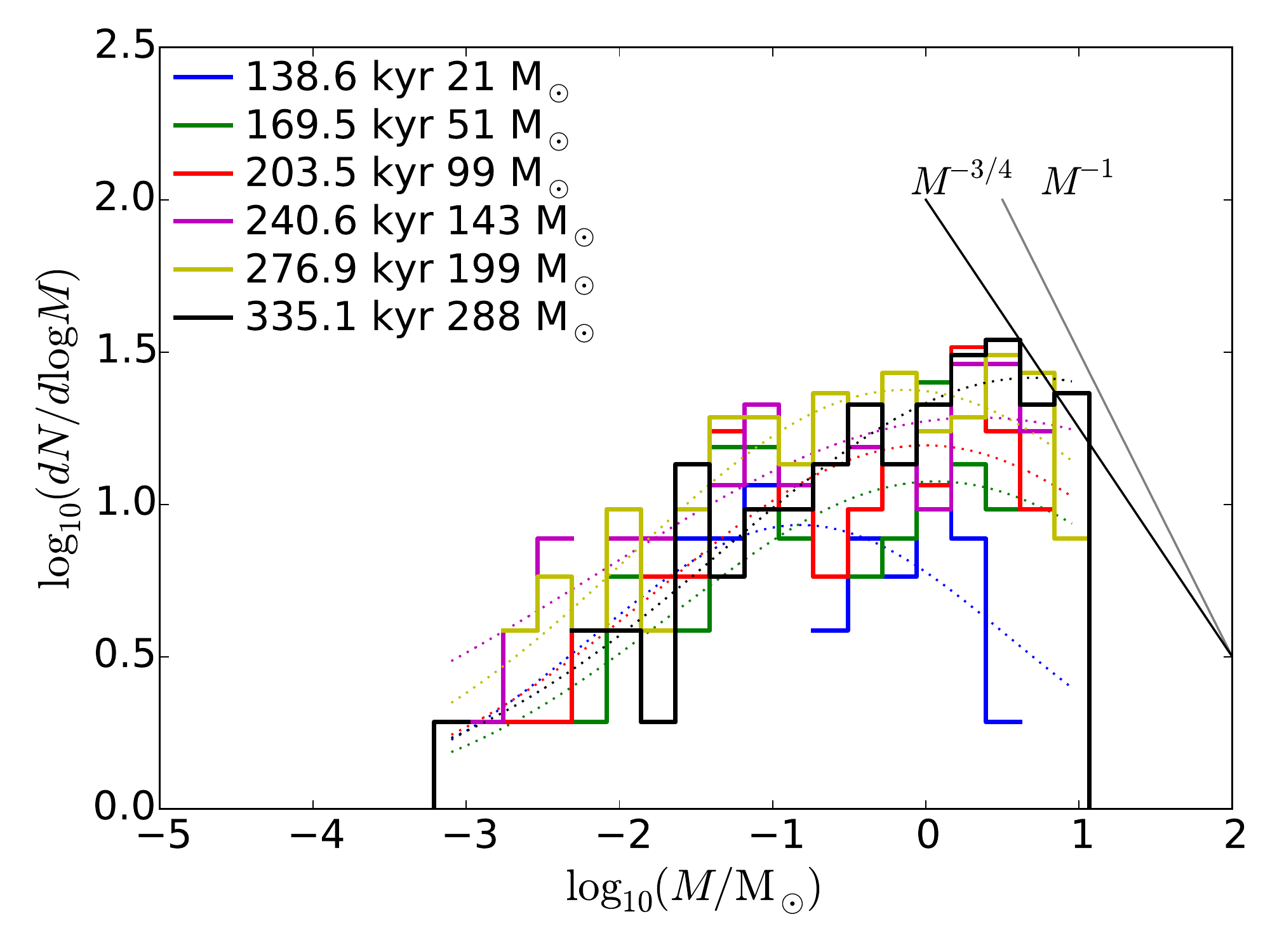}}
\put(5,8.4){A0}
\put(5,3.9){A1d}
\put(11,3.9){A2}
\put(17,3.9){A2d}
\end{picture}
\caption{Mass spectra of runs A0, A1d, A2, and A2d, with the
same color coding as in Fig. \ref{fig_SMF_density}.  The cloud
with a radius of 0.75 pc is evolved without gravity on levels 8 or 9 during 0, 0.3, and 0.6 turbulence-crossing times. With longer relaxation, more massive stars develop, and the IMF is slightly more top-heavy.}
\label{fig_IMF_A}
\end{figure*}

\setlength{\unitlength}{1cm}
\begin{figure*}[]
\begin{picture} (0,4.5)
\put(12,0){\includegraphics[width=6cm]{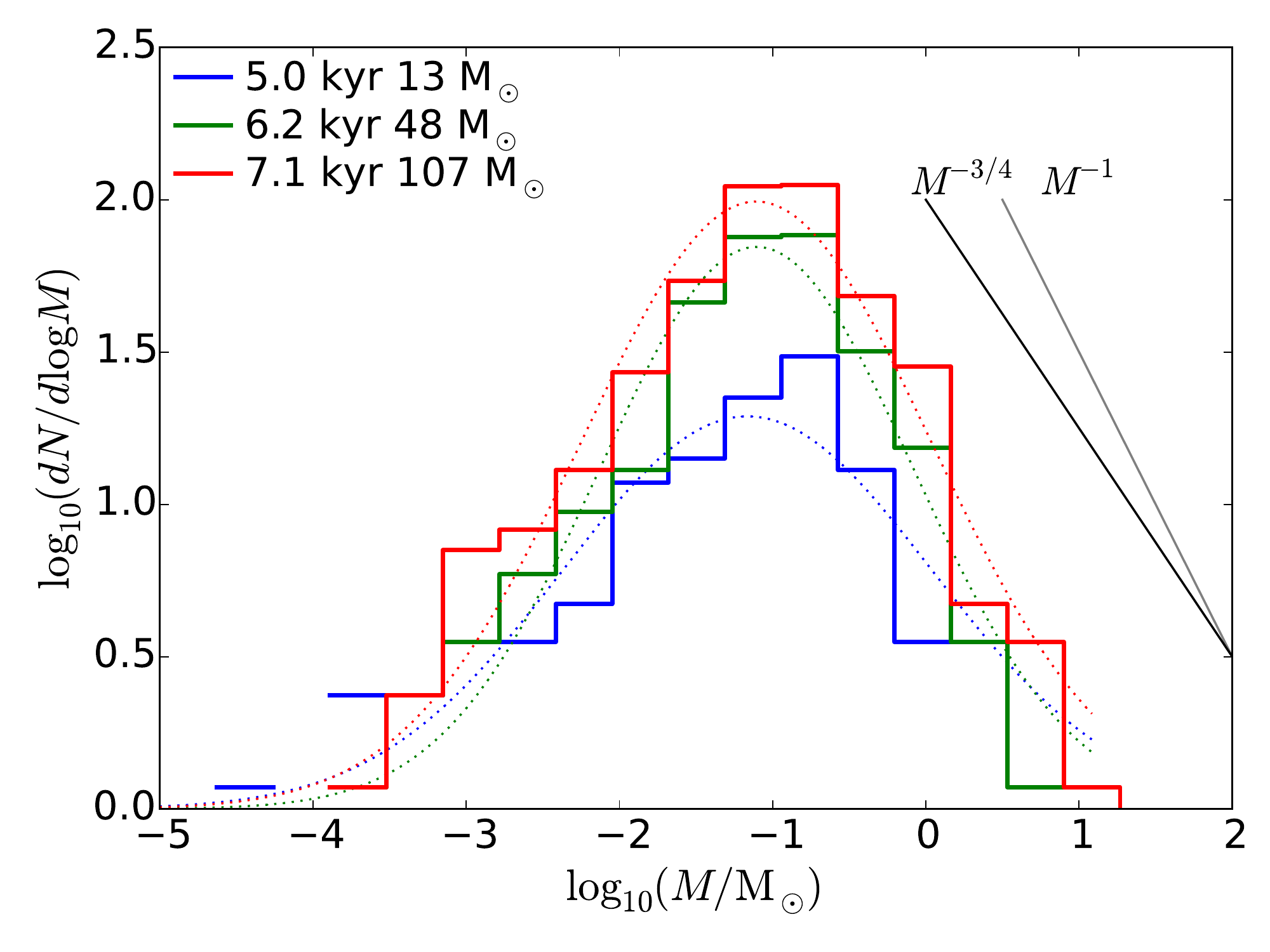}}
\put(6,0){\includegraphics[width=6cm]{C1_14_SMF.pdf}}
\put(0,0){\includegraphics[width=6cm]{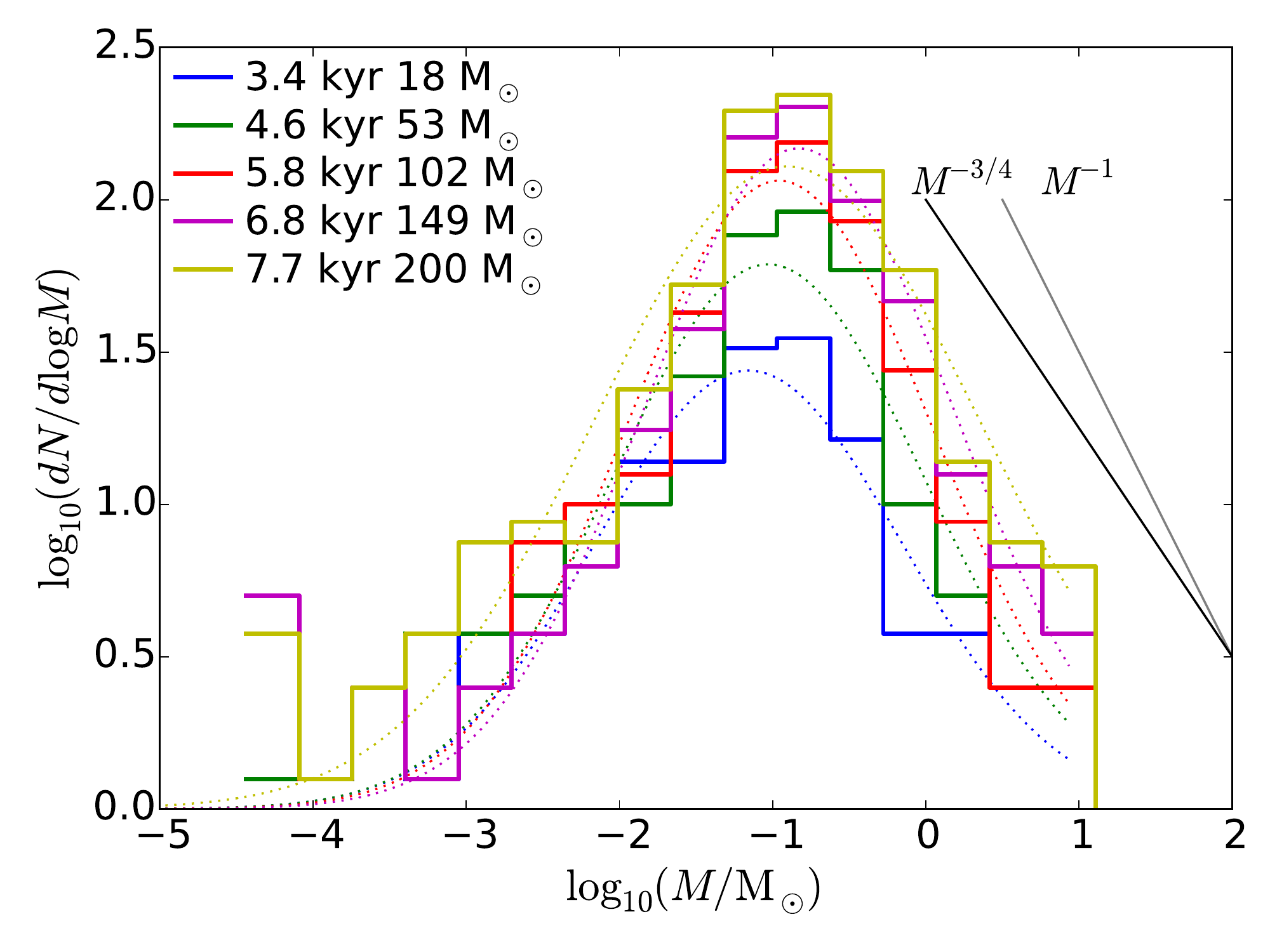}}
\put(5,3.9){C0}
\put(11,3.9){C1}
\put(17,3.9){C2}
\end{picture}
\caption{Mass spectra of runs C0, C1, and C2 with the same color coding as in Fig. \ref{fig_SMF_density}. The cloud with a radius
of 0.084 pc evolved directly without relaxation or evolved without gravity on levels 8 and 9 during 0.3 turbulence-crossing time. There is no significant dependence of the IMF on the initial density fluctuations.}
\label{fig_IMF_C}
\end{figure*}

\begin{table}[]
\caption{Parameters of the study on influences of relaxation time and refinement: label, relaxation level, relaxation time, and the maximum resolution level. The clouds initially have $1000~\Ms$ in mass and 0.75/0.084 pc radius in run A/C. The numbers in the label column denote the levels on which the relaxation run is performed, and `d' denotes the double relaxation time. }
\label{table_relax}
\centering
\begin{tabular}{l r r r}
\hline\hline
Label    & \makecell{relaxation\\level}  & \makecell{relaxation\\time (kyr) } &  $l_\mathrm{max}$ \vspace{.5mm}\\
\hline
A0      &  -  &  -     & 14   \\%& -- &\\
A1d    & 8  &  180 & 14     \\%& -- &\\
A2      & 9  &  90 & 14     \\%& -- &\\
A2d    & 9  &  180 & 14     \\% & -- &\\
C0      &  -  &  -     & 14   \\%& -- &\\
C1      & 8  &  3    & 14    \\%& -- &\\
C2      & 9  &  3    & 14     \\%& -- &\\
\hline
\end{tabular}
\end{table}

One conventional method to initialize a molecular cloud in simulations is to give a density profile, usually smooth, and introduce randomly seeded turbulence, 
in which case the cloud is artificially constructed such that the velocity and density fields are not self-consistent. 
We generate relaxed initial conditions by evolving the cloud with the hydrodynamic equations without considering self-gravity during a fraction of the system turbulence-crossing time. 
This relaxation generates local density fluctuations that create
a self-consistent correlation between density and velocity fields.  
Since it is important to verify that the results do not depend on the initial fluctuations and on this procedure to set up the initial conditions,
 we have explored the dependence on the initial conditions using virialized clouds of 0.75 and 0.084 pc radius (see parameters of run series A and C in Table \ref{table_relax}).

We performed simulations after three different relaxation times, 0, 0.3 or 0.6 $t_\mathrm{tc}$. 
To verify that the initial fluctuations are not determinant, we also varied the initial resolutions by using a configuration
with either $256^3$ (level 8) or $512^3$ (level 9). 
The relaxation was made on the uniform base grid, and the AMR refinement was activated only for the second phase and was allowed down to level 14. 
The resulting mass spectra for runs A are shown in Fig.~\ref{fig_IMF_A} and runs C in Fig.~\ref{fig_IMF_C}. 
There are no noticeable trends for either runs A or C with varied resolution of relaxation, except for runs A1d and A2d, which are relaxed during 0.6 $t_\mathrm{tc}$ and have probably lost too much kinetic energy and thus yield fewer low-mass stars.

%------------------GAUSSIAN FITS------------------------
\section{Gaussian fit to the IMF}\label{appen_g}
The IMFs in Figs. \ref{fig_SMF_density} and \ref{fig_SMF_turbulence} are fitted to a Gaussian distribution. 
The purpose of these fits is to provide a systematic way to estimate the peak position 
and width of the distribution. We note that while reasonable, these fits are not always 
very good. In particular, the fitted distribution tends to underestimate the number of objects around the peak.   
In Table \ref{table_gaussian}, the peak mass and the width of the Gaussian fit are listed. 
\begin{table*}[]
\caption{Gaussian fits: the peak mass ($\Ms$) followed by the width (dex) of the Gaussian in parentheses.}
\label{table_gaussian}
\centering
\begin{tabular}{l c c c c c c }
\hline\hline
Label   & $20~\Ms$  & $50~\Ms$   &  $100~\Ms$ &  $150~\Ms$ &  $200~\Ms$ &  $300~\Ms$ \\
\hline
A1    & 0.44(3.03) & 0.49(1.61) & 0.93(2.42) & 0.61(1.68) & 0.79(1.82) & 0.92(2.23) \\
A1+  & 0.39(1.42) & 0.30(1.63) & 0.36(1.84) & 0.71(2.19) & 0.48(1.87) & 0.65(1.86) \\
A1++  & 0.31(4.73) & 0.14(1.73) & 0.17(1.84) & 0.16 (1.92) & 0.18(2.23) & 0.12(2.52) \\      
\hline
B1    & 0.52(1.26) & 0.23(2.72) & 0.75(2.24) & 0.36(2.22) & 0.63(1.75) & 0.41(2.37) \\
B1+  &  0.18(1.29) & 0.16(1.43) & 0.25(1.58) & 0.30(1.53) & 0.31(1.57) & \\
B1++  &  0.21(0.86) & 0.15(1.31) & 0.13(1.47) & 0.13 (1.52) & 0.13(1.58) &\\
\hline
C1    & 0.06(1.77) & 0.08(1.89) & 0.12(1.64) & 0.11(1.78) & 0.14(1.83) &\\
C1+     &  0.04(1.21) & 0.07(1.55) & 0.10(1.55) & 0.15(1.55) & 0.12(1.57) & 0.15(1.75) \\
C1--     & 0.07(1.35) & 0.10(1.46) & 0.10(1.48) & 0.13(1.43) & 0.15(1.46) & 0.14(1.68) \\
C1-- --  & 0.12(2.03) & 0.19(1.38) & 0.14(1.56) & 0.16(1.65) & 0.20(1.64) & \\
C1t15   & 0.06(1.33) & 0.08(1.49) & 0.10(1.41) & 0.10(1.58) & 0.14(159) &\\
C1t05   & 0.09(1.05) & 0.13(1.09) &&&&\\
C1t03   & 0.16(1.14) & 0.17(1.31) & 0.21(1.34) & 0.17(1.53) &&\\
C1t01   & 5.22(0.24) & 0.29(1.27) & 0.19(1.23) & 0.17(1.19) &&\\
\hline
D1    & 0.12(1.36) & 0.11(1.72) & 0.10(1.58) & 0.09(1.76) & 0.11(1.75) &\\
\hline
\end{tabular}
\end{table*}

\bibliographystyle{aa}
\bibliography{lars}

\begin{thebibliography}{67}
\expandafter\ifx\csname natexlab\endcsname\relax\def\natexlab#1{#1}\fi

\bibitem[{{Alves de Oliveira} {et~al.}(2013){Alves de Oliveira}, {Moraux},
  {Bouvier}, {Duch{\^e}ne}, {Bouy}, {Maschberger}, \& {Hudelot}}]{alves2013}
{Alves de Oliveira}, C., {Moraux}, E., {Bouvier}, J., {et~al.} 2013, \aap, 549,
  A123

\bibitem[{{Ballesteros-Paredes} {et~al.}(2015){Ballesteros-Paredes},
  {Hartmann}, {P{\'e}rez-Goytia}, \& {Kuznetsova}}]{BallesterosParedes15}
{Ballesteros-Paredes}, J., {Hartmann}, L.~W., {P{\'e}rez-Goytia}, N., \&
  {Kuznetsova}, A. 2015, \mnras, 452, 566

\bibitem[{{Bastian} {et~al.}(2010){Bastian}, {Covey}, \& {Meyer}}]{Bastian10}
{Bastian}, N., {Covey}, K.~R., \& {Meyer}, M.~R. 2010, \araa, 48, 339

\bibitem[{{Bate}(2005)}]{Bate05b}
{Bate}, M.~R. 2005, \mnras, 363, 363

\bibitem[{{Bate}(2009{\natexlab{a}})}]{Bate09a}
{Bate}, M.~R. 2009{\natexlab{a}}, \mnras, 392, 590

\bibitem[{{Bate}(2009{\natexlab{b}})}]{Bate09b}
{Bate}, M.~R. 2009{\natexlab{b}}, \mnras, 392, 1363

\bibitem[{{Bate}(2012)}]{Bate12}
{Bate}, M.~R. 2012, \mnras, 419, 3115

\bibitem[{{Bate} \& {Bonnell}(2005)}]{Bate05a}
{Bate}, M.~R. \& {Bonnell}, I.~A. 2005, \mnras, 356, 1201

\bibitem[{{Bate} {et~al.}(2003){Bate}, {Bonnell}, \& {Bromm}}]{Bate03}
{Bate}, M.~R., {Bonnell}, I.~A., \& {Bromm}, V. 2003, \mnras, 339, 577

\bibitem[{{Bertelli Motta} {et~al.}(2016){Bertelli Motta}, {Clark}, {Glover},
  {Klessen}, \& {Pasquali}}]{bertelli2016}
{Bertelli Motta}, C., {Clark}, P.~C., {Glover}, S.~C.~O., {Klessen}, R.~S., \&
  {Pasquali}, A. 2016, \mnras, 462, 4171

\bibitem[{{Bleuler} \& {Teyssier}(2014)}]{Bleuler14}
{Bleuler}, A. \& {Teyssier}, R. 2014, \mnras, 445, 4015

\bibitem[{{Bonnell} {et~al.}(2001){Bonnell}, {Bate}, {Clarke}, \&
  {Pringle}}]{bonnell2001}
{Bonnell}, I.~A., {Bate}, M.~R., {Clarke}, C.~J., \& {Pringle}, J.~E. 2001,
  \mnras, 323, 785

\bibitem[{{Bonnell} {et~al.}(2003){Bonnell}, {Bate}, \& {Vine}}]{Bonnell03}
{Bonnell}, I.~A., {Bate}, M.~R., \& {Vine}, S.~G. 2003, \mnras, 343, 413

\bibitem[{{Bonnell} {et~al.}(2008){Bonnell}, {Clark}, \& {Bate}}]{Bonnell08}
{Bonnell}, I.~A., {Clark}, P., \& {Bate}, M.~R. 2008, \mnras, 389, 1556

\bibitem[{{Bonnell} {et~al.}(2011){Bonnell}, {Smith}, {Clark}, \&
  {Bate}}]{Bonnell11}
{Bonnell}, I.~A., {Smith}, R.~J., {Clark}, P.~C., \& {Bate}, M.~R. 2011,
  \mnras, 410, 2339

\bibitem[{{Bonnell} {et~al.}(2004){Bonnell}, {Vine}, \& {Bate}}]{bonnell2004}
{Bonnell}, I.~A., {Vine}, S.~G., \& {Bate}, M.~R. 2004, \mnras, 349, 735

\bibitem[{{Breslau} {et~al.}(2014){Breslau}, {Steinhausen}, {Vincke}, \&
  {Pfalzner}}]{breslau2014}
{Breslau}, A., {Steinhausen}, M., {Vincke}, K., \& {Pfalzner}, S. 2014, \aap,
  565, A130

\bibitem[{{Chabrier}(2003)}]{chabrier2003}
{Chabrier}, G. 2003, \pasp, 115, 763

\bibitem[{{Clark} {et~al.}(2008){Clark}, {Bonnell}, \& {Klessen}}]{Clark08}
{Clark}, P.~C., {Bonnell}, I.~A., \& {Klessen}, R.~S. 2008, \mnras, 386, 3

\bibitem[{{Clarke} \& {Pringle}(1993)}]{clarke1993}
{Clarke}, C.~J. \& {Pringle}, J.~E. 1993, \mnras, 261, 190

\bibitem[{{Commer{\c c}on} {et~al.}(2010){Commer{\c c}on}, {Hennebelle},
  {Audit}, {Chabrier}, \& {Teyssier}}]{commercon2010}
{Commer{\c c}on}, B., {Hennebelle}, P., {Audit}, E., {Chabrier}, G., \&
  {Teyssier}, R. 2010, \aap, 510, L3

\bibitem[{{Commer{\c c}on} {et~al.}(2011){Commer{\c c}on}, {Hennebelle}, \&
  {Henning}}]{commercon2011}
{Commer{\c c}on}, B., {Hennebelle}, P., \& {Henning}, T. 2011, \apjl, 742, L9

\bibitem[{{Federrath} {et~al.}(2010{\natexlab{a}}){Federrath}, {Banerjee},
  {Clark}, \& {Klessen}}]{federrath2010}
{Federrath}, C., {Banerjee}, R., {Clark}, P.~C., \& {Klessen}, R.~S.
  2010{\natexlab{a}}, \apj, 713, 269

\bibitem[{{Federrath} {et~al.}(2008){Federrath}, {Klessen}, \&
  {Schmidt}}]{Federrath08}
{Federrath}, C., {Klessen}, R.~S., \& {Schmidt}, W. 2008, \apjl, 688, L79

\bibitem[{{Federrath} {et~al.}(2010{\natexlab{b}}){Federrath}, {Roman-Duval},
  {Klessen}, {Schmidt}, \& {Mac Low}}]{Federrath10}
{Federrath}, C., {Roman-Duval}, J., {Klessen}, R.~S., {Schmidt}, W., \& {Mac
  Low}, M.-M. 2010{\natexlab{b}}, \aap, 512, A81

\bibitem[{{Fromang} {et~al.}(2006){Fromang}, {Hennebelle}, \&
  {Teyssier}}]{Fromang06}
{Fromang}, S., {Hennebelle}, P., \& {Teyssier}, R. 2006, \aap, 457, 371

\bibitem[{{Girichidis} {et~al.}(2011){Girichidis}, {Federrath}, {Banerjee}, \&
  {Klessen}}]{Girichidis11}
{Girichidis}, P., {Federrath}, C., {Banerjee}, R., \& {Klessen}, R.~S. 2011,
  \mnras, 413, 2741

\bibitem[{{Girichidis} {et~al.}(2014){Girichidis}, {Konstandin}, {Whitworth},
  \& {Klessen}}]{girichidis2014}
{Girichidis}, P., {Konstandin}, L., {Whitworth}, A.~P., \& {Klessen}, R.~S.
  2014, \apj, 781, 91

\bibitem[{{Hennebelle} \& {Chabrier}(2008)}]{HC08}
{Hennebelle}, P. \& {Chabrier}, G. 2008, \apj, 684, 395

\bibitem[{{Hennebelle} \& {Chabrier}(2009)}]{HC09}
{Hennebelle}, P. \& {Chabrier}, G. 2009, \apj, 702, 1428

\bibitem[{{Hennebelle} \& {Chabrier}(2013)}]{HC13}
{Hennebelle}, P. \& {Chabrier}, G. 2013, \apj, 770, 150

\bibitem[{{Hennebelle} {et~al.}(2016){Hennebelle}, {Commer{\c c}on},
  {Chabrier}, \& {Marchand}}]{hennebelle2016}
{Hennebelle}, P., {Commer{\c c}on}, B., {Chabrier}, G., \& {Marchand}, P. 2016,
  \apjl, 830, L8

\bibitem[{{Hennebelle} {et~al.}(2011){Hennebelle}, {Commer{\c c}on}, {Joos},
  {Klessen}, {Krumholz}, {Tan}, \& {Teyssier}}]{hennebelle2011}
{Hennebelle}, P., {Commer{\c c}on}, B., {Joos}, M., {et~al.} 2011, \aap, 528,
  A72

\bibitem[{{Hennebelle} \& {Falgarone}(2012)}]{HF12}
{Hennebelle}, P. \& {Falgarone}, E. 2012, \aapr, 20, 55

\bibitem[{{Hennebelle} \& {Teyssier}(2008)}]{hennebelle2008}
{Hennebelle}, P. \& {Teyssier}, R. 2008, \aap, 477, 25

\bibitem[{{Hillenbrand}(2004)}]{hillenbrand2004}
{Hillenbrand}, L.~A. 2004, in The Dense Interstellar Medium in Galaxies, ed.
  S.~{Pfalzner}, C.~{Kramer}, C.~{Staubmeier}, \& A.~{Heithausen}, Vol.~91, 601

\bibitem[{{Hopkins}(2012)}]{hopkins2012}
{Hopkins}, P.~F. 2012, \mnras, 423, 2037

\bibitem[{{Jappsen} {et~al.}(2005){Jappsen}, {Klessen}, {Larson}, {Li}, \& {Mac
  Low}}]{Jappsen05}
{Jappsen}, A.-K., {Klessen}, R.~S., {Larson}, R.~B., {Li}, Y., \& {Mac Low},
  M.-M. 2005, \aap, 435, 611

\bibitem[{{J{\'{\i}}lkov{\'a}} {et~al.}(2016){J{\'{\i}}lkov{\'a}}, {Hamers},
  {Hammer}, \& {Portegies Zwart}}]{jilkova2016}
{J{\'{\i}}lkov{\'a}}, L., {Hamers}, A.~S., {Hammer}, M., \& {Portegies Zwart},
  S. 2016, \mnras, 457, 4218

\bibitem[{{Kritsuk} {et~al.}(2007){Kritsuk}, {Norman}, {Padoan}, \&
  {Wagner}}]{Kritsuk07}
{Kritsuk}, A.~G., {Norman}, M.~L., {Padoan}, P., \& {Wagner}, R. 2007, \apj,
  665, 416

\bibitem[{{Kritsuk} {et~al.}(2011){Kritsuk}, {Norman}, \& {Wagner}}]{kritsuk11}
{Kritsuk}, A.~G., {Norman}, M.~L., \& {Wagner}, R. 2011, \apjl, 727, L20

\bibitem[{{Kroupa}(2001)}]{kroupa2001}
{Kroupa}, P. 2001, \mnras, 322, 231

\bibitem[{{Krumholz} {et~al.}(2011){Krumholz}, {Klein}, \&
  {McKee}}]{Krumholz11}
{Krumholz}, M.~R., {Klein}, R.~I., \& {McKee}, C.~F. 2011, \apj, 740, 74

\bibitem[{{Krumholz} {et~al.}(2004){Krumholz}, {McKee}, \&
  {Klein}}]{Krumholz04}
{Krumholz}, M.~R., {McKee}, C.~F., \& {Klein}, R.~I. 2004, \apj, 611, 399

\bibitem[{{Krumholz} {et~al.}(2007){Krumholz}, {Stone}, \&
  {Gardiner}}]{krumholz07}
{Krumholz}, M.~R., {Stone}, J.~M., \& {Gardiner}, T.~A. 2007, \apj, 671, 518

\bibitem[{{Lada} \& {Lada}(2003)}]{Lada03}
{Lada}, C.~J. \& {Lada}, E.~A. 2003, \araa, 41, 57

\bibitem[{{Liptai} {et~al.}(2017){Liptai}, {Price}, {Wurster}, \&
  {Bate}}]{liptai2017}
{Liptai}, D., {Price}, D.~J., {Wurster}, J., \& {Bate}, M.~R. 2017, \mnras,
  465, 105

\bibitem[{{Machida} {et~al.}(2008){Machida}, {Tomisaka}, {Matsumoto}, \&
  {Inutsuka}}]{machida2008}
{Machida}, M.~N., {Tomisaka}, K., {Matsumoto}, T., \& {Inutsuka}, S.-i. 2008,
  \apj, 677, 327

\bibitem[{{Maschberger} {et~al.}(2014){Maschberger}, {Bonnell}, {Clarke}, \&
  {Moraux}}]{Maschberger14}
{Maschberger}, T., {Bonnell}, I.~A., {Clarke}, C.~J., \& {Moraux}, E. 2014,
  \mnras, 439, 234

\bibitem[{{Maschberger} {et~al.}(2010){Maschberger}, {Clarke}, {Bonnell}, \&
  {Kroupa}}]{Maschberger10}
{Maschberger}, T., {Clarke}, C.~J., {Bonnell}, I.~A., \& {Kroupa}, P. 2010,
  \mnras, 404, 1061

\bibitem[{{Masson} {et~al.}(2016){Masson}, {Chabrier}, {Hennebelle}, {Vaytet},
  \& {Commer{\c c}on}}]{masson2016}
{Masson}, J., {Chabrier}, G., {Hennebelle}, P., {Vaytet}, N., \& {Commer{\c
  c}on}, B. 2016, \aap, 587, A32

\bibitem[{{Masunaga} \& {Inutsuka}(1999)}]{Masunaga99}
{Masunaga}, H. \& {Inutsuka}, S.-i. 1999, \apj, 510, 822

\bibitem[{{Moraux} {et~al.}(2007){Moraux}, {Bouvier}, {Stauffer}, {Barrado y
  Navascu{\'e}s}, \& {Cuillandre}}]{moraux2007}
{Moraux}, E., {Bouvier}, J., {Stauffer}, J.~R., {Barrado y Navascu{\'e}s}, D.,
  \& {Cuillandre}, J.-C. 2007, \aap, 471, 499

\bibitem[{{Myers} {et~al.}(2013){Myers}, {McKee}, {Cunningham}, {Klein}, \&
  {Krumholz}}]{myers2013}
{Myers}, A.~T., {McKee}, C.~F., {Cunningham}, A.~J., {Klein}, R.~I., \&
  {Krumholz}, M.~R. 2013, \apj, 766, 97

\bibitem[{{Offner} {et~al.}(2008){Offner}, {Klein}, \& {McKee}}]{Offner08}
{Offner}, S.~S.~R., {Klein}, R.~I., \& {McKee}, C.~F. 2008, \apj, 686, 1174

\bibitem[{{Offner} {et~al.}(2009){Offner}, {Klein}, {McKee}, \&
  {Krumholz}}]{Offner09}
{Offner}, S.~S.~R., {Klein}, R.~I., {McKee}, C.~F., \& {Krumholz}, M.~R. 2009,
  \apj, 703, 131

\bibitem[{{Padoan} {et~al.}(1997){Padoan}, {Nordlund}, \& {Jones}}]{padoan1997}
{Padoan}, P., {Nordlund}, A., \& {Jones}, B.~J.~T. 1997, \mnras, 288, 145

\bibitem[{{Peters} {et~al.}(2011){Peters}, {Banerjee}, {Klessen}, \& {Mac
  Low}}]{peters2011}
{Peters}, T., {Banerjee}, R., {Klessen}, R.~S., \& {Mac Low}, M.-M. 2011, \apj,
  729, 72

\bibitem[{{Salpeter}(1955)}]{Salpeter55}
{Salpeter}, E.~E. 1955, \apj, 121, 161

\bibitem[{{Schmidt} {et~al.}(2010){Schmidt}, {Kern}, {Federrath}, \&
  {Klessen}}]{schmidt2010}
{Schmidt}, W., {Kern}, S.~A.~W., {Federrath}, C., \& {Klessen}, R.~S. 2010,
  \aap, 516, A25

\bibitem[{{Shu}(1977)}]{shu77}
{Shu}, F.~H. 1977, \apj, 214, 488

\bibitem[{{Stacy} \& {Bromm}(2013)}]{stacy2013}
{Stacy}, A. \& {Bromm}, V. 2013, \mnras, 433, 1094

\bibitem[{{Stacy} {et~al.}(2016){Stacy}, {Bromm}, \& {Lee}}]{stacy2016}
{Stacy}, A., {Bromm}, V., \& {Lee}, A.~T. 2016, \mnras, 462, 1307

\bibitem[{{Stamatellos} {et~al.}(2012){Stamatellos}, {Whitworth}, \&
  {Hubber}}]{stamatellos2012}
{Stamatellos}, D., {Whitworth}, A.~P., \& {Hubber}, D.~A. 2012, \mnras, 427,
  1182

\bibitem[{{Teyssier}(2002)}]{Teyssier02}
{Teyssier}, R. 2002, \aap, 385, 337

\bibitem[{{Vázquez-Semadeni}(1994)}]{vazquez94}
{Vázquez-Semadeni}, E. 1994, \apj, 423, 681

\bibitem[{{Wurster} {et~al.}(2017){Wurster}, {Price}, \& {Bate}}]{wurster2017}
{Wurster}, J., {Price}, D.~J., \& {Bate}, M.~R. 2017, \mnras, 466, 1788

\end{thebibliography}

\end{document}